%% file: lht.tex
\documentclass[11pt,a4paper]{article} 
\usepackage{jheppub}

\usepackage{graphicx}
\usepackage{amsmath,amssymb,amsfonts}
\usepackage{multirow}
\usepackage{hyperref,enumitem}
\usepackage{bbm}
\usepackage{xspace}
\usepackage{verbatim}
\usepackage{bm}
\usepackage{color}
\usepackage{cancel}
\usepackage{mathtools}
\usepackage[T1]{fontenc} 
\usepackage{slashed}
\usepackage{soul}
\usepackage{booktabs}
\usepackage{enumitem}
\usepackage{booktabs}
\usepackage{units}
\usepackage{subfigure}

\usepackage{makecell, tabularx}
\setcellgapes{3pt}

\newcommand{\lam}{\lambda}
\newcommand{\Lam}{\Lambda}
\newcommand{\eps}{\epsilon}

\newcommand{\whizard}{\texttt{WHIZARD}\xspace}

\newcommand{\Checkmate}{\texttt{CheckMATE}\xspace}

\definecolor{weddingpurple}{rgb}{0.6,0, 0.6}

\newcommand{\mgfive}{\texttt{MG5\_aMC@NLO}}
\newcommand{\checkmate}{\texttt{CheckMATE}}

\newcommand{\AddrHH}{
II. Institut f\"ur Theoretische Physik,
Universität Hamburg,
Luruper Chaussee 149,
22761 Hamburg, Germany
 }
\newcommand{\AddrDESY}{
DESY, 
Notkestra{\ss}e 85, 
D-22607 Hamburg, Germany
 }
\begin{document}

\title{The fate of the Littlest Higgs Model with $T$-parity under 13 TeV
  LHC Data} 

\author[a]{Daniel Dercks,} 
\author[a]{Gudrid Moortgat-Pick,}
\author[b]{Jürgen Reuter,} 
\author[b]{and So Young Shim}

\affiliation[a]{\AddrHH}
\affiliation[b]{\AddrDESY}

\emailAdd{daniel.dercks@desy.de}
\emailAdd{gudrid.moortgat-pick@desy.de}
\emailAdd{juergen.reuter@desy.de}
\emailAdd{soyoung.shim@desy.de}

\abstract{
  We exploit all LHC available Run 2 data at center-of-mass
  energies of 8 and 13 TeV for searches for physics beyond the
  Standard Model. We scrutinize the allowed parameter space of Little
  Higgs models with the concrete symmetry of $T$-parity by providing
  comprehensive analyses of all relevant production channels of
  heavy vectors, top partners, heavy quarks and heavy leptons
  and all phenomenologically relevant decay channels. Constraints on
  the model will be derived from the signatures of jets and missing
  energy or leptons and missing energy. Besides the symmetric case, we
  also study the case of T-parity violation. Furthermore, we give an
  extrapolation to the LHC high-luminosity phase at 14 TeV as well.}


\arxivnumber{1801.XXXXX}

\begin{flushright}
  \normalsize{} DESY 17--192
  \vspace{+1em}
\end{flushright}

\maketitle

\input{sections/01_introduction}
\input{sections/02_littlehiggstheory}
\input{sections/03_ewpohiggsconstraints}
\input{sections/04_toolsetup}
\input{sections/05_checkmateresults}
\input{sections/06_summary}

\section*{Acknowledgments}

The authors like to thank Marco Tonini for helping in the validation
of our results. SYS would like to thank Pedro Schwaller for
clarifications regarding the $T$-parity violating decays. JRR likes to
thank Maxim Perelstein for helpful discussions. DD and JRR acknowledge
funding and support from the Collaborative Research Unit (SFB) 676 of
the Deutsche Forschungsgemeinschaft (DFG), projects B1 and B11. 

\medskip

\appendix
\input{sections/07_additionalplots}

\bibliography{lht}

\end{document}

%% file: sections/01_introduction.tex
\section{Introduction}

The main legacy of the 7/8 and 13 TeV runs of the Large Hadron
Collider (LHC) is the discovery of a 125 GeV Higgs
boson~\cite{Aad:2012tfa,Chatrchyan:2012xdj} as well as the absence of
any signals for other new particles. This is in accordance with the
measurements of electroweak precision observables (EWPO) which agree
very well with a Standard Model (SM) containing a light Higgs boson and no
further degrees of freedom in the range up to a TeV. Besides the EWPO from
the pre-LHC era and flavor physics observables, both direct searches
and the ever more precise measurements of the properties of the Higgs
boson (as well as the top quark and weak gauge bosons) are the tools
to search for physics beyond the Standard Model (BSM) at the
LHC. These are used to constrain any type of BSM model.

In this paper we study the Littlest Higgs Model with $T$-parity
(LHT). This is an attractive representative of Little Higgs
models~\cite{ArkaniHamed:2001nc,ArkaniHamed:2002pa} since  fine tuning
problems in the Higgs potential can be avoided via a discrete global
$Z_2$ symmetry. Little Higgs models in general regard a naturally light
Higgs boson as a pseudo--Nambu-Goldstone boson (pNGB) arising from a
(new) global symmetry at high scale, see
e.g.\ Ref.~\cite{Georgi:1975tz,Kaplan:1983fs}. However, such a mechanism
would require new strong interactions to tie the constituents of the
Higgs boson together, which unavoidably would show up in electroweak
precision observables. In order to avoid such strong
constraints from EWPO, the mechanism of so-called collective symmetry
breaking has been applied, i.e. interweaving several
global symmetries which all have to broken in order to give mass to the
pNGBs charged under them. 
This means that the Higgs mass achieves only a logarithmic
sensitivity to the cutoff scale at one-loop , while a quadratic sensitivity only
arises at the two-loop level, thereby shifting the
strongly-interacting UV completion scale from the multi-TeV to
the multi-10 TeV-region. Note, however, that Little Higgs model are
effective field theories (with new degrees of freedom beyond the SM
like heavy vectors, scalars and quarks) that not necessarily have a
direct strongly coupled UV completion, but could also have
weakly-coupled sectors at the next scale~\cite{Kaplan:2004cr}. 

In this paper we consider the LHT model just as sucg an effective
(low-energy) field theory consisting of the SM degrees of freedom
augmented by ($T$-odd) heavy vector bosons, heavy quarks (and leptons)
as well as additional heavy pNGBs (which turn out to be irrelevant for
the phenomenology of that model). All of these particles just have the
SM gauge interactions as well as generalizations of the SM Yukawa
couplings, which reflect the implementation of both the Little Higgs
collective symmetries as well as $T$-parity. We consider all
phenomenologically relevant production mechanisms for the heavy new
particles, including all relevant decays in order to compare the
predictions within the LHT model with the LHC 13 TeV data from Run
2. In addition, we reproduce the constraints from 
the EWPO. For completeness, we review the
status from the 8 TeV Run 1 data. Because of the possibility
of $T$-parity breaking in a strongly coupled UV completion of the LHT,
as well as tensions from dark matter (DM) constraints, we also take
signatures and limits from a scenario with $T$-parity breaking into
account which is different than the
Littlest Higgs model without $T$-parity. We also give prospects for
the upcoming high-luminosity runs at the LHC at 14 TeV.

The outline of the paper is as follows: in order to make the paper
self-contained, in Sec.~\ref{sec:littlehiggstheory} we briefly
summarize the model-building setup of the Littlest Higgs model (with
$T$-parity) needed to understand the phenomenological analyses later
on. In Sec.~\ref{sec:EWPO} we review the existing limits from EWPO on
the LHT model. In the next section, Sec.~\ref{sec:toolsetup}, we
discuss the tool chain for generating events and recasting the LHC
analyses. We then collect the relevant collider topologies along with
cross sections and branching ratios for different regions of parameter
space in Sec.~\ref{sec:topologies}. Our main collider results are
collected in Sec.~\ref{sec:results}, and compared to the sensitivity
from electroweak precision data in Sec.~\ref{sec:comp}. Finally, we
give a summary and outlook in Sec.~\ref{sec:summary}. 

%% file: sections/02_littlehiggstheory.tex
\section{Little Higgs Models with $T$-Parity}
\label{sec:littlehiggstheory}

The Littlest Higgs model~\cite{ArkaniHamed:2002qy} is based on a
non-linear sigma model with a single field $\Sigma$ parameterizing a
$SU(5)/SO(5)$ symmetry breaking structure.\footnote{For different
  implementations of Little Higgs models in terms of product group and
  simple group models and a way to distinguish them,
  cf. e.g.~\cite{Kilian:2004pp,Perelstein:2005ka,Kilian:2006eh}.}
The vacuum expectation value (vev) causing the breaking from $SU(5)$ to
$SO(5)$, $\Sigma_0$, can be cast into the form of the $5\times5$ matrix 
\begin{equation}
  \label{eq:vev}
  \Sigma_0 =  ~ \left( \begin{array}{ccc}
    &  & {\mathbf{1}}_{2 \times 2} \\
    & 1 &  \\
    {\mathbf{1}}_{2 \times 2} &  & 
  \end{array}\right).
\end{equation}

The gauge group of the Littlest Higgs is $G_1\times G_2 ~=~
(SU(2)_1\times U(1)_1) \times (SU(2)_2\times U(1)_2)$ embedded in
$SU(5)$ as a subgroup such that the vev in Eq.~(\ref{eq:vev}) above
breaks it down into the diagonal subgroup $SU(2)_L\times U(1)_Y$ which
is identified with the SM electroweak group. 
The kinetic term for the non-linear sigma model field is
\begin{equation}
  \label{eq:kinterm}
  \mathcal{L}_{kin}~=~\frac{f^2}{4} {\rm{Tr}}|D_\mu \Sigma|^2
\end{equation}
with 
\begin{equation}
  \Sigma ~=~ e^{i\Pi/f}~\Sigma_0~e^{i\Pi^T/f}~=~ e^{2i\Pi/f}~\Sigma_0 \;
\end{equation}
where $f$ is the Nambu-Goldstone-Boson (NGB) decay constant of the model. At this scale the
symmetry breakings $SU(5) \to SO(5)$ and $G_1 \times G_2 \to SU(2)_L
\times U(1)_Y$ take place. The covariant derivative
in Eq.~(\ref{eq:kinterm}) is given by
  \begin{equation}
   D_\mu \Sigma ~=~ \partial_\mu \Sigma - i \sum_j [ g_j W_j^a(Q_j^a
     \Sigma + \Sigma Q_j^{a^T}) + g_j'B_j(Y_j\Sigma + \Sigma Y_j) ] 
  \end{equation} 
with the generators
\begin{align}
  Q_1^a =~ \frac{1}{2}\left( \begin{array}{cc}
    \sigma^a  &  {\mathbf{0}}_{2 \times 3} \\
          {\mathbf{0}}_{3 \times 2} &  {\mathbf{0}}_{3 \times 3} 
  \end{array}\right), \quad & \quad
  Q_2^a =~ \frac{1}{2}\left( \begin{array}{cc}
    {\mathbf{0}}_{3 \times 3} & {\mathbf{0}}_{2 \times 3} \\
    {\mathbf{0}}_{3 \times 2} & - (\sigma^a)^*
  \end{array}\right) \\
  Y_1 =~ \frac{1}{10} \,\text{diag} \left( -3, -3, 2, 2, 2 \right), \quad & \quad  
  Y_2 =~ \frac{1}{10} \,\text{diag} \left( -2, -2, -2, 3, 3 \right)
  \quad, 
\end{align}
where $\sigma^a$ are Pauli matrices. The $SU(5)\to SO(5)$ symmetry
breaking generates a total of 14 NGBs $\Pi^a$ which decompose under
the unbroken EW group $SU(2)_L \times U(1)_Y$ as $\mathbf{1_0\oplus
3_0\oplus 2_{\pm \frac{1}{2}}\oplus 3_{\pm1}}$. Four of these NGBs
are eaten by the extra gauge bosons,  $Z_H$, $W_H$ and $A_H$, which
get masses of the order $f$. The remaining ten physical (p)NGBs
decompose into the complex Higgs doublet and a hypercharge one complex
triplet. The latter is phenomenologically irrelevant as the production 
cross section for these particles is negligibly small, cf. Ref.~\cite{Reuter:2013iya}.

Like many other BSM models, the Littlest Higgs model suffers from
constraints by electroweak precision observables, particularly as the
heavy hypercharge boson, $A_H$, has an accidentally small prefactor,
cf. the right hand side of Eq.~(\ref{eq:heavygaugebosonmass}). To
alleviate these constraints, a discrete symmetry, TeV- or short
$T$-parity has been added~\cite{Cheng:2003ju,Cheng:2004yc}, which
phenomenologically plays a similar role as $R$-parity in supersymmetry
(SUSY). $T$-parity is an involutary automorphism that exchanges the
sets of the two different gauge algebras $G_1$ and $G_2$, or
alternatively, their gauge bosons:
\begin{equation}
  W_{1\mu}^a ~ \xleftrightarrow{T} ~W_{2\mu}^a, \qquad\qquad
  B_{1\mu}~ \xleftrightarrow{T} ~B_{2\mu}.
\end{equation}
This fixes the gauge coupling constants of the two different $SU_{1,2}(2)$
and $U_{1,2}(1)$ to be equal:
 \begin{align}
 g_1&=~g_2~=~\sqrt{2}g, \\
 g_1'&=~g_2'~=~\sqrt{2}g'.
 \end{align}
The mass eigenstates are then just the (normalized) sum and difference
of the two gauge fields, respectively, with mixing angles of
${\pi}/{4}$. This results in the mass terms of the heavy gauge bosons
\begin{subequations}
  \label{eq:heavygaugebosonmass}
  \begin{align} 
    m_{W_H}=m_{Z_H} &= g f, 
    \label{eq:heavygaugebosonmass1}\\
    m_{A_H} &= \frac{g'f}{\sqrt{5}}.
    \label{eq:heavygaugebosonmass2}
  \end{align}
\end{subequations}

In order to implement collective symmetry breaking in the fermion
fields, a partner state to the third generation quark doublet has to
be introduced, forming an incomplete $SU(5)$ multiplet $\Psi$ and its
$T$-parity partner $\Psi'$
\begin{equation}
  \Psi = \left( \begin{array}{c}
  i b_L \\
  -it_{1L} \\
  t_{2L} \\
  \mathbf{0}_{2\times1}
  \end{array}\right) 
  = \left( \begin{array}{c}
  q_L \\
  t_{2L} \\
  \mathbf{0}_{2\times1}
  \end{array}\right), \qquad
  \Psi' = \left( \begin{array}{c}
  \mathbf{0}_{2\times1}\\
  t'_{2L} \\
  i b'_L \\
  -i t'_{1L} 
  \end{array}\right) 
  = \left( \begin{array}{c}
  \mathbf{0}_{2\times1}\\
  t'_{2L} \\
  q'_L 
  \end{array}\right) \;,
\end{equation}
which are related via
\begin{align}
 \Psi \xleftrightarrow{T} - \Sigma_0 \Psi'
\end{align}
Here, $q_L$ denotes the quark doublet of the SM following the
conventions in~\cite{ArkaniHamed:2002qy}, while
$q_L'$ and $t_2'$ are the $T$-parity partner fermions needed to
reconcile both $T$-parity and the collective symmetry breaking
mechanism. The $T$-parity invariant Lagrangian then reads as
\begin{equation}
 \mathcal{L}_Y \, \supset \; \frac{\lam_1 f}{2 \sqrt{2}}\eps_{ijk}
 \eps_{xy}  (\overline{\Psi}_i \Sigma_{jx} \Sigma_{ky} - (\overline{\Psi'})_i
 \tilde{\Sigma}_{jx} \tilde{\Sigma}_{ky} )t_{1R}+ \lam_2 f
 (\overline{t}_{2L}t_{2R} + \overline{t'}_{2L} t_{2R} ) + h.c., 
\end{equation}
where $\lambda_{1,2}$ denote the top-quark Yukawa couplings, respectively.
The $T$-parity eigenstates are now the (normalized) differences (even
states) and sums (odd states) of the primed and unprimed fermion
fields $t_+ = (t_{1L,+},t_{R})$, $t_ = (t_{1L,-},t_{1R,-})$, $T^- =
(t_{2L,-},t_{2R,-})$ and $T'_+ = (t_{2L,+},t_{2R,+})$. Diagonalizing
the left-handed $T$-even fermions yields the (SM) top quark and the
heavy $T$-even top quark, $T^+$. The $t_-$ gets a mass with the help
of the so-called mirror fermions, cf. below for the first and second
generation fermions, while the masses for the SM top quark and the
other top partners are given by, 
\begin{align}
  m_{t_{SM}} &= m_{t+} =\; \frac{\lam_2 R}{\sqrt{1+ R^2}}v,
  \label{eq:topmass} \\ 
  m_{T^-} &= \lam_2 f =\; \frac{m_{t_+}}{v}\frac{f\sqrt{1+R^2}}{R}
   \\ 
  m_{T^+} &=\; \frac{m_{t_+}}{v}\frac{f (1+R^2)}{R} =\; m_{T-}\sqrt{1+
    R^2} \quad .  
  \label{eq:heavytopmass} 
\end{align}
$R$ is defined as the ratio between the Yukawa coefficients of the two 
different possible terms, $R=\lam_1/\lam_2$ and is one of the
parameters used for investigating the parameter space in this paper.

Up-type quarks for the first and second generations have a similar
Lagrangian than the top quark except for the vector-like quark, which
is not present as there is no need to cancel the contribution from
light quarks to Higgs self energies:
\begin{align}
 \mathcal{L}_{Y} \supset \frac{i\lam_d
   f}{2\sqrt{2}}\eps_{ij}\eps_{xyz} ( \overline{\Psi'}_x\Sigma_{jy}
 \Sigma_{jz} X - (\overline{\Psi} \Sigma_0)_x
 \tilde{\Sigma}_{iy}\tilde{\Sigma_{jz}}\tilde{X}) d_R 
\end{align}
The $SU(2)_{1,2}$ singlet $X$ with $U(1)_{1,2}$ charges 
$(Y_1, Y_2)=(1/10, -1/10)$ renders the term gauge invariant. There
are two different $X$ embeddings as $(3,3)$ component into the NGB
multiplet, namely $X=(\Sigma_{33})^{-1/4}$ [\emph{Case A}] and
$X=(\Sigma_{33})^{1/4}$ [\emph{Case B}]. These cases do not differ
in the context of BSM collider phenomenology which is why we choose
Case A in this study. Differences only arise in the discussion of
constraints from the Higgs sector and electroweak precision 
observables and more details can be found in Ref.~\cite{Reuter:2013iya}.

To give rise to mass terms for the $T$-odd fermions without
introducing any anomalies, another $SO(5)$ multiplet $\Psi_c$ is
introduced as 
\begin{align}
   \Psi_c = \left( i d_c, \, -iu_{c}, \, \chi_{c}, \, i \tilde{d}_c, \,
  -i \tilde{u}_c \right)^T = \left( q_c, \, \chi_{c}, \, \tilde{q}_c
  \right)^T\,, \qquad \Psi_c \xleftrightarrow{T} - \Psi_c \;.
\end{align}
The $q_c$ fields are called mirror fermion. 

The $T$-parity invariant Lagrangian for the light fermions is
\begin{align}
 \mathcal{L}_\kappa = 
 -\kappa f (\overline{\Psi'}\xi\Psi_c +\overline{\Psi}\Sigma_0 \Omega
 \xi^\dagger\Omega\Psi_c ) + h.c. . 
\end{align}
This Lagrangian not only adds the $T$-odd mass terms but also 
imposes new interactions between Higgs boson and up-type partners.
\begin{align}
 \mathcal{L}_\kappa \supset &-\sqrt{2}\kappa f (
 \overline{d}_{L-}\tilde{d}_{c} + \frac{1+c_\xi}{2} \overline{u}_{L-}
 \tilde{u}_c -\frac{s_\xi}{\sqrt{2}}\overline{u}_{L-} \chi_c
 -\frac{1-c_\xi}{2}\overline{u}_{L-} u_c ) 
 + h.c. + \cdots
\end{align}
where $c_\xi = \cos((v+h)/\sqrt{2}f)$, $s_\xi=\sin((v+h)/\sqrt{2}f)$. 

The parameter $\kappa$ characterizing the coupling between the Higgs
and the $T$-odd fermions is another degree of freedom in the model
parameter space we investigated. We will distinguish between 
$\kappa_q$ for the light quarks and $\kappa_l$ for the leptons.

The mass spectrum for heavy $T$-odd fermions is given (at order
$\mathcal{O}(v^2/f^2) $) by
\begin{align}
 m_{u,-}    &=\; \sqrt{2}\,\kappa_q f
                 \Big(1-\frac{1}{8}\frac{v^2}{f^2}\Big), \\ 
 ~~m_{d,-}  &=\; \sqrt{2}\,\kappa_q f \label{eq:qHmasses}  \\
 m_{\ell,-} &=\; \sqrt{2}\,\kappa_l f
\end{align}

% \dacomment{- heavy lepton masses, differentiation between kappaq and kappal}

\subsection{$T$-parity Violation}
\label{sec:tpv}

For the phenomenology of the LHT model, we will also consider
$T$-parity violation. There are two reasons for that: first, in the
context of strongly interacting UV completions $T$-parity violation
can naturally occur via an anomalous Wess-Zumino-Witten
term,~\cite{Hill:2007zv,Hill:2007nz}, secondly, there is a certain
tension for the case that the lightest $T$-odd particle, the heavy
photon $A_H$ is absolutely stable from relic density calculations and
direct detection dark matter
experiments~\cite{Hubisz:2004ft,Wang:2013yba}. In order to avoid any 
constraints from dark matter bounds, one can assume that the $A_H$
only has a microscopic lifetime and that dark matter instead is made
up of an axion-like particle in the strongly interacting UV completion
of the Little Higgs model.

As has been studied in \cite{Hill:2007zv,Freitas:2008mq}, $T$-parity
violation generates decays the heavy photon partner $A_H$ into the
electroweak gauge bosons $WW$ and $ZZ$ similar to the decay of the
pion into two photons. Above the kinematic threshold for these $A_H$
decays, the partial width is given by:
\begin{align} \label{eq:TPVwidthzz}
  \Gamma(A_H \rightarrow ZZ) &= \left(
  \frac{Ng'}{80\sqrt{3}\pi^3} \right)^2\frac{M_{A_H}^3
    m_Z^2}{f^4}
  \left( 1-\frac{4 m_Z^2}{M_{A_H}^2} \right)^{\frac{5}{2}},  \\
  \label{eq:TPVwidthww} 
  \Gamma(A_H \rightarrow W^+ W^-)& = \left(
  \frac{Ng'}{40\sqrt{3}\pi^3} \right)^2 \frac{M_{A_H}^3
    m_W^2}{f^4}%\nn\\ 
  \left( 1-\frac{4 m_W^2}{M_{A_H}^2} \right)^{\frac{5}{2}} . 
\end{align}
Here, the integer $N$ depends on the UV completion of the theory. 
As we are only interested in branching ratios, shown in a later section,
 the precise choice of this number does not matter for our analysis.

If the mass of $A_H$ is below the $WW$ and $ZZ$ thresholds, it will
decay into the SM fermions via $WW$- and $ZZ$-induced triangle loops
leading to the partial widths:
\begin{align}
 \Gamma(A_H \rightarrow ff) &= 
 \left( \frac{N_{C,f}M_{A_H}}{48\pi}\right)  \left[
 c_-^2\left(1-\frac{4m_f^2}{M_{A_H}^2}\right) +
 c_+^2\left(1+\frac{2m_f^2}{M_{A_H}^2}\right) \right]
 \left( 1-\frac{4m_f^2}{M_{A_H}^2} \right)^{\frac{1}{2}} ,
 &\label{eq:TPVwidthff} 
\end{align}
with $c_\pm := c_R \pm c_L$, and $c_L$ and $c_R$ being the
left- and right-handed fermion couplings, shown in
Table~\ref{tab:TPVcoeffi}. $N_{C,f}$ is the number of colors of the
final state fermions. For the range $f~\sim ~\unit[1-10]{TeV}$ and $N = \mathcal{O}(1)$, the
total $A_H$ width $\Gamma_{A_H}$ ranges between \unit[0.01-1]{eV} which
corresponds to a lifetime of order \unit[10$^{-17}$]{s}. This excludes
$A_H$ from being a viable dark matter candidate. On the other hand, it
leads to a mean free path of approximately \unit[10]{nm}, resulting in
nearly prompt decays which do not produce observable displaced vertices in the LHC detectors.

\begin{table*}
\renewcommand{\tabularxcolumn}[1]{>{\centering\arraybackslash}p{#1}}
\footnotesize
\begin{tabularx}{\textwidth}{lXX}
\toprule
% Particles & $c_{L,\eps}^f$ & $c_{R,\eps}^f$ \\
Particles & $c_{L}^f$ & $c_{R}^f$ \\
\midrule
$A_H e^+ e^-$  & $\frac{9 \hat{N}}{160\pi^2} \frac{v^2}{f^2} g^4 g' (4
+ (c_w^{-2}-2t_w^2)^2)$ & $-\frac{9\hat{N}}{40\pi^2}\frac{v^2}{f^2}
g'^5$  \\[4pt] 
$A_H \overline{\nu}\nu$ & $\frac{9 \hat{N}}{160\pi^2} \frac{v^2}{f^2}
g^4 g' (4 + c_w^{-4})$ & 0  \\[4pt] 
$A_H \overline{u}_a u_b$  & $-\frac{\hat{N}}{160\pi^2} \frac{v^2}{f^2}
g^4 g' (36 + (3c_w^{-2}-4t_w^2)^2)\delta_{ab}$ &
$-\frac{\hat{N}}{10\pi^2}\frac{v^2}{f^2} g'^5 \delta_{ab}$  \\[4pt]  
$A_H \overline{d}_a d_b$  & $-\frac{\hat{N}}{160\pi^2} \frac{v^2}{f^2}
g^4 g' (36 + (3c_w^{-2}-2t_w^2)^2)\delta_{ab}$ &
$-\frac{\hat{N}}{40\pi^2}\frac{v^2}{f^2} g'^5 \delta_{ab}$  \\[4pt]
\midrule
\bottomrule
\end{tabularx}
\caption{Coefficients for the $A_H$ TPV decays, cf. Eq.~(\ref{eq:TPVwidthff}). The indices $a, b$ refer to the color of the respective quarks and  we use $\hat N = N/48 \pi^2, c_W = \cos \theta_W, t_W = \tan \theta_W$.}
\label{tab:TPVcoeffi}
\end{table*}

\subsection{Naturalness and Fine Tuning}
\label{sec:naturalness}

Together with the model setup, we discuss in this section the
definition of fine tuning, that is sometimes used as a guideline for
the naturalness of a model or of certain regions of parameter space. 
The naturalness is generally tied to the radiative corrections to the
scalar potential in quantum field theories. In order for a model to be
considered natural, those corrections should be of the same order as
the scalar mass term from the mechanism that originally created that
mass term (the explicit breaking of the global symmetries in Little
Higgs models). A fine-tuning measure usually compares the size of the
radiative corrections to this bare mass term. In the absence of a
special cancellation mechanism, this measure depends quadratically on
the typical scale of these corrections; cancellation by means of a
symmetry turns this into a logarithmic dependence, or even zero if the
symmetry is mighty enough like exact supersymmetry or conformal
symmetries. 

In Little Higgs models, the cancellation comes from SM partner
particles of like statistics by means of nonlinearly realized global
symmetries. The most severe SM radiative corrections from the top
quark are cancelled by the $T$-odd and even top partners, $T^\pm$,
followed by the cancellations of the EW gauge bosons due to the heavy
new gauge bosons, $A_H$, $Z_H$, and $W_H$. In this paper, we adopt the
fine-tuning measure defined in~\cite{ArkaniHamed:2002qy}, which only
accounts for the top partners, and neglects the contributions from the
gauge boson partners as well as from the heavy pNGBs and the light
fermion partners. The fine tuning is then defined as the ratio of the
experimentally measured Higgs mass squared and the absolute value of
the radiative corrections from the top partners to the Higgs quadratic
operator:
\begin{align}
  \label{eq:finetuning}
  \Delta~=~\frac{\mu_{exp}^2}{|\delta \mu^2|} , \qquad
  \delta \mu^2~=~ -\frac{3 \lam_t M_T^2}{8\pi^2}\log\frac{\Lam^2}{M_T^2}.
\end{align}
Here $\Lam=4\pi f$ is the cut-off scale of the LHT model, i.e. the
equivalent to $\Lambda_{\text{QCD}}$ in a strongly-interacting
embedding of the LHT, $\lam_t$ is the SM top Yukawa coupling and $M_T$
is a generic mass scale of the top partner sector. Note that this
definition of the fine-tuning measure leads to the fact that smaller
values of that measure (provided in per cent in general) constitute a
higher amount of fine tuning, hence a more finely tuned point of
parameter space. While the LHC Run 1 datasets at 7 and 8 TeV together
with electroweak precision observables still allowed parameter space
with $\mathcal{O}(1\%)$~\cite{Reuter:2013iya}, we will see in this paper that
the fine tuning including LHC Run 2 data is now everywhere around one
per cent or even in the sub-per cent regime. This is still comparable
with or better than the amount of fine tuning in generic parameter
regions of the minimal supersymmetric SM (MSSM), and it is generically
(much) better than the fine tuning for Composite Higgs models.

%% file: sections/03_ewpohiggsconstraints.tex
\section{Electroweak Precision Constraints}
\label{sec:EWPO}

Even before the start of data taking at the LHC, Little Higgs models
were already grossly constrained by comparing their predictions to
precise measurements in the electroweak sector, the so-called
electroweak precision observables
(EWPO)~\cite{Csaki:2002qg,Hewett:2002px,Han:2003wu,Kilian:2003xt}. Additional 
constraints come from flavor data (in the $K$, $D$ and $B$ sector), as
well as for the models with $T$ parity and stable massive particles
from dark matter searches. We will not discuss the first point here as
this has been studied elsewhere~\cite{Blanke:2006eb,Blanke:2015wba},
and the second point has been addressed in the last section.

EWPO mainly contain a list of measurements from $e^+e^-$ colliders
like LEP1, LEP2, SLC, and TRISTAN, and a few selected measurements
from hadron colliders where the precision has superseded that from
lepton colliders, like the $W$ mass, or was only possible there, like the
Higgs mass and couplings. In
Refs.~\cite{Reuter:2012sd,Reuter:2013iya,Tonini:2014dza}, both the
EWPO as well as the latest Higgs data have been scrutinized in order
to give the then best constraints on the parameter space of the
LHT model.

We will not repeat the complete table of the EWPO fit of the LHT model
from~\cite{Reuter:2013iya} here, but just remind that the two main
observables with the highest pull in the fit giving the highest
constraint are the total hadronic cross section at the $Z$ pole as
well as the left-right asymmetry on the $b$ quarks,
$A^{(b)}_{LR}$. Higgs observables in general do not give any further
constraints beyond that as EWPO already drive the Little Higgs scale
$f$ in a region where the deviations of the Higgs couplings are well
within the LHC experimental uncertainties. The only exception to this
statement comes from the case when the decay $H \to A_HA_H$ is
possible which is ruled out by the LHC limits on Higgs invisible
branching ratios and excludes $m_{A_H} < \unit[62.5]{GeV}$, i.e.\ $f
< \unit[480]{GeV}$ \cite{Reuter:2013iya}.

The first EWPO constraints that have been applied to Little Higgs
models came from oblique corrections, the so-called Peskin-Takeuchi
$\Delta S$, $\Delta T$ and $\Delta U$
parameters~\cite{Peskin:1990zt,Peskin:1991sw}. These 
parameterize corrections to the self energies of EW gauge bosons, that
are measured in two-(and four-) fermion processes at lepton
colliders. $T$-parity was specifically introduced to minimize the
contributions from Little Higgs heavy particles to the oblique
parameters as far as possible, as no $T$-odd particle can contribute
to them at tree level. However, at loop-level there are contributions
from $T$-odd heavy quarks, the $T$-even top quark, the mirror fermions
and the heavy gauge bosons. These have been calculated
in~\cite{Hubisz:2005tx,Berger:2012ec}.

One interesting feature derived in~\cite{Reuter:2012sd,Reuter:2013iya}
from the contribution of the heavy top partners to the $\Delta T$ parameter, 
is the exclusion limit from EWPO as a function of the parameter $R$,
the ratio of the two different Yukawa couplings $\lambda_1$ and
$\lambda_2$ in the top sector. There is an accidental cancellation to
the EWPO in terms of $R$ for the value of $R$=1. This gives an only
relatively weak exclusion limit for $f \gtrsim 405$ GeV at 95\% confidence
level from EWPO only. For $R \ll 1$ this bound goes up to roughly
750 GeV while for large $R \sim 3$ the bound from EWPO goes up to 1.3
TeV. 

For our discussion in this paper and the motivation into which regions
of parameter space to look at, even more relevant are the
contributions from the mirror fermions: 
\begin{equation}
  \label{eq:delta_t_mirror}
  \Delta T_{q_H,\ell_H} = - \sum_{q_H,\ell_H}
  \frac{\kappa_{q,\ell}^2}{192\pi^2 \alpha_w} \frac{v^2}{f^2} \quad . 
\end{equation}
These expressions come from box diagrams contributing to
four-fermion operators with heavy quark and lepton mirror fermions
running in the loop:
\begin{equation}
  \label{eq:4ferm-op}
  \mathcal{O}_{4-\text{ferm.}} = -
  \frac{\kappa^2_{q,\ell}}{128\pi^2f^2} \Bigl( \overline{\psi}_L
  \gamma^\mu \psi_L\Bigr) \Bigl(\overline{\psi}^\prime_L \gamma_\mu
  \psi^\prime_L \Bigr)
\end{equation}
Here, $\psi$ and $\psi^\prime$ are any combinations of different SM
fermions. These four-fermion operators can be reinterpreted in terms
of a contribution to the oblique $\Delta T$ parameter. The peculiar feature
about them is that they increase with the mass of the mirror fermions
for fixed scale $f$. This is clear from the fact that in that case the
Yukawa-type coupling which enters the box diagrams has to be enlarged
leading to a larger contribution from the box diagrams. The $\kappa$
is usually assumed to be a diagonal matrix in flavor space or even
proportional to the unit matrix. In this paper, we do not lift the
degeneracy in generation space, however, we investigate different
values for the $\kappa$ couplings for mirror quarks and mirror
leptons.

As was shown in~\cite{Reuter:2012sd,Reuter:2013iya,Tonini:2014dza}, the
end of LHC Run 1 was sort of a turning point where limits from direct
searches of heavy particles in Little Higgs models started to become
competitive with EWPO, and now with Run 2 even superseded them. As
the only relevant EWPO result is Eq.~(\ref{eq:4ferm-op}) and the $R$
dependence from the top partner contributions to the $\Delta T$
parameter, we do not discuss EWPO any further here, and
take Eq.~(\ref{eq:4ferm-op}) as a motivation to look into different
scenarios of combinations of all-light degenerate mirror fermions,
heavy mirror quarks, as well as light mirror leptons and decoupled
quarks and vice versa. 

%% file: sections/04_toolsetup.tex
\section{Tool Framework and Scan Setup}
\label{sec:toolsetup}

The main goal of this paper is to derive limits on the LHT model from
all available LHC run II data. In this section we describe the
framework that we used in order to derive numerically the current LHC
bounds on the LHT model. 

\subsection{Used Software}

To be able to generate Monte-Carlo events for our model, we make use
of the FeynRules implementation of the LHT model as in
Ref.~\cite{Reuter:2012sd,Reuter:2013iya,Tonini:2014dza}. We slightly
extended the model definition such that the heavy fermion Yukawa
couplings $\kappa$ are transformed into independent coupling constants
$\kappa_\ell$ and $\kappa_q$.  We then exported the LHT model to the
event generators \mgfive~\cite{Alwall:2011uj} and
\whizard~\cite{Kilian:2007gr,Moretti:2001zz,Nejad:2014sqa,Kilian:2011ka}~\footnote{\whizard
  recently also has been extended towards next-to-leading order functionality,
  cf.~\cite{Kilian:2006cj,Robens:2008sa,Binoth:2009rv,Greiner:2011mp,
    Nejad:2016bci,Bach:2017ggt}.}
via the UFO file format.\footnote{The model file is available on demand
from the authors.}

The collider phenomenology of the LHT model studied in this paper
depends on the mass scale $f$, the two Yukawa coupling parameters
$\kappa_\ell$ and $\kappa_q$, as well as the ratio of top Yukawa
couplings $R$. For these four parameters we derive the corresponding
masses according to 
Eqs.~(\ref{eq:heavygaugebosonmass}), (\ref{eq:heavytopmass}),
(\ref{eq:qHmasses}) and store these in a spectrum file which follows the
definitions of the UFO model. The branching ratios and corresponding
decay tables for all LHT particles are calculated analytically using
the formulae in the above linked model file. These include all 2-body
decays for all relevant particles. Note that within the parameter
space that we analyze, no 3-body decays need to to be considered as
there is always a dominating 2-body final state. The only difference
is the anomaly-mediated decay of $A_H$ in the case of $T$-parity
violation, see Sec.~\ref{sec:tpv}. For this, we use the branching
ratios as functions of $f$ taken from Ref.~\cite{Freitas:2008mq}
which will be shown later in this work. For decays into gauge bosons,
we assume that for $m(A_H) > \unit[185]{GeV}$, i.e.\ for $f \gtrsim
\unit[1080]{GeV}$, $A_H$ decays via 2-body decays into $WW$ and
$ZZ$. For smaller masses, we formulate 3-body decays for the decay
table as follows: we consider all possible decay modes of the $W$ or
$Z$, replace one of the final state gauge bosons with the
corresponding decay products and multiply the branching ratio
accordingly.  

For the main tasks of this numerical study, we make use of the
collider analysis tool \Checkmate{}
\cite{Drees:2013wra,Kim:2015wza,Dercks:2016npn}. This program is
useful to test a given BSM model in an automatized way. It makes again
use of the aforementioned  generator \mgfive{} to simulate partonic
events. By making use of the \texttt{UFO} model description file
format, \mgfive{} or \whizard{} are able to simulate partonic events
for a given BSM model which was implemented in a model building
framework like
\texttt{FeynRules}~\cite{Christensen:2008py,Alloul:2013bka} or
\texttt{SARAH}~\cite{Staub:2013tta}, e.g. via the 
\whizard-\texttt{FeynRules} interface~\cite{Christensen:2010wz}. The
showering and hadronization of these events is subsequently performed
by \texttt{Pythia8}~\cite{Sjostrand:2007gs}, followed by the fast
detector simulation \texttt{Delphes}~\cite{deFavereau:2013fsa} which
considers the effect of measurement uncertainties, finite
reconstruction efficiencies and the jet clustering of the observed
final state objects. These detector events are then quantified by
various analyses from both \texttt{ATLAS} and \texttt{CMS} at
center-of-mass energies of 8 and 13 TeV (more details below). Events
are categorized in different signal regions and \Checkmate{}
determines which signal region provides the strongest expected
limit. If the input model predicts more signal events than are allowed
by the observed limit of that signal region, \Checkmate{} concludes
that the model is \emph{excluded} at the 95\% confidence level,
otherwise the model is \emph{allowed}. For more details on the inner
functionality of \Checkmate, we refer to the 
manual papers in Refs.~\cite{Drees:2013wra,Kim:2015wza,Dercks:2016npn}.  

\subsection{Details on Event Generation}
\label{sec:tools:evtgen}
For the event generation, we consider the production of all relevant
two-body final states. In the following, we use $q_H$ for all heavy
fermion squarks $\{d_H, u_H, s_H, c_H, b_H, t_H$\}, $\ell_H$ for all
other heavy fermions $\{e_H, \mu_H, \tau_H, \nu_{e H}, \nu_{\mu H},
\nu_{\tau H}\}$, $V_H$ for all heavy gauge bosons $\{W_H, Z_H$ and
$A_H\}$ and $T^\pm$ for the additional heavy $T$-even/odd top partner,
respectively. We analyzed the following processes for the LHC (cf.
also~\cite{Reuter:2012sd,Reuter:2013iya})
\begin{equation}
  \label{eq:process_list}
  \arraycolsep=1.4pt\def\arraystretch{1.6}  
  \begin{array}{rl}
    1. \quad & \quad p p \rightarrow q_H q_H, q_H \bar{q}_H, \bar{q}_H
    \bar{q}_H \\
    2. \quad & \quad p p \rightarrow q_H V_H \\
    3. \quad & \quad p p \rightarrow \ell_H \bar{\ell}_H \\
    4. \quad & \quad p p \rightarrow V_H V_H \\
    5. \quad & \quad p p \rightarrow T^+ \bar{T}^+, T^- \bar{T}^- \\
    6. \quad & \quad p p \rightarrow T_+ \bar{q}, \bar{T}_+ q, T_+ W^\pm,
    \bar{T}_+ W^\pm
  \end{array}
\end{equation}
At this stage we give some remarks on the choice of these final states.
\begin{itemize}
\item If $T$-parity is conserved, $T$-odd particles need to be
  produced in pairs. Therefore, the $T$-even top partner $T^+$ is the
  only LHT particle which can be produced in association with Standard
  Model particles. This rule  
 also holds in case of anomaly-triggered $T$-parity violation as the
 corresponding TPV couplings  $A_H-V-V$ are too small to result in
 another $T$-odd final state with experimentally accessible cross
 section. 
\item We focus our discussion on certain benchmark scenarios and
  within these scenarios, some processes are expected to be negligible
  compared to others. We give more details on this when we discuss the
  individual scan setups below. 
\item While the production of color-charged objects is expected to be
  dominant at the LHC in case $q_H$ and $V_H$ have similar masses,
  heavy gauge boson production can become dominant in regions of
  parameter space where the heavy gauge bosons are significantly
  lighter than the heavy quarks (i.e. for large $\kappa$). We discuss
  the parameter dependence of the respective cross sections below. 
\item Processes with additional hard radiation in the final state,
  e.g. the process $p p \rightarrow q_H q_H j$, are not considered
  here. They are expected to be relevant in regions with strong mass
  degeneracy between the produced particle and the stable particle it
  decays into as in such a case the process $p p \rightarrow q_H q_H$
  produces too soft jets to be observed. By requiring an additional
  hard jet in the event, $p p \rightarrow q_H q_H j$, the additional
  jet can boost the $q_H q_H$ system and create a new, potentially
  observable multijet topology (see
  e.g.\ Ref.~\cite{Dreiner:2012gx}). However, in our case the gauge
  bosons $W_H$ and $Z_H$ are always predicted to be at least 100 GeV  
  heavier than the $A_H$, cf.\ Eqs.~(\ref{eq:heavygaugebosonmass1}),
  (\ref{eq:heavygaugebosonmass2}).   
  Similarly, the $q_H-A_H$ and $T^--A_H$ mass splittings are always
  large enough in the studied parameter regions. Therefore, we do not
  need to look at these peculiar topologies which have a significantly
  smaller cross section than our discussed two-body final states. 

\item 
  Another interesting final state is $p p \rightarrow A_H A_H j$ whose
  analysis is motivated because of the distinct and typical monojet
  signature as generally expected in models with a dark matter
  candidate (see e.g.\ Ref.~\cite{Aad:2015zva}). However, since our
  dark matter candidate has a mass of the  order of about 100~GeV and
  since it couples directly to quarks via the $A_H-q-q_H$ vertex, we
  do not expect LHC searches for dark matter final states to be more 
  constraining than existing bounds from direct detection
  searches. Furthermore, the consideration of this final state is
  technically involved as double-counting with the decay topology $p p
  \rightarrow q_H A_H, q_H \rightarrow q A_H$ could occur in specific
  parts of the parameter region and needs to be under precise control
  within the simulation. The detailed discussion of such a decay
  topology is postponed to a forthcoming study. 
\end{itemize}

All simulations have been done automatically by\Checkmate{} using the
event generator \mgfive{} and have been cross-checked with \whizard.

\subsection{Scan Benchmark Scenarios}
The LHT model 
---as already described earlier--- 
depends on the following four parameters
\begin{enumerate}
\item the symmetry breaking scale $f$ which affects the masses of all
  $q_H, \ell_H$, $V_H$ and $T^\pm$. 
\item the Yukawa parameter $\kappa_q$ which affects the masses of the
  heavy quarks $q_H$, 
\item the Yukawa parameter $\kappa_\ell$ which affects the masses of
  the color-neutral heavy fermions $\ell_H$ and 
\item the Yukawa parameter $R$ which affects the masses of the heavy
  top partners $T^+, T^-$. 
\end{enumerate}
Furthermore we distinguish models in which a) $T$-parity is exactly
conserved and b) models where gauge anomalies introduce the $T$-parity
violating couplings $A_H-W-W$ and $A_H-Z-Z$. 

In order to reduce the number of free parameters we focus on
particular benchmark scenarios with different theoretical and/or
phenomenological motivation and with different assumptions on the
fermion sector, the heavy top sector and the validity of
$T$-parity. These scenarios result in $3  \times 2 \times 2 = 12$
different benchmark cases, summarized in Tab.~\ref{tab:benchmarks}.

\begin{table*}
\footnotesize
\begin{tabularx}{\textwidth}{lllXl}
\toprule
Sector & Model & Constraint & Phenomenology & Considered Topology \\
\midrule
\midrule
\multirow{6}{*}{$f_H$} & \multirow{2}{*}{\emph{Fermion
    Universality}} & \multirow{2}{*}{$\kappa_l = \kappa_q$} &
$\bullet$ mass degeneracy of $q_H, \ell_H$ & \multirow{2}{*}{Exclude
  process 3}  \\ 
&                                          &  & $\bullet$ $\ell_H$
production negligible & \\[4pt] 
& \multirow{2}{*}{\emph{Heavy $q_H$}} & \multirow{2}{*}{$\kappa_q
  = 3.0$}  & $\bullet$  $q_H$ decoupled & \multirow{2}{*}{Exclude
  processes 1, 2} \\ 
&            &                    & $\bullet$ $\ell_H$ production
relevant & \\[4pt] 
& \multirow{2}{*}{\emph{Light $\ell_H$}} &
\multirow{2}{*}{$\kappa_l = 0.2$} & $\bullet$ $\ell_H$ very light &
\multirow{2}{*}{Exclude process 3} \\ 
&                               & & $\bullet$ $V_H$ branching ratios
change & \\ 
\midrule
\multirow{2}{*}{$T^\pm$} & \emph{Light $T^\pm$} & $R = 1.0$ &
$\bullet$ $T^\pm$ are light/accessible & Include process 4, 5 \\[4pt] 
                 & \emph{Heavy $T^\pm$} & $R = 0.2$ & $\bullet$
$T^\pm$ are heavy/inaccessible & Exclude process 4, 5 \\ 
\midrule 
\multirow{2}{*}{$A_H$} & \emph{TPC} & No TPV & $\bullet$ $A_H$ is
stable and invisible & $A_H$ stable \\[4pt] 
& \emph{TPV} & With TPV & $\bullet$ $A_H$ is unstable  & $A_H
\rightarrow V V$ decays \\ 
\bottomrule
\end{tabularx}
\caption{Definitions of the considered benchmark models of this
  study. In this work we consider all $3 \times 2 \times 2$
  combinations of the options given in this table. The process numbers
  refer to the list in Eq.~(\ref{eq:process_list}).} 
\label{tab:benchmarks}
\end{table*}

\paragraph{Heavy Fermion Sector:}
We first discuss the different assumptions on the heavy fermion
sector. In the \emph{Fermion Universality}  model we set the two  
%G: maybe more explicit:
coefficients $\kappa_q=\kappa_l$ 
equal and hence get a mass degeneracy in the heavy fermion sector.
Due to their color charge, the production cross sections for processes
involving heavy quarks are significantly higher than the respective
cross sections for final states with color-neutral heavy
fermions. Hence, we do not consider process 3 of our list
in~\ref{eq:process_list}.

The masses of the heavy fermions have two important consequences for
the phenomenology: they affect their production cross sections and
they change the branching ratios of the heavy gauge bosons $V_H
\rightarrow \ell_H^{(*)} \ell^\prime$. To get an understanding which
role this plays when setting bounds on the model we choose two further
benchmark cases, each taking into account one of these effects.  

In the \emph{Heavy $q_H$}  model we decouple the heavy quarks from the
model by fixing $\kappa_q=3.0$. This raises the heavy quark masses to
the multi-TeV-scale and hence makes them experimentally
inaccessible. Therefore, we do not consider production modes which
involve $q_H$, i.e.\ processes 1 and 2 of~\ref{eq:process_list},
but take into account $\ell_H$ pair production, process 3,
instead. The results of this benchmark scenario should give insight to
which degree the LHC sensitivity relies on the presence of the
color-charged objects and which limits can be determined from searches
looking for color-neutral particles only.  

The \emph{Light $\ell_H$} benchmark is also designed to lift the
degeneracy of the color-charged and color-neutral objects. Here, by
fixing $\kappa_\ell$ to a small value of 0.2, the latter are light
enough for the heavy gauge bosons to decay into them. We are
interested to see how this change in the expected decay patterns
affects the bounds compared to the \emph{Fermion Universality}
model. Note that even though the $\ell_H$ are light we do not take
into account the bounds from $\ell_H$ production as we are interested
in how only a change in the decay pattern affects the resulting
bounds. The bounds resulting from direct $\ell_H$ production are
determined in the previously discussed \emph{Heavy $q_H$} benchmark.  

The results of these three benchmark cases should be sufficient to
qualitatively determine the resulting bounds for other
$\kappa_q-\kappa_\ell$ combinations and to avoid a full 3D parameter
scan in the $f-\kappa_q-\kappa_\ell$ plane.

\paragraph{Heavy Top Partner Sector:}

The main phenomenological difference between the heavy top partners
$T^\pm$ and the other heavy fermions $q_H$ is that their mass depends
on $R$ instead of $\kappa$. We choose two benchmark values for this
parameter in such a way that one results in experimentally accessible
top partners ($R=1.0$) while the other ($R=0.2$) does not. The value
$R=1.0$ also corresponds to a case where minimal fine-tuning can be
achieved, see~\cite{Reuter:2012sd,Reuter:2013iya}, and thus this
benchmark case tests the natural regions of parameter space of the LHT
model. In the \emph{Heavy $T^\pm$} scenario we ignore any processes
which involve these particles as they are too heavy to result in an
LHC exclusion. The comparison of the two bounds at $R=1.0$ and $R=0.2$
gives insight to which degree the masses of the particles in this
sector are relevant for the overall sensitivity.  

\paragraph{$T$-Parity Violation:}
As discussed in Sec.~\ref{sec:tpv}, gauge anomalies in the heavy
sector can result in anomalous $T$-parity violating $A_H-W-W$ and
$A_H-Z-Z$ couplings. The presence of these operators may drastically
change the expected collider phenomenology as the final state not
necessarily contains an invisible particle any more. Supersymmetry
motivated searches are however still expected to be sensitive as the
leptonic decays of the $W$ and the invisible decays of the $Z$ boson
can still produce a significant amount of missing energy. We are
interested to see by how much the bounds derived for the $T$-parity
conserving case are changed due to these anomaly-mediated decays. For
that reason we analyze each of the above discussed benchmark scenarios
once with a stable $A_H$ and once with enabling $A_H \rightarrow V V$
decays.

%% file: sections/05_checkmateresults.tex
\newcommand{\etmiss}{$\slashed{E}_T$}
\section{Collider Topologies}
\label{sec:topologies}

For the discussion of the LHC results, it is useful to understand both
the values of the production cross sections for all the processes we
listed in the last section and the dominant branching ratios of the
relevant final state BSM particles. Collider bounds are expected to be
set by processes with a large production cross section times a decay
topology with only a small Standard Model contamination. In this
section we review the parameter dependence of these observables in
order to determine the theoretically expected collider topologies of
our LHT benchmark scenarios. Many of them are relevant for the 
discussion of the exclusion bounds that we determine with \checkmate{}
in the upcoming section.

\subsection{Cross Sections}
\label{sec:xsects}
\begin{figure*}
\centering
\includegraphics[width=0.45\textwidth]
                {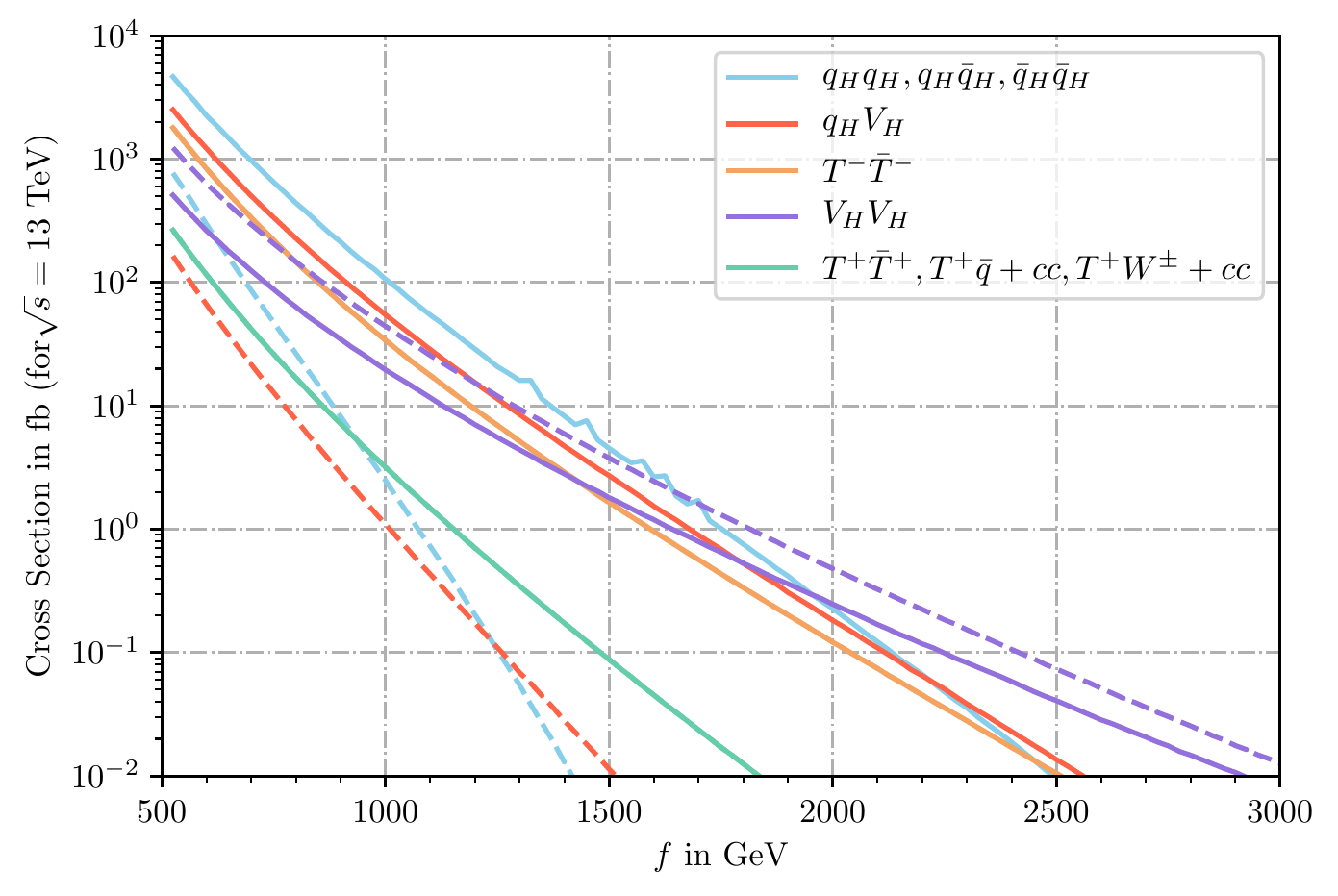}  \quad
\includegraphics[width=0.45\textwidth]
                {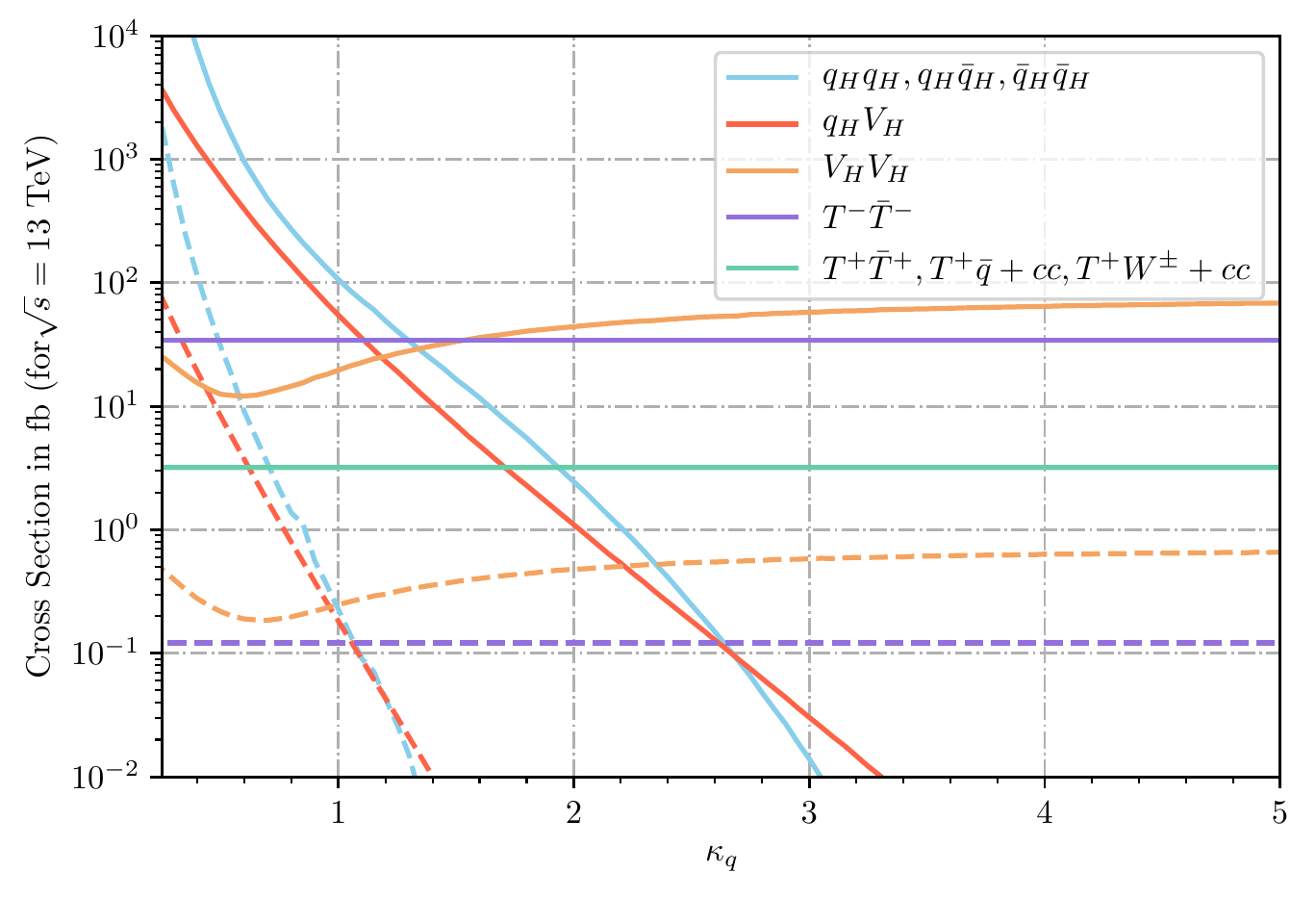}
\caption{LHC production cross sections ($\sqrt{s} = \unit[13]{TeV}$)
  for benchmark models \emph{Fermion Universality/Light $\ell_H$} +
  \emph{Light $T^\pm$}. Left: Dependence on $f$ for fixed $\kappa=1.0$
  (solid), $\kappa=2.0$ (dashed). Right: Dependence on $\kappa$ for
  fixed $f=\unit[1]{TeV}$ (solid), $f=\unit[2]{TeV}$ (dashed).  Labels
  in the legend appear in decreasing order of the respective maximum
  value of the solid lines.} 
\label{fig:xs:univ}
\includegraphics[width=0.45\textwidth]
                {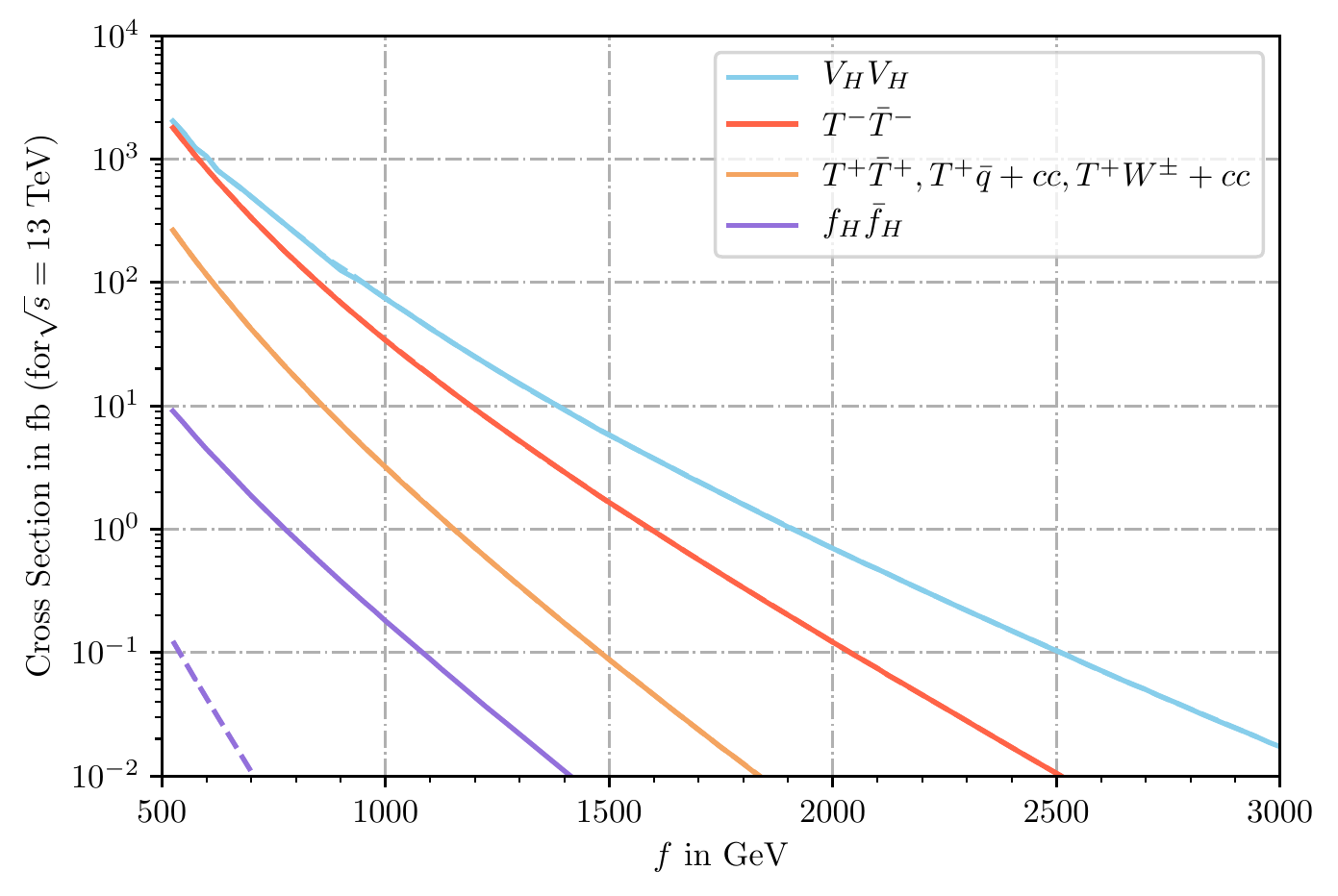} \quad
\includegraphics[width=0.45\textwidth]
                {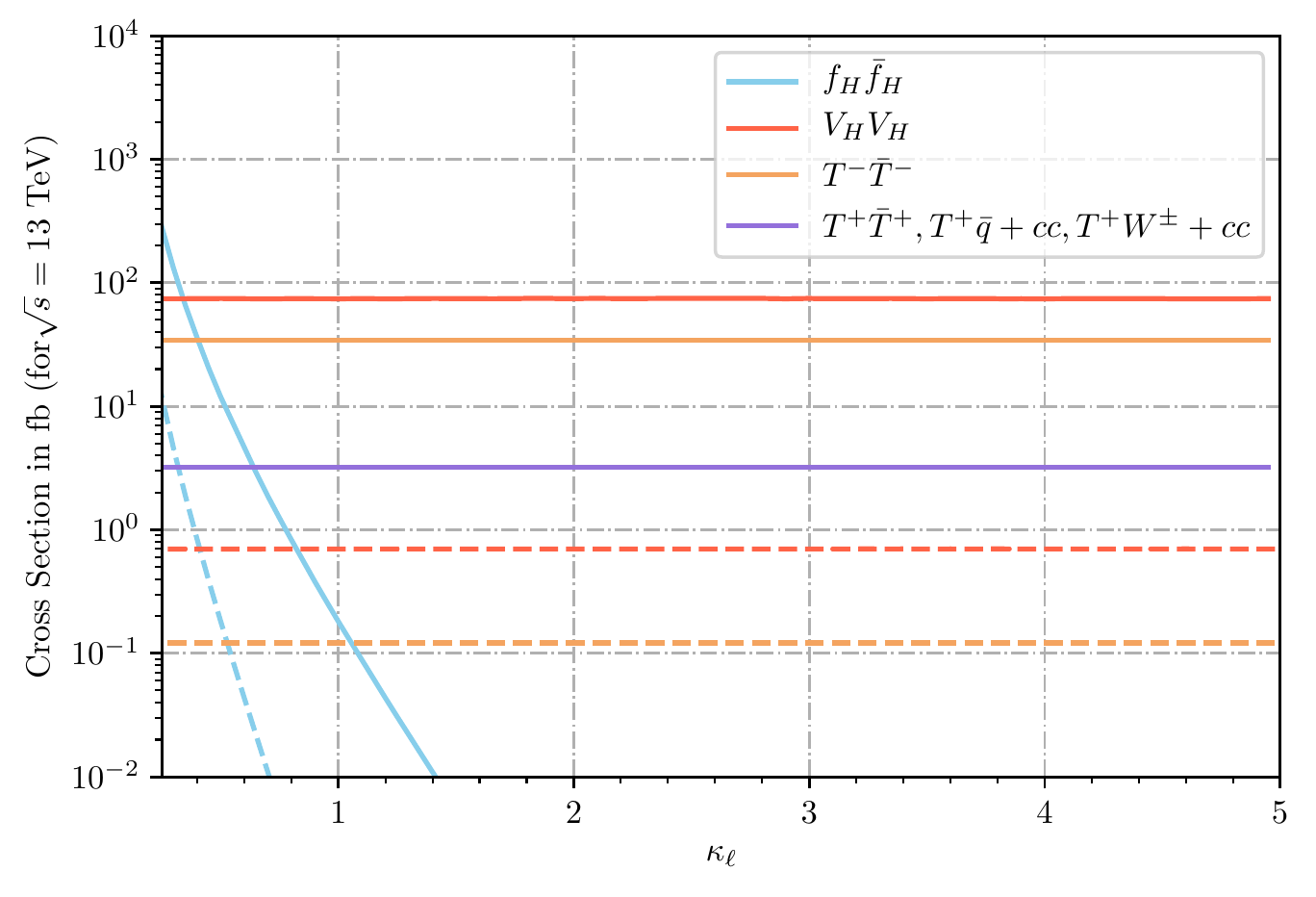}
\caption{Same as Fig.~\ref{fig:xs:univ} for benchmark model
  \emph{Heavy $q_H$} + \emph{Light $T^\pm$}.}\label{fig:xs:kq4} 
\end{figure*}
We start with a discussion of the production cross sections for all
the process sets listed in Sec.~\ref{sec:tools:evtgen}. In
Figs.~\ref{fig:xs:univ},\ref{fig:xs:kq4} we show the cross sections
for $\sqrt{s} = \unit[13]{TeV}$ as a function of the symmetry breaking
parameter $f$ with fixed $\kappa$ and vice versa. As the benchmark
case \emph{Light $\ell_H$} does not affect any production mode, the
cross sections are identical to those in the \emph{Fermion
  Universality} benchmark. In all cases we show the results in the
\emph{Light $T^\pm$} subscenario for which the $T^\pm$ are
kinematically accessible and the cross sections would nearly vanish in
the case of \emph{Heavy $T^\pm$}. Note that $\kappa$ refers to
$\kappa_q = \kappa_\ell$ in the \emph{Fermion Universality} case and
to $\kappa_q (\neq \kappa_\ell)$ in the \emph{Light $\ell_H$}
scenario. $T$-parity violation does not play a role in the discussion
of LHT particle production which is why we do not distinguish
\emph{TPC} and \emph{TPV} here. Results for center-of-mass energies
of 8 and 14 TeV are provided in App.~\ref{app:figures}.  

Since the mass of all heavy sector particles increases with $f$, the
cross sections for all processes drop with increasing
$f$.\footnote{Small fluctuations in the $f$-dependent $q_H q_H$
  production cross section are caused by numerical noise.}
Similarly, since the mass of the heavy fermions depends  linearly on
$\kappa$, the cross sections for producing these particles becomes
smaller for larger values of this parameter. As both mass and
couplings of the $T^\pm$ only depend on $f$ and the fixed parameter
$R$, no dependence on $\kappa$ can be seen.  

Interestingly, even though the mass of the vector bosons $V_H$ also
depends on $f$ only, their production cross sections show a small
$\kappa$-dependence in the \emph{Fermion Universality} scenario. This
is due to contributions of $t$-channel $q_H$ which interfere
destructively with the $s$-channel vector-boson
diagrams. Since all masses scale linearly with $f$, this effect
appears nearly independently of $f$ at the position $\kappa \approx
0.5$. As a result, the cross section for $V_H$ pair production is
roughly a factor 5 smaller for small $\kappa \approx 0.5$ than for
large values $\kappa \gtrsim 4$ when the heavy fermions are
decoupled. As the $q_H$ are by construction decoupled in the
\emph{Heavy $q_H$} benchmark scenario, the $\kappa$ dependence of the
$V_H V_H$ production cross section vanishes in the resulting
distribution shown in Fig.~\ref{fig:xs:kq4}.  

The production cross sections can reach values up to $10^3$ fb and we
thus expect the $\sqrt{s} = \unit[13]{TeV}$ LHC to be sensitive to
large regions of the parameter space we considered. Even for values of
$f \approx \unit[3]{TeV}$, cross sections of order \unit[$10^{-1}$]{fb}
and thus detectable event rates can be expected which improves results
from LHC Run 1 which were insensitive to values of the symmetry
breaking scale above
\unit[2]{TeV}~\cite{Reuter:2012sd,Reuter:2013iya}. Comparing the
results of both the $f$-$\sigma$ and the $\kappa$-$\sigma$ planes, it
becomes clear that there is no dominant process with a universally
largest cross section. The cross sections have very different
dependencies on $\kappa$ and $f$ and thus different regions in
parameter space are expected to have different dominating final
states. 

Generally, regions with small values of $\kappa$ and thus with light
$q_H, \ell_H$ predict a large rate of produced heavy fermions. As
expected for a hadron collider, the $q_H$ production is about two to
three orders of magnitude larger than the production of heavy leptons
$\ell_H$ and the latter appear only to be relevant for small values $f
\lesssim \unit[1]{TeV}, \kappa \lesssim 0.5$. In regions with larger
values of $\kappa$, the production of heavy vector bosons becomes more
important as their mass is independent of $\kappa$. If heavy top
partners $T^\pm$ are accessible, they are produced with comparable
abundance as the heavy vector bosons.\footnote{Note that this
  statement in general depends on the specific value of the additional
  parameter $R$ which we fixed to $1.0$ in our benchmark scenario.}
Since the $T^-$ is always lighter than the $T^+$, the production of
the latter appears to be negligible in comparision.  

\begin{figure*}
\centering
\includegraphics[width=0.45\textwidth]
                {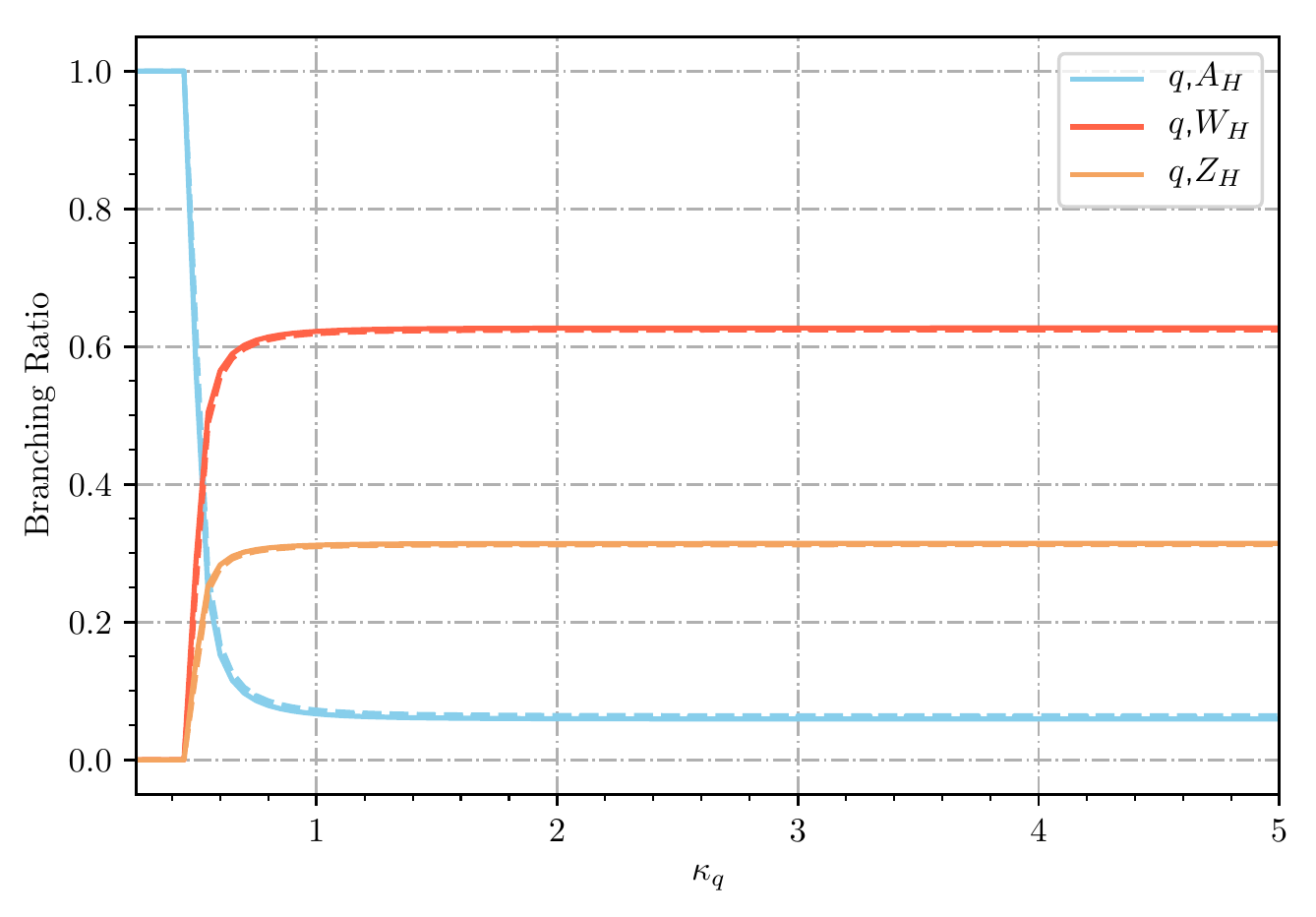} \quad
\includegraphics[width=0.45\textwidth]
                {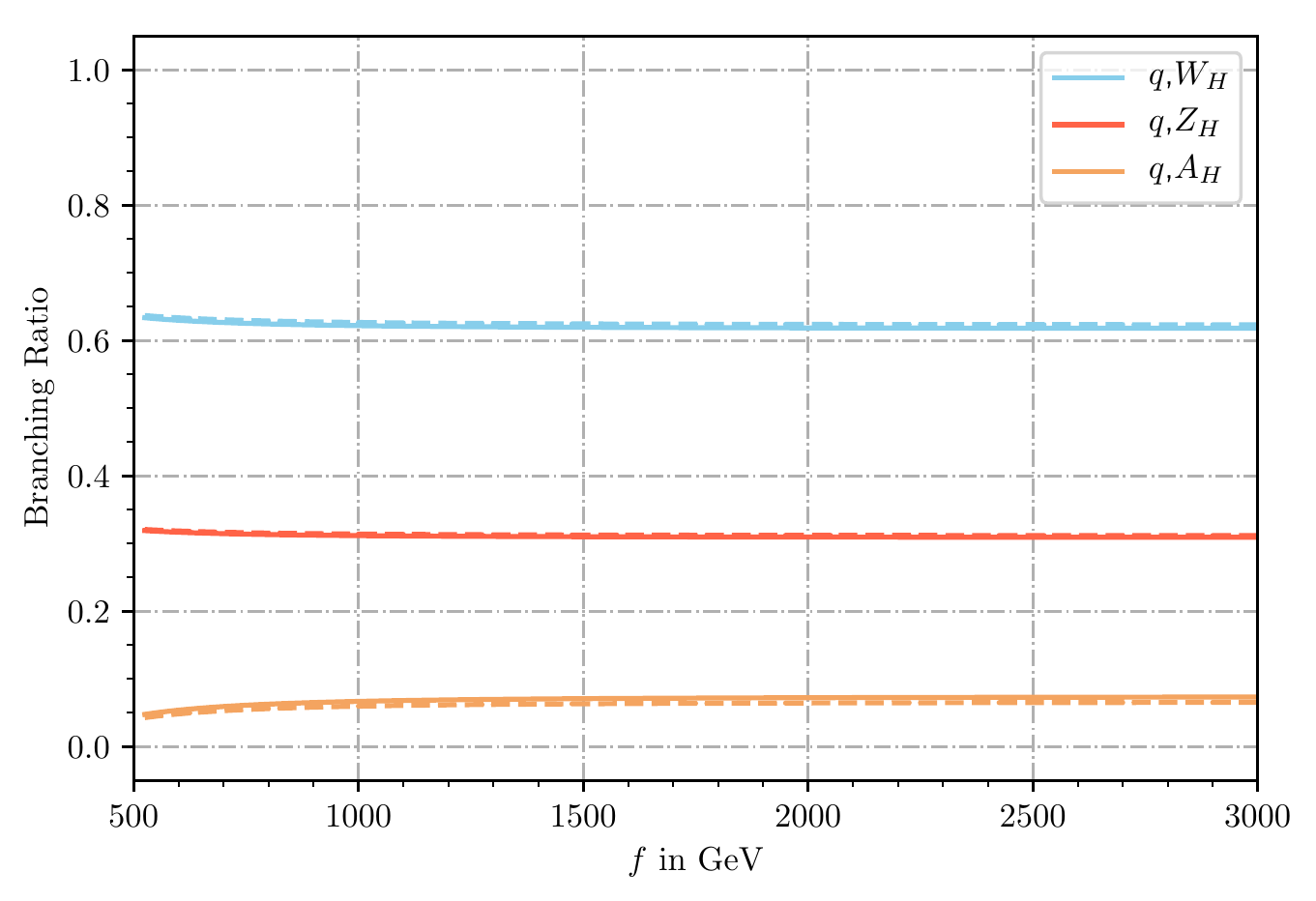} \quad
\caption{Branching ratios of $d_H$ in the \emph{Fermion
    Universality/Light $\ell_H$} model. Items in legend appear in
  decreasing order of the maximum value of the respective curve. Left:
  Fixed $f = \unit[1]{TeV}$ (solid), $f = \unit[2]{TeV}$
  (dashed). Right: Fixed $\kappa = 1$ (solid), $\kappa = 2$ (dashed).} 
\label{fig:cm:br1}
\end{figure*}
\begin{figure*}
\centering
\includegraphics[width=0.45\textwidth]
                {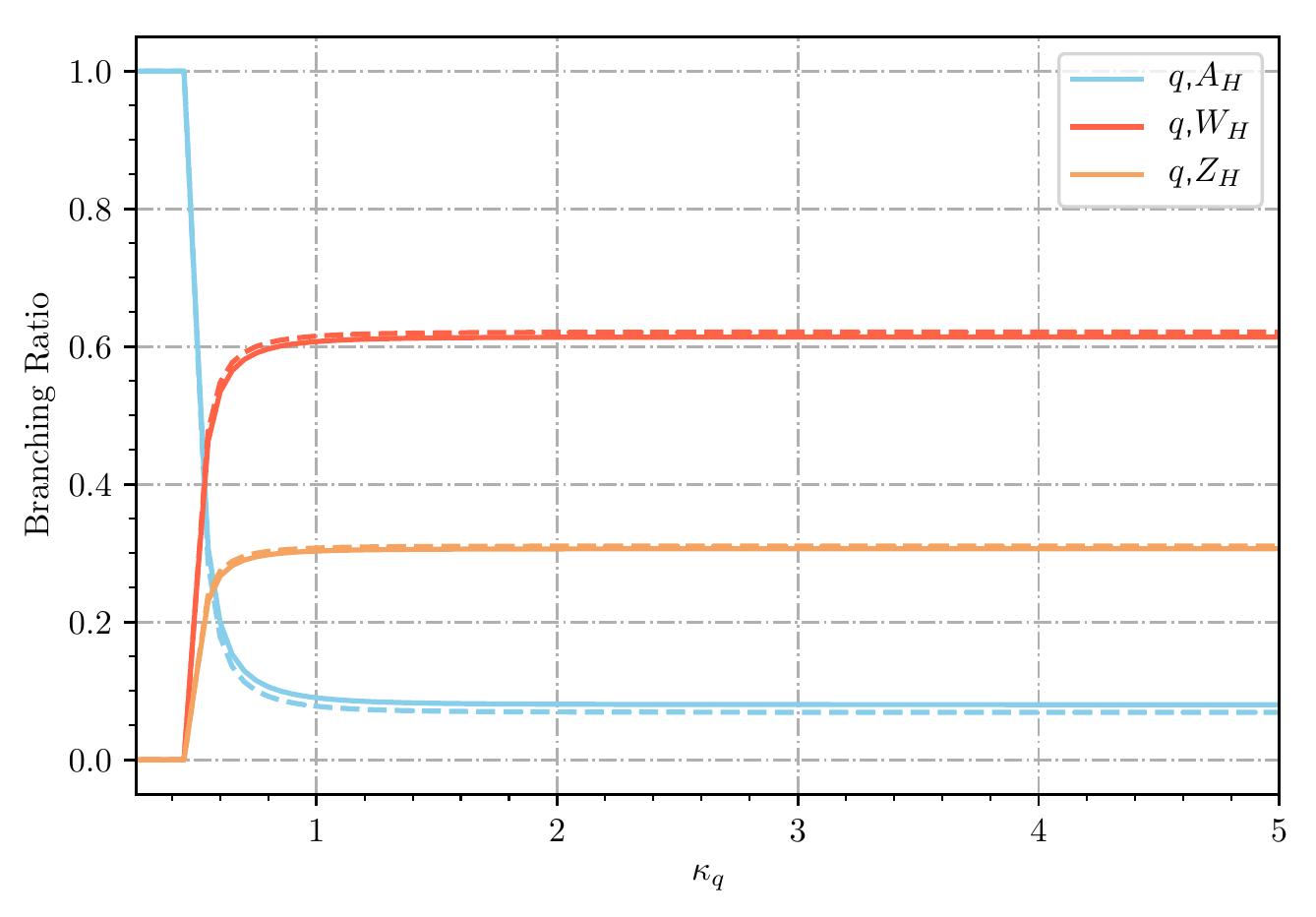} \quad
\includegraphics[width=0.45\textwidth]
                {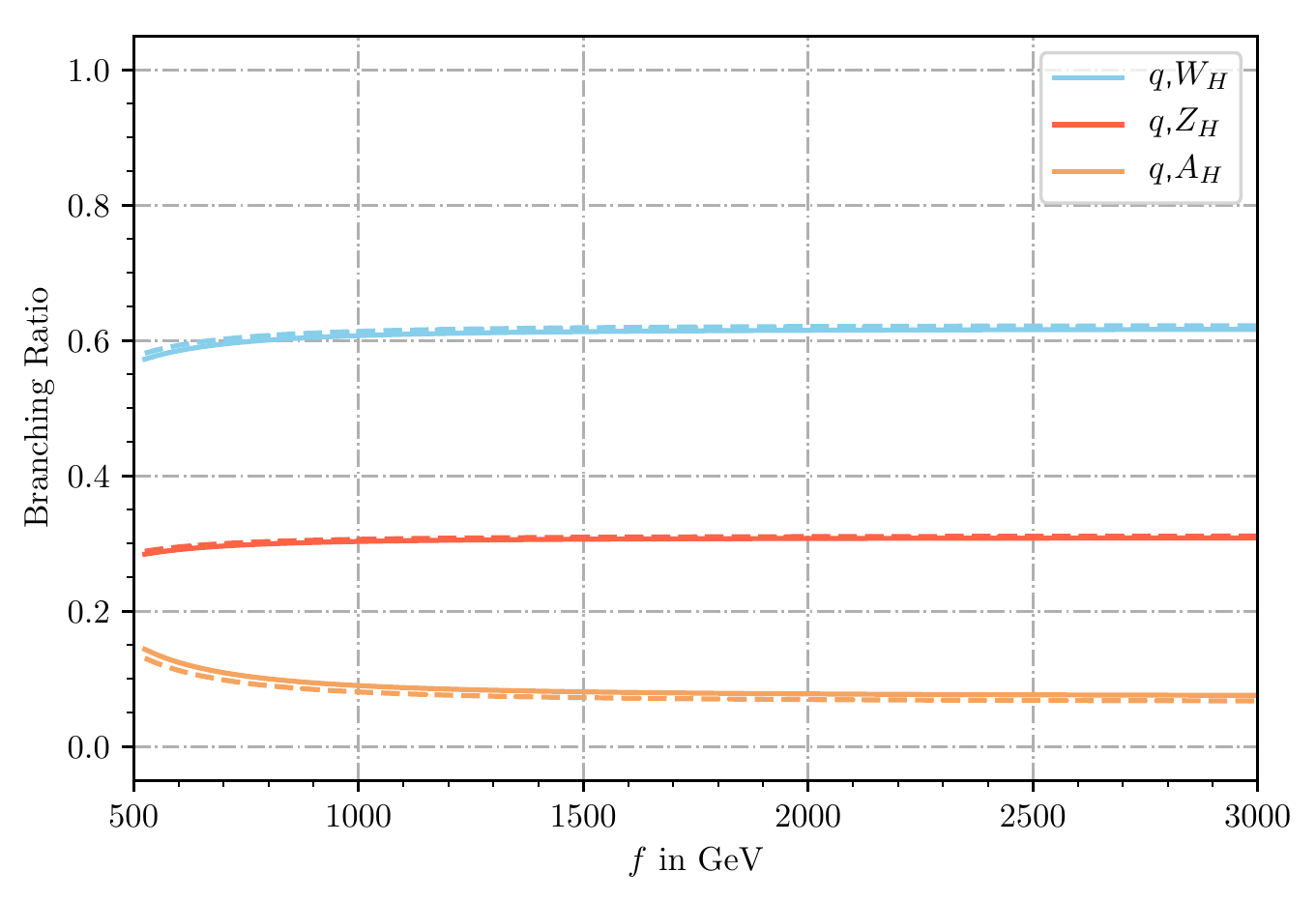} \quad
\caption{Branching ratios of $u_H$ in the \emph{Fermion
    Universality/Light $\ell_H$} model. Parameters as in
  Fig.~\ref{fig:cm:br1}.} 
\label{fig:cm:br2}
\end{figure*}

\begin{figure*}
\centering
\includegraphics[width=0.45\textwidth]
                {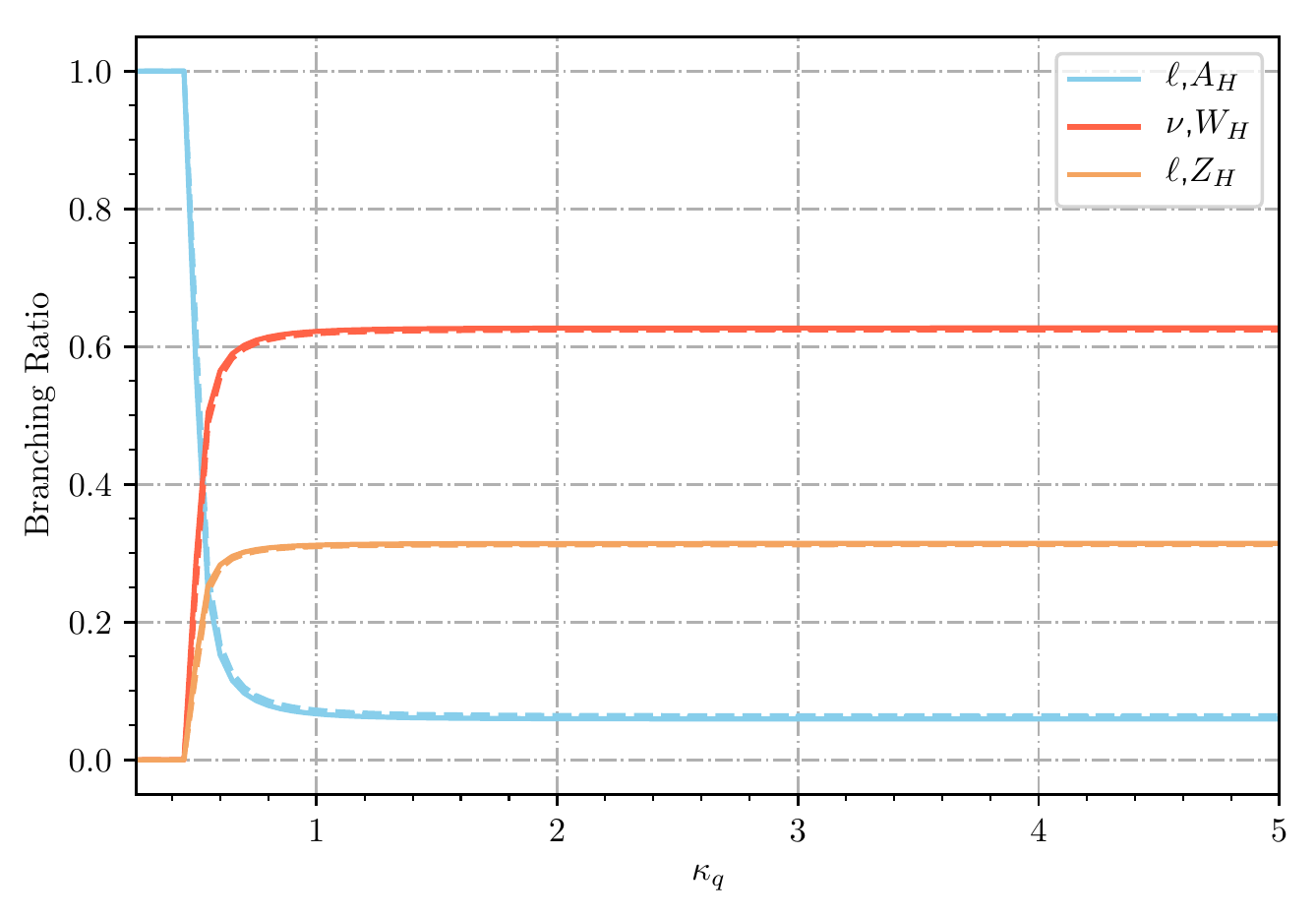} \quad
\includegraphics[width=0.45\textwidth]
                {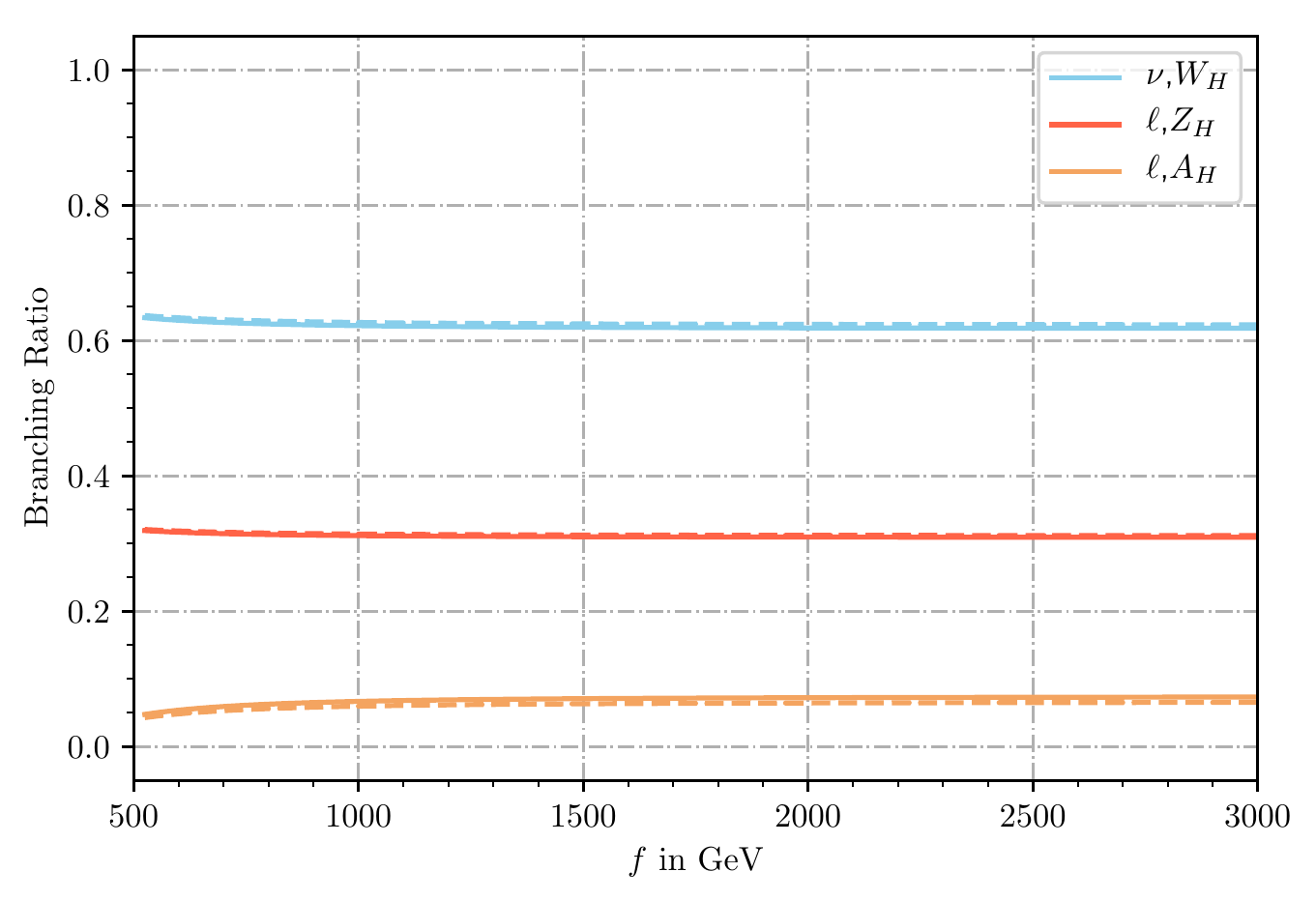} \quad
\caption{Branching ratios of $\ell_H$ in the \emph{Fermion Universality} model.}
\label{fig:cm:br11}
\end{figure*}

\begin{figure*}
\centering
\includegraphics[width=0.45\textwidth]
                {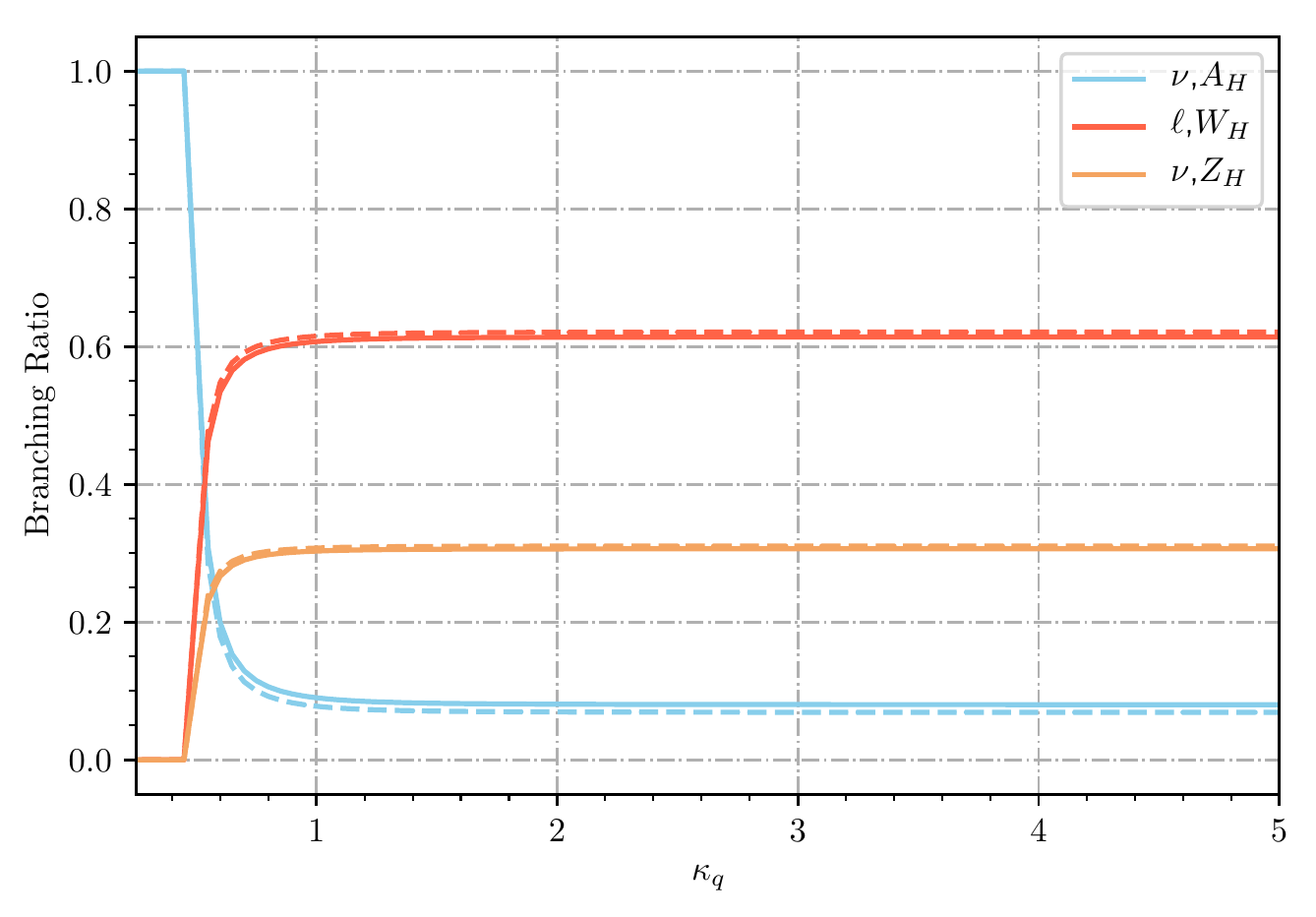} \quad
\includegraphics[width=0.45\textwidth]
                {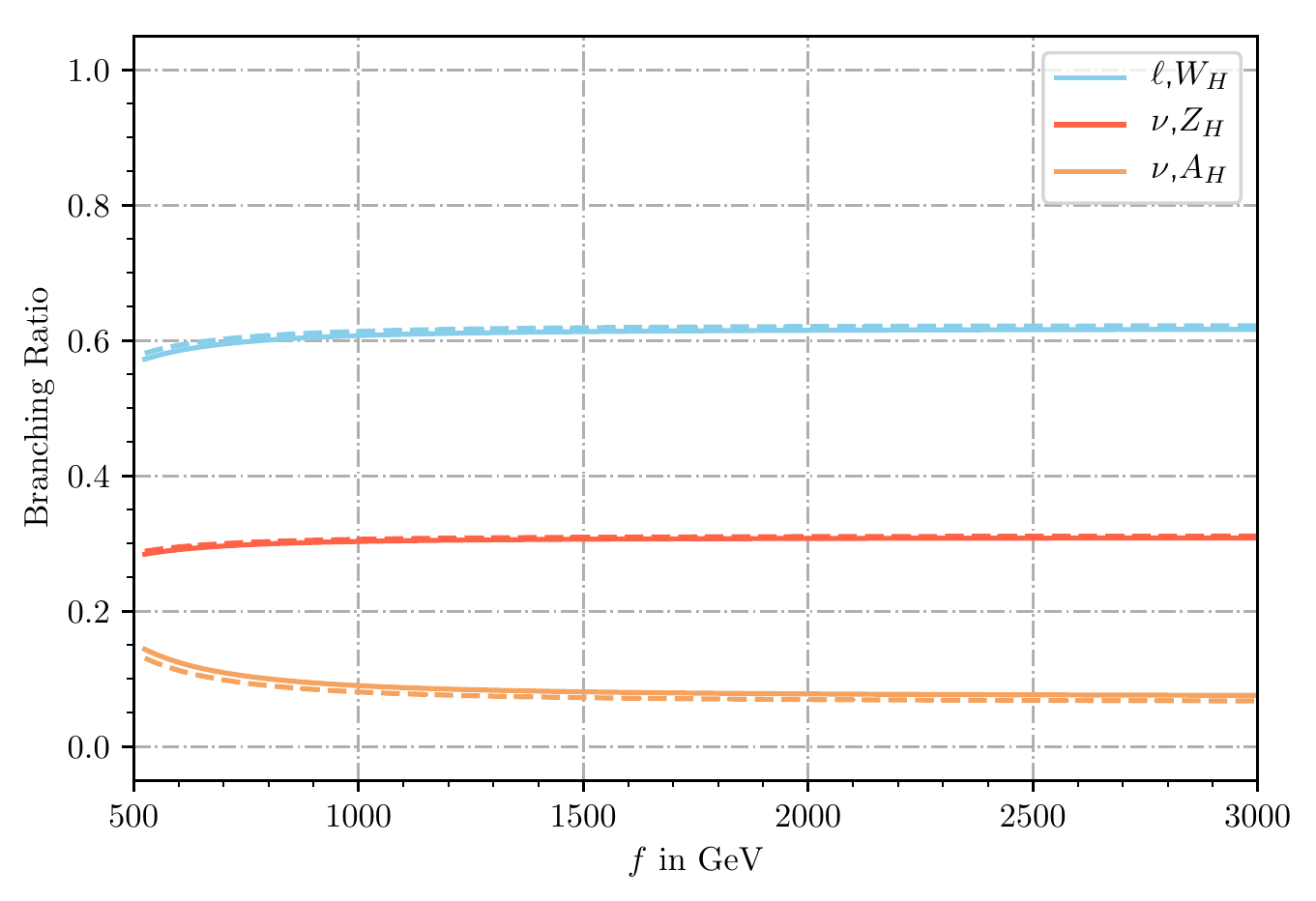} \quad
\caption{Branching ratios of $\nu_{e,H}$ in the \emph{Fermion Universality} model.}
\label{fig:cm:br12}
\end{figure*}

\begin{figure*}
\centering
\includegraphics[width=0.45\textwidth]
                {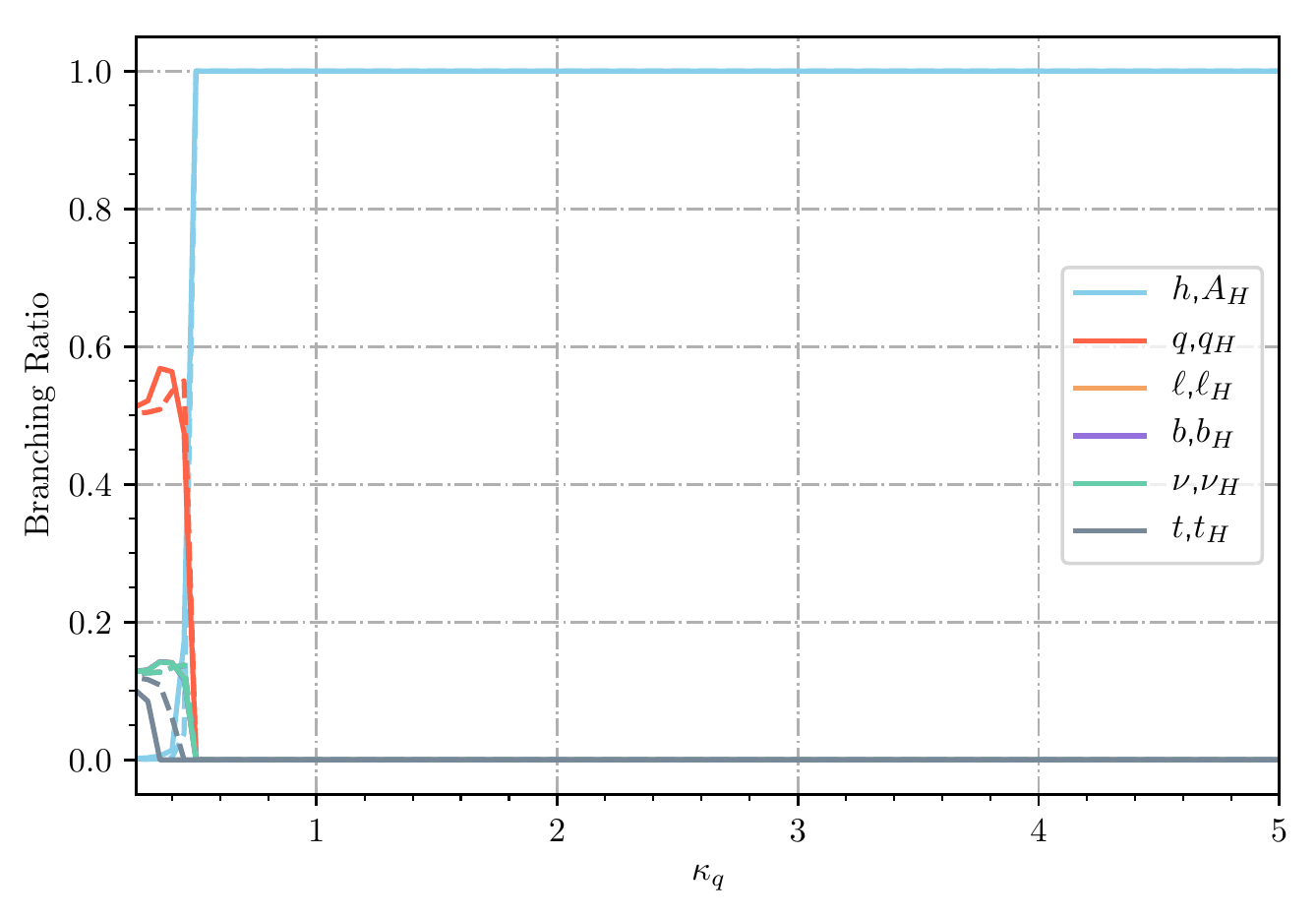} \quad
\includegraphics[width=0.45\textwidth]
                {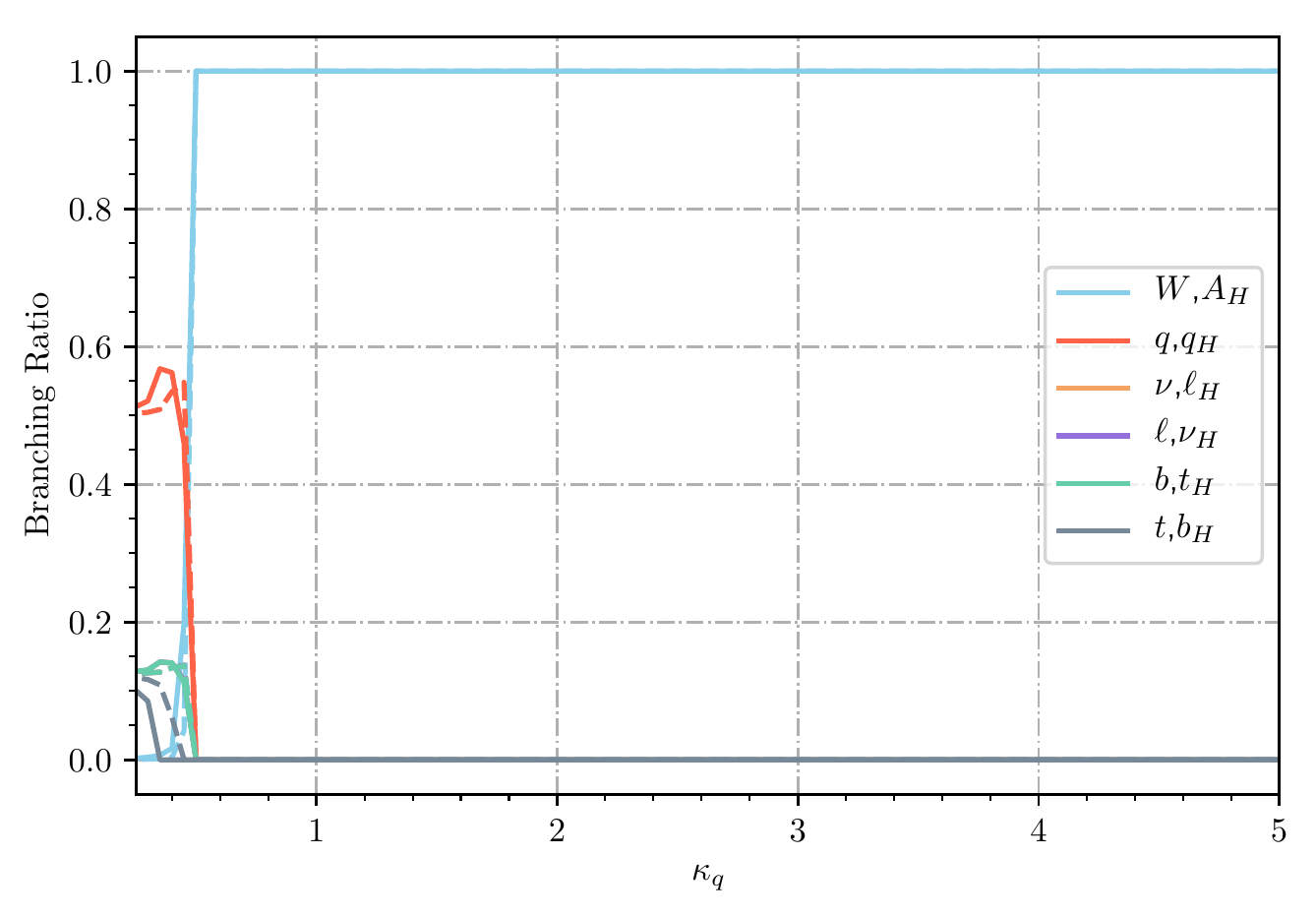}
\caption{Branching ratios of $Z_H$ (left) and $W_H$ (right)  in the
  \emph{Fermion Universality} (The \emph{Heavy $q_H$} scenario is very
  similar, cf. text). Parameters as in Fig.~\ref{fig:cm:br1}. In both
  plots, curves corresponding to decays with $\nu, \ell$ or $b$ are
  nearly identical.} 
\label{fig:cm:br2324}
\end{figure*}

\begin{figure*}
\centering
\includegraphics[width=0.45\textwidth]
                {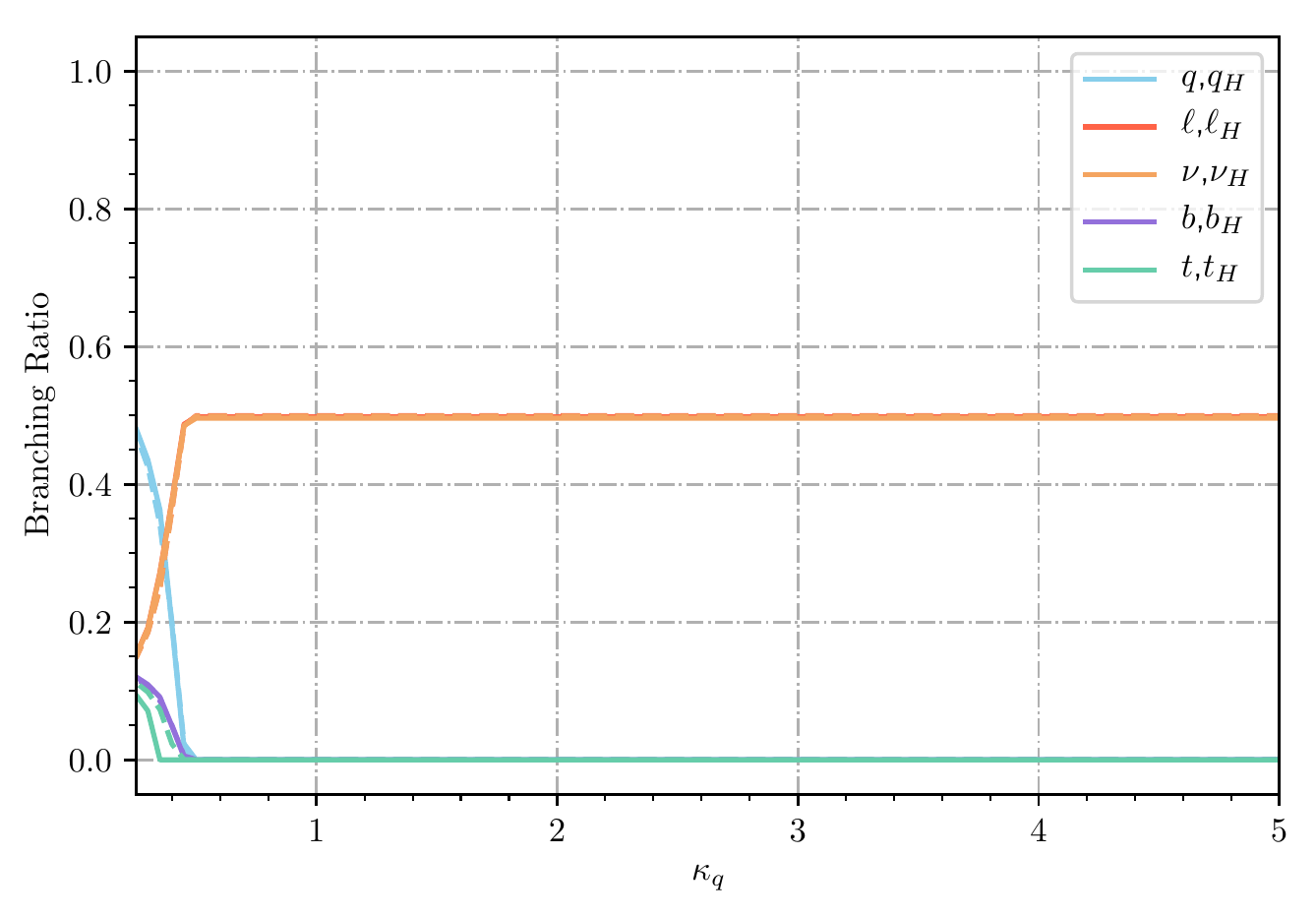} \quad
\includegraphics[width=0.45\textwidth]
                {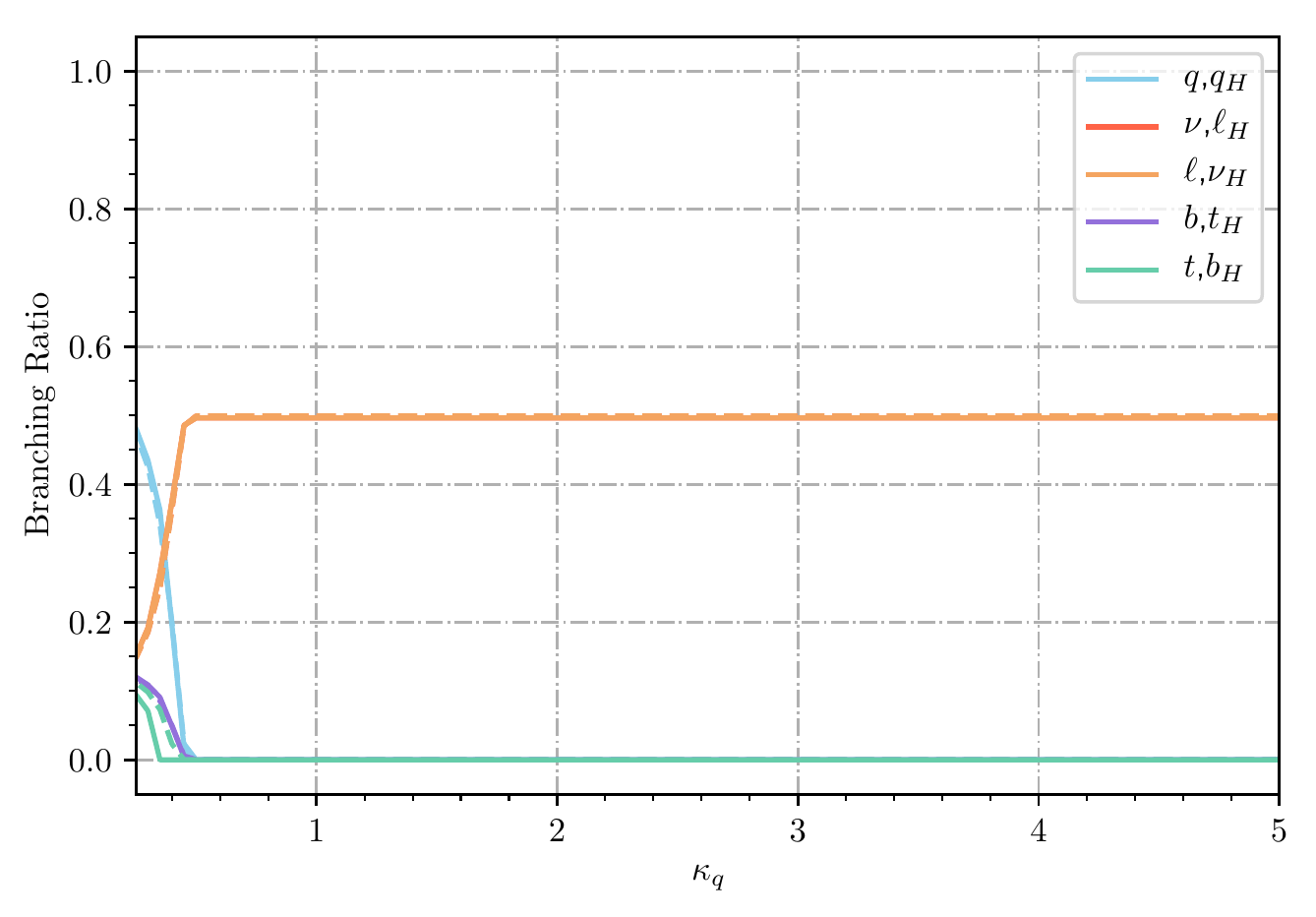}
\caption{Branching ratios of $Z_H$ (left) and $W_H$ (right)  in the
  \emph{Light $\ell_H$} model. $f$ is chosen as in
  Fig.~\ref{fig:cm:br1}, left. In both plots, the curves corresponding
  to decays with $\nu$ or $\ell$ are nearly identical.} 
\label{fig:cm:br2324kl}
\end{figure*}

\begin{figure}
\begin{center}
  \includegraphics[width=0.45\textwidth]
                  {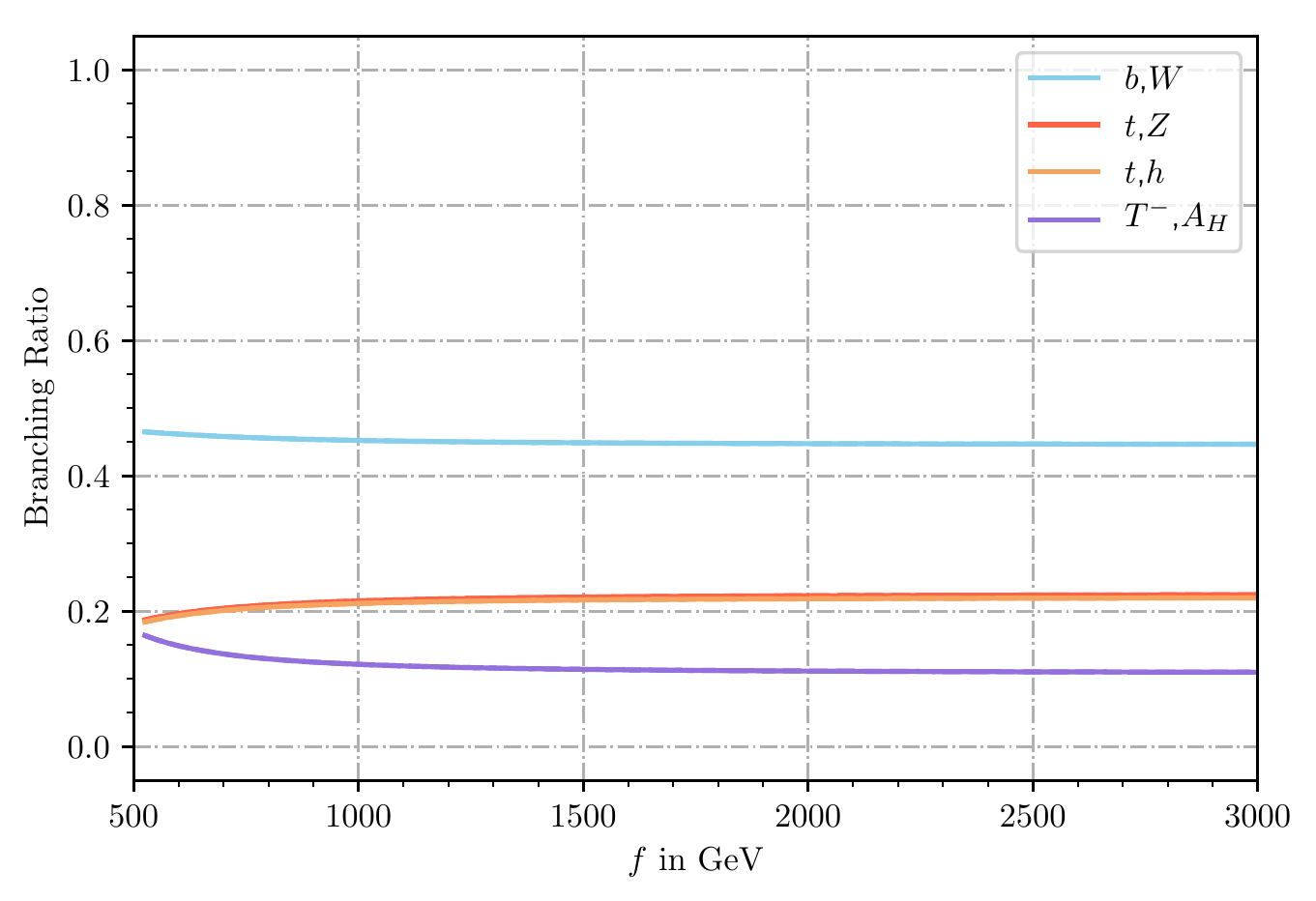} 
\end{center}
\caption{Branching ratios of $T^+$ in the \emph{Light $T^\pm$} scenario.}
\label{fig:cm:br7}
\end{figure}
\begin{figure}
\begin{center}
  \includegraphics[width=0.45\textwidth]
                  {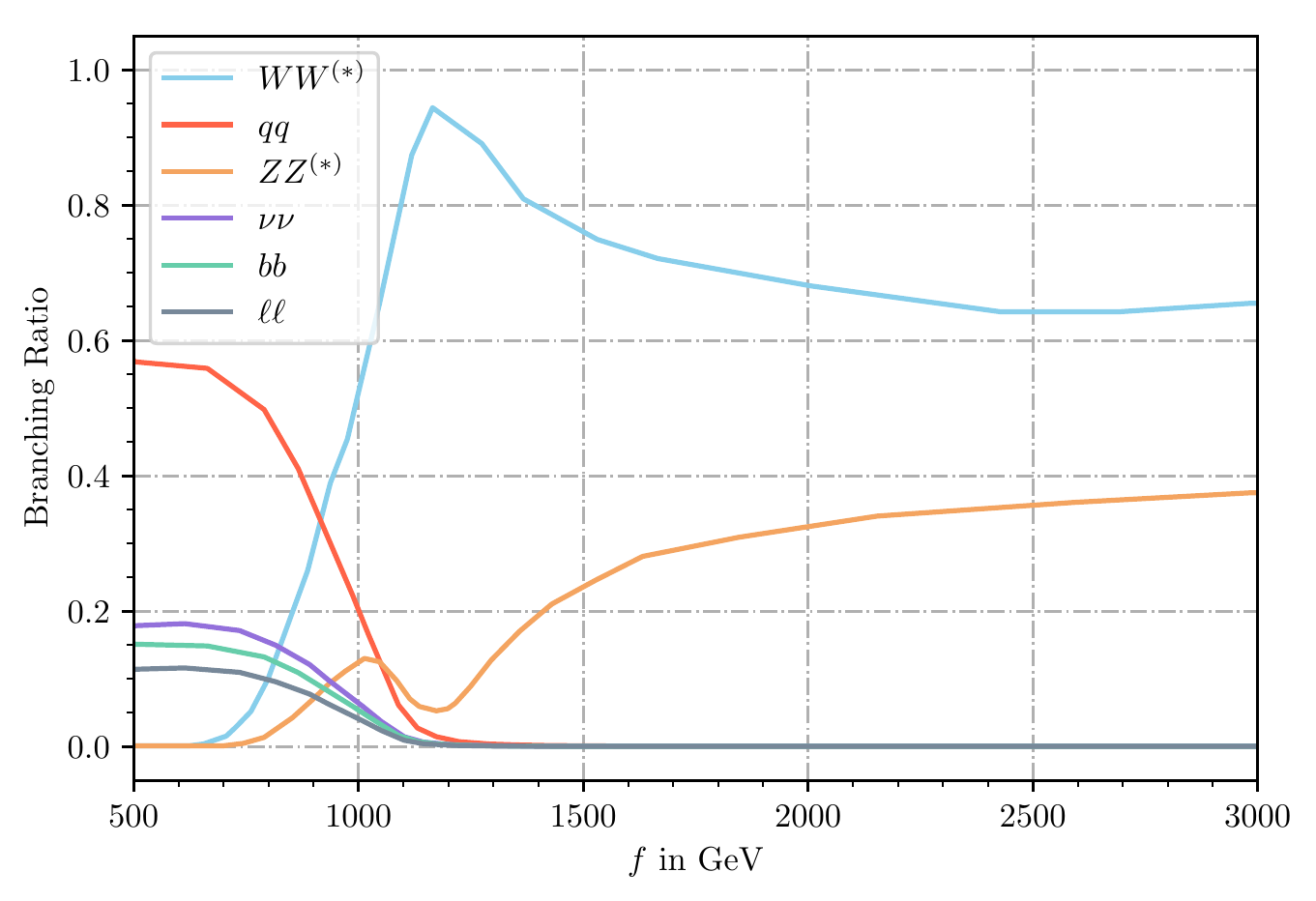}
\end{center}
\caption{Branching ratios of $A_H$ in the \emph{TPV} benchmark.}
\label{fig:cm:br22}
\end{figure}

\subsection{Branching Ratios}
\label{sec:brs}

We now continue with a discussion of the branching ratios for the
relevant partner particles within the given benchmark cases. Note that
we combine phenomenologically similar  branching ratios which involve
$q :=u, d, c, s$, (so we particularly do not distinguish heavy up- and
down-type quarks here) $\ell = e, \mu, \tau$, $\nu := \nu_e, \nu_\mu,
\nu_\tau$ and their respective heavy partner fermions.\footnote{It is
  only in this section where we distinguish between the charged heavy
  fermion $\ell_H$ and the neutral particle $\nu_H$. In the rest of
  this work, $\ell_H$ refers to both heavy charged and heavy neutral
  leptons.} Also, we only discuss those decays with a branching ratio
of at least 1\ \% anywhere in the discussed parameter
space. Though we do not show it in the plots, we analytically
calculated all decay widths and considered all kinematically allowed
$2$-body final states in the decay tables used in our scans in order
to get correct values for the branching ratios. We mainly discuss
results for the \emph{Fermion Universality} and the \emph{Light
$\ell_H$} scenarios as the \emph{Heavy $q_H$} scenario does not show
any differences in the observable decay pattern - except for one
difference which we mention along the way. Obviously, it is only the
decay of the $A_H$ which shows different behaviour in the benchmark
cases \emph{TPV} and \emph{TPC}. These two benchmarks are hence not
disinguished in the discussion regarding the decays for the other
particles.

Within the parameter ranges that we focus on, the particles $T^-,
\ell_H$ and $\nu_H$ each only have one decay mode in some scenarios: 
\begin{align}
\text{\emph{Light $T^-$}:}&\qquad \text{BR}(T^- \rightarrow t A_H) = 1
\\ 
\text{\emph{Light $\ell_H$}:}&\qquad \text{BR}(\ell_H/\nu_H
\rightarrow \ell/\nu A_H) = 1 
\end{align}
For other particles and/or other scenarios there is more than one
decay mode and the branching ratios depend on the values of $f$ and/or
$\kappa$.  

In Figs.~\ref{fig:cm:br1},\ref{fig:cm:br2} we show the dominating
branching ratios of the heavy quark partners $d_H, u_H$, respectively,
in the \emph{Fermion Universality/Light $\ell_H$} models which show
identical results in this regard. As before, we show curves as
functions of both $\kappa$ and $f$. For both up- and down-type heavy
quark partners, the decay into a heavy $W_H$ boson and a quark is the
most important decay with a branching ratio of nearly 60 \% ---
whenever it is kinematically allowed. They are followed by decays into
$Z_H q$ of order 30 \% and to $A_H q$ of order 10 \%. A small
variation with $f$ becomes visible which is caused by a subdominant
dependence of the respective coupling constants on $v/f$ (see
e.g.~\cite{Hubisz:2005tx}). This dependence differs between up-
and down-type quarks and thus the variation with $f$ differs for these
two flavors. Note that very small values of $\kappa_q \lesssim 0.5$
lead to $m(q_H) < m(W_H),m(Z_H)$ and thus forbids decays $q_H
\rightarrow (W/Z)_H + X$. All $q_H$ therefore decay to the light $A_H$
in this region of parameter space.

Note that due to the overall mass degeneracy and the identical quantum
numbers within the \emph{Fermion Universality} model, the decay
signatures of all other heavy fermions, except for the $T^\pm$, are
identical after replacing the corresponding up- and down-type
components of the respective SU(2) doublets. For example, the
branching ratio for $\nu_{eH} \rightarrow W_H e$ is identical to the
branching ratio $u_H \rightarrow W_H d$, see
Figs.~\ref{fig:cm:br1}-\ref{fig:cm:br11}.

Next, we discuss the decays of the heavy gauge bosons $W_H$ and $Z_H$
for the \emph{Fermion Universality} model in Fig.~\ref{fig:cm:br2324}
and for the \emph{Light $\ell_H$} model in
Fig.~\ref{fig:cm:br2324kl}. We only show results depending on
$\kappa$ as there is no $f$ dependence for the two standard benchmark
values $\kappa = 1.0, 2.0$ which we considered. In case of
\emph{Fermion Universality}, the decay $V_H \rightarrow f_H f^\prime$
into a heavy fermion partner is only allowed for $\kappa \lesssim 0.5$
and in this region decays into heavy quarks dominate. For larger
values of $\kappa$, the only available decays are $W_H \rightarrow W
A_H$ and $Z_H \rightarrow h A_H$. In the \emph{Light $\ell_H$}
scenario, this picture changes by construction: the $\ell_H$ are fixed
to light masses and thus for $\kappa_q \gtrsim 0.5$ both heavy gauge
bosons decay to \unit[50]{\%} into $\ell_H \ell$ and $\nu_H
\nu$. Again, for smaller values of $\kappa_q$ decays into $q_H$ are
kinematically accessible and have a dominant branching ratio. The
branching ratio curve for the benchmark scenario \emph{Heavy $q_H$}
corresponds to the one for \emph{Fermion Universality} with the only
exception that the decay $V_H \rightarrow q_H q$ disappears for
$\kappa < 0.5$ and the branching ratios for the other modes scale up
accordingly.  

In Fig.~\ref{fig:cm:br7} we show the branching ratios of the heavy top
partner $T^+$ (note that $T^-$ always decays to $t A_H$ as listed
above) in the \emph{Light Top} benchmark, i.e.\ for $R=1.0$. As $T^+$
is a $T$-parity even particle it must decay into pairs of $T$-odd
particles or purely into SM particles. This results in four main decay
scenarios. The SM decays follow mainly the pattern of a $SU(2)_L$
singlet top partner (cf.~e.g.~\cite{Reuter:2014iya}) of 50 \%
branching ratio into $b W^+$ and equally a quarter into $th$ and
$tZ$. This is only slightly modified by the only accessible $T$-odd
particle decay, namely roughly 15 \% branching ratio into $T^- A_H$.
The changes the top-like decay into $b W^+$ into nearly 45 \%
branching ratio, while $th$ and $tZ$ have roughly 20 \% branching
ratio each. These branching ratios have no dependence on $\kappa$
and only little dependence on $f$ which originates from the
$f$-dependence of the $T^\pm$ and $A_H$ masses. 

We finish the discussion with the branching ratios of the $A_H$ in the
\emph{TPV} scenario shown in Fig.~\ref{fig:cm:br22}, which only depend
on $f$. The information shown in this figure has been taken from a
detailed calculation performed in Ref.~\cite{Freitas:2008mq}. One
observes that for $f > \unit[1200]{GeV}$, decays into on-shell
Standard Model gauge boson pairs dominate. For smaller values of $f$,
the $A_H$ mass drops below \unit[180]{GeV}, the partial decay widths
into gauge bosons decrease due to kinematic suppression and the
loop-induced decays into Standard Model leptons become equally
relevant. For $f \lesssim \unit[900]{GeV}$, $A_H$ decays predominantly
into SM quark pairs.

\subsection{Expected Final State Topologies and Correspondence to
  Supersymmetric Searches} 
\label{subsec:topologies}

In this section we combine the information of the preceding one with
the list of dominant production processes given in
Sec.~\ref{sec:tools:evtgen} in order to find the following expected 
final state signatures. Comparing them to the specialized analyses of
the experimental collaborations for supersymmetry, we can make the
following classification of the signatures and their applicability to
the LHT model:
\begin{itemize}
\item In general -- if $T$-parity is conserved -- all $T$-odd particles
  produce decay chains  with a stable $A_H$ as the lightest $T$-odd
  particle at the end. This particle is experimentally invisible and
  thus produces missing transverse momentum \etmiss in the event. This
  is in close analogy to $R$-parity conserving supersymmetry which
  produces decay chains with the lightest neutralino at the end which
  similarly produces \etmiss. Therefore, many searches looking for
  $R$-parity conserving supersymmetry require \etmiss{} in the event
  and thus are sensitive to our model.  
\item Final states with heavy gauge bosons $W_H, Z_H$ behave
  differently in the main benchmark cases. In the \emph{Fermion
    Universality} model, where $W_H$ decays produce $W$ bosons which
  either contribute with further jets in their hadronic decays or with
  further hard leptons in their leptonic decays, the heavy $Z_H$ adds
  Higgs bosons in the final state which mainly lead to additional
  $b$-jets in the event. This final state topology is thus similar to
  supersymmetric electroweakino production $\tilde \chi^\pm_1 \tilde
  \chi^0_2 + \tilde \chi^\pm_1 \tilde \chi^\pm_1 + \tilde \chi^0_2
  \tilde \chi^0_2$ with a Wino-like chargino and a Higgsino-like
  neutralino.  

  In the \emph{Light Leptons} model, the heavy gauge bosons almost
  always decay into a lepton and the corresponding heavy lepton partner
  which itself always decays into a lepton and $A_H$. This behavior
  corresponds to a supersymmetry model with very light scalar leptons
  for which there exist specific signal regions in experimental searches
  for electroweakinos.
\item Final states with heavy $q_H$ always produce quarks and $A_H$ in
  their decays and hence result in final states with jets and missing
  transverse momentum. In most cases these decays produce further
  heavy gauge bosons $V_H$ which, as explained above, add more
  leptons, $b$-jets or normal jets to the event. This topology is very
  similar to supersymmetric scalar quark production with either direct
  decays into the lightest supersymmetric particle or with decay
  chains producing further neutralinos and/or charginos in the final
  state. 
\item Final states with the $T$-odd $T^-$ produce final states with SM
  tops and missing transverse momentum, a typical signature of natural
  supersymmetry with a light scalar top.  
\item Final states with the $T$-even $T^+$ not necessarily produce
  missing transverse momentum but instead decay top-like into $b
  W^+$, hence are expected to affect SM top measurements, or decay into
  top + Higgs/gauge boson final states which is a typical feature 
  of models with an extended quark sector. Since processes involving
  $T^+$ have a reduced production cross section, see our earlier
  discussion, and since our searches mostly focus on SUSY-like final
  states, we do not expect these particles to be of great relevance
  for our results. 
\item If $T$-parity is violated by small couplings, we still expect
  the same production and decay topologies as in the $T$-parity
  conserving case which typically produce 2 $A_H$ and the same hard
  final state objects which we listed in the previous
  discussion. However, as now each of these decays into pairs of
  Standard Model particles, many more final state topologies
  appear. Especially if $f \gtrsim \unit[1.2]{TeV}$ we expect four
  Standard Model vector bosons in the final state and as each of these
  can decay hadronically or leptonically, a plethora of possible final
  state exists with various combinations of additional jets and
  leptons. These can be covered by analyses which target very large
  final state multiplicities for which the Standard Model background
  is very small. Furthermore, as both $Z$ and $W$ have sizable decay
  rates into final states with neutrinos, the final states may even
  have a significant amount of missing transverse momentum and thus
  may still be covered by the same supersymmetry-based analysis
  strategies as mentioned for the $T$-parity conserving case. 
\end{itemize}

All in all we expect various final states which are very similar to
those expected in typical supersymmetric models and we expect that
this model can be strongly constrained by applying LHC searches
originally designed to find supersymmetric particles. Even though
theoretically expected, some of these topologies not necessarily will
result in a large enough signal event rate to produce a sensible bound
and/or various topologies appear simultaneously and it is difficult to
say \emph{a priori} which of these topologies is expected to result in
the strongest sensitivity. Fortunately, as many of these searches are
implemented in the tool \Checkmate{}, we expect this tool to perform
very well in our scenarios and determine the respectively strongest
bounds for each benchmark case conveniently. However, not all
conceivable topologies mentioned here have a matching analysis
implemented in \Checkmate{} and therefore most sensitive topologies
determined below might not necessarily correspond to what we 
theoretically expect at this stage. Our previous discussion should
hence be understood as a more general summary of interesting LHC
topologies worthwhile investigating at the Large Hadron Collider out
of which we cover a large fraction with our following \Checkmate{}
analysis. 

\newcommand{\alert}[1]{#1}\begin{table}
\footnotesize
    \begin{tabularx}{\columnwidth}{XXXl}
      \toprule \midrule
      CM identifier & Final State & Designed for & Ref.\\
\midrule
      \texttt{atlas\_conf\_2016\_096} &       \etmiss{} + 2-3 $\ell$  & $\tilde \chi^\pm, \tilde \chi^0, \tilde \ell$ & \cite{ATLAS:2016uwq} \\               
      \texttt{atlas\_conf\_2016\_054} &    \etmiss{} + 1 $\ell$ + (b)-j           & $\tilde q, \tilde g$ & \cite{ATLAS:2016lsr} \\
      \texttt{atlas\_conf\_2017\_022} &    \etmiss{} + 0 $\ell$ + 2-6 j          & $\tilde q, \tilde g$ & \cite{ATLAS:2017cjl} \\
      \texttt{atlas\_conf\_2017\_039} &   \etmiss{} + 2-3 $\ell$               & $\tilde \chi^\pm, \tilde \chi^0, \tilde \ell$ &\cite{ATLAS:2017uun}\\
%      \texttt{atlas\_conf\_2017\_019} &     search for stops with Higgs or Z  (coming soon)   &   \color{green}13\color{black}  &     \alert{36.1}  & 6 \\
%      \texttt{atlas\_conf\_2017\_020} &    search for stop pair production, 0 $\ell$  (coming soon)  &   \color{green}13\color{black} & \alert{36.1}   &   14  \\                       
%      \texttt{cms\_pas\_sus\_15\_011} &                SUSY with 2 leptons + jets + \etmiss{}   &    \color{green}13\color{black} &                           2.2  &   47 \\
      \bottomrule
    \end{tabularx}
\caption{Small summary of all $\sqrt{s} = \unit[13]{TeV}$ analyses
  which appear in the discussion of our results. More details, also on
  other tested analyses, are given in Tab.~\ref{tab:app:analyses} in
  the appendix.} 
\label{tab:analysisleg}
\end{table}

%%%%%

\section{Collider Results from CheckMATE}
\label{sec:results}

\begin{figure*}
\centering
\includegraphics[width=0.45\textwidth]
                {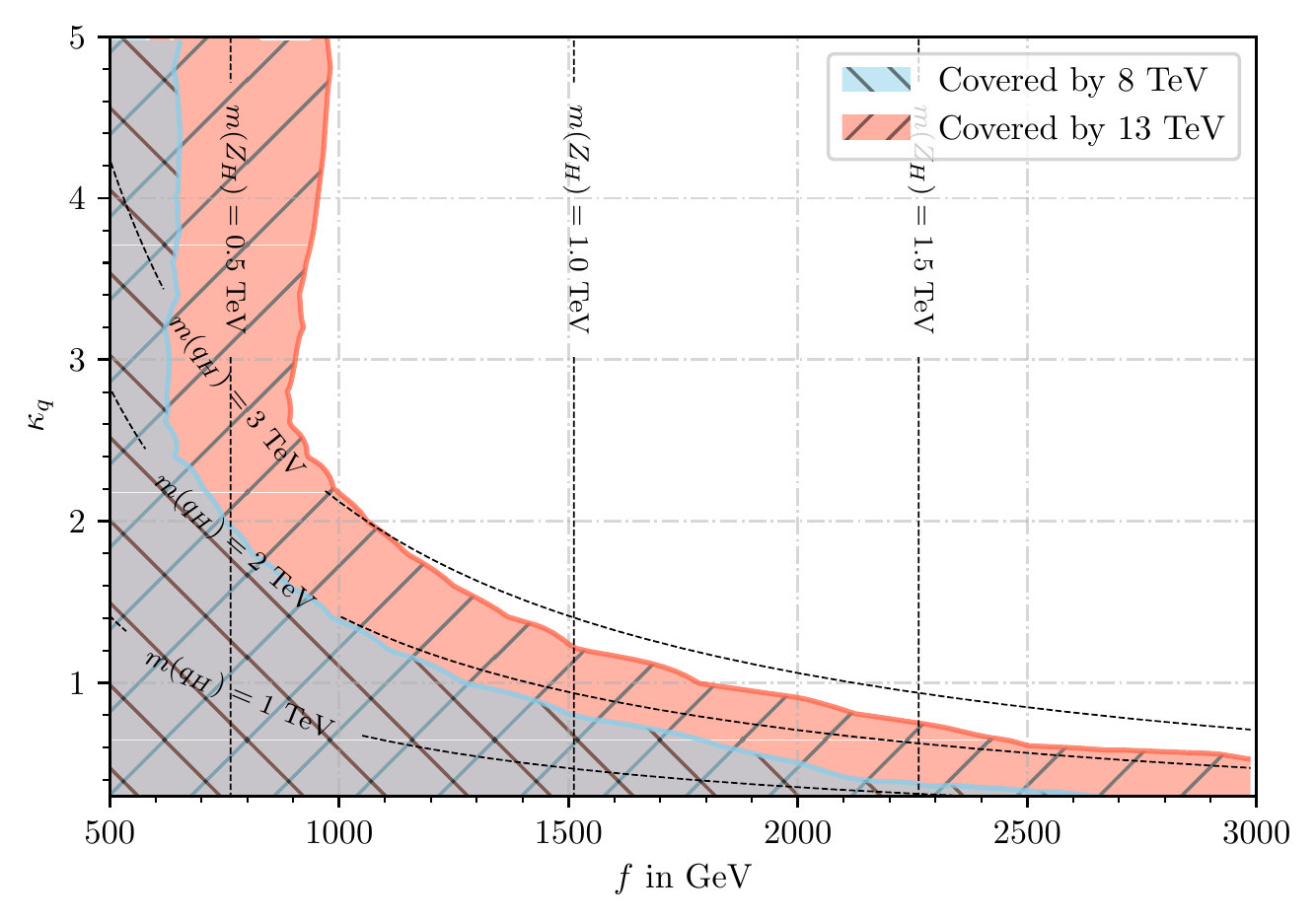} 
\includegraphics[width=0.45\textwidth]
                {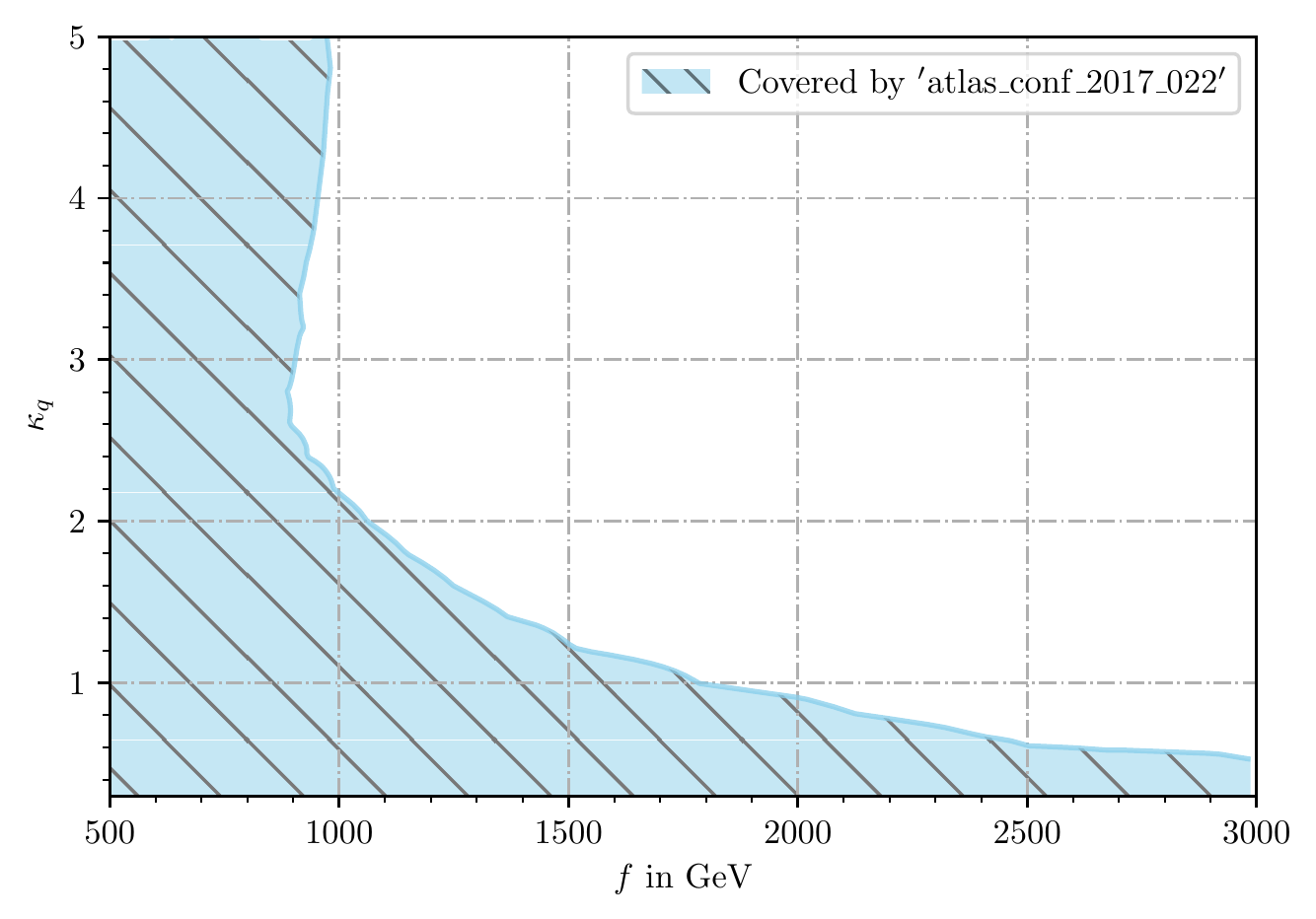}
\caption{Results for scenario (\emph{Fermion
    Universality})$\times$(\emph{Heavy $T^\pm$})$\times$(\emph{TPC})} 
\label{fig:cmresults:univtpcnotop}
\includegraphics[width=0.45\textwidth]
                {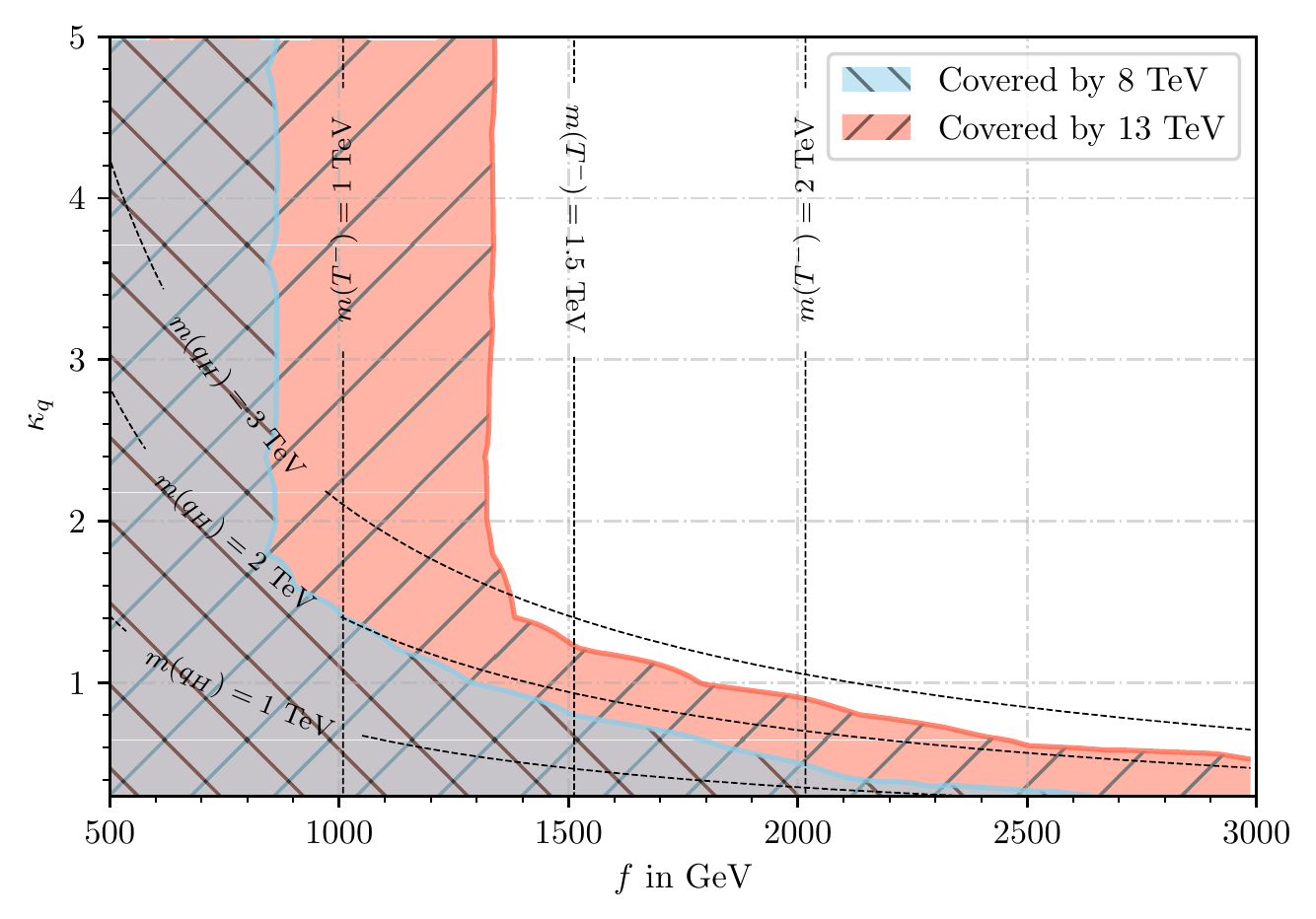} 
\includegraphics[width=0.45\textwidth]
                {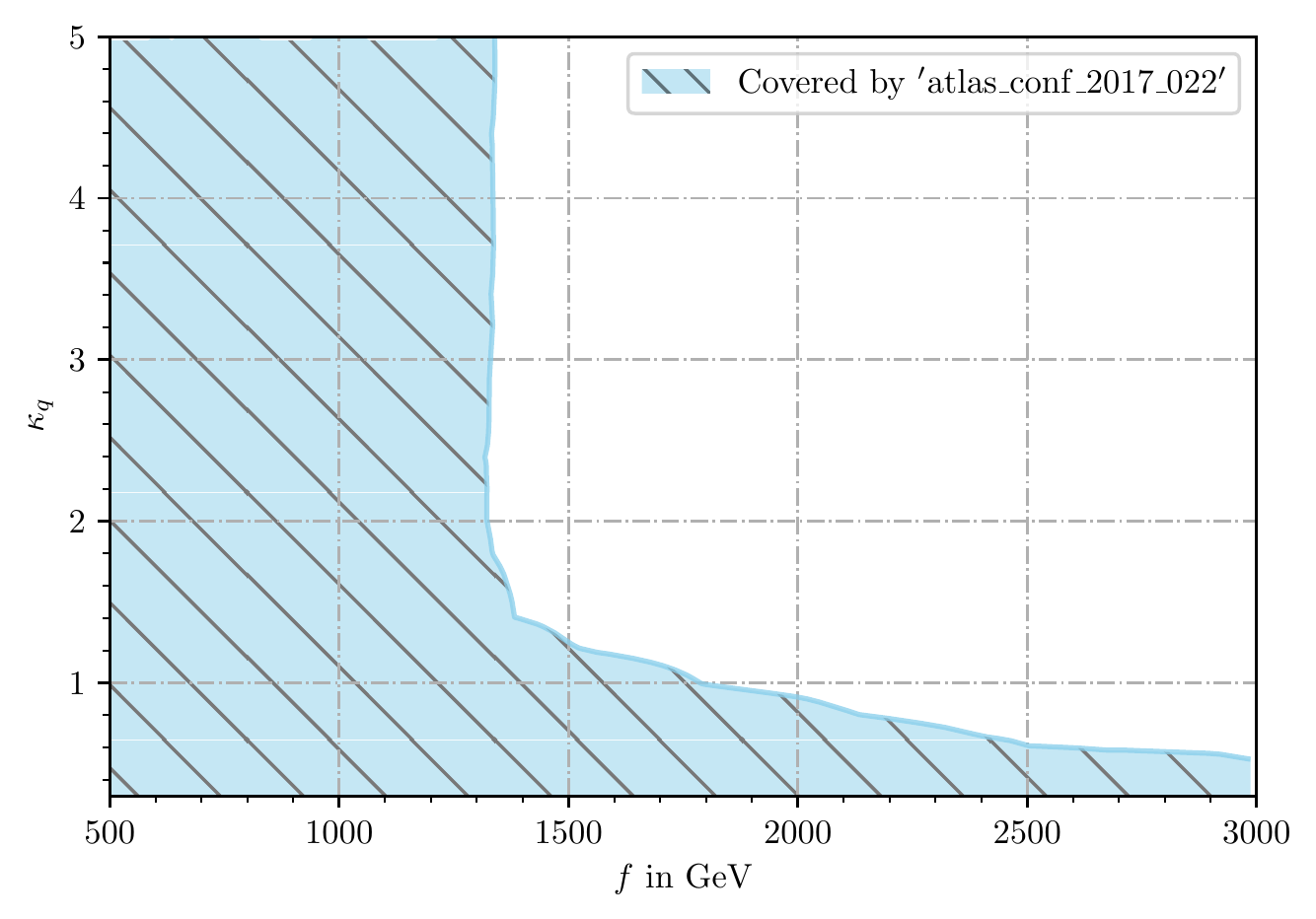}
\caption{Results for scenario (\emph{Fermion
    Universality})$\times$(\emph{Light $T^\pm$})$\times$(\emph{TPC})} 
\label{fig:cmresults:univtpctop}
\includegraphics[width=0.45\textwidth]
                {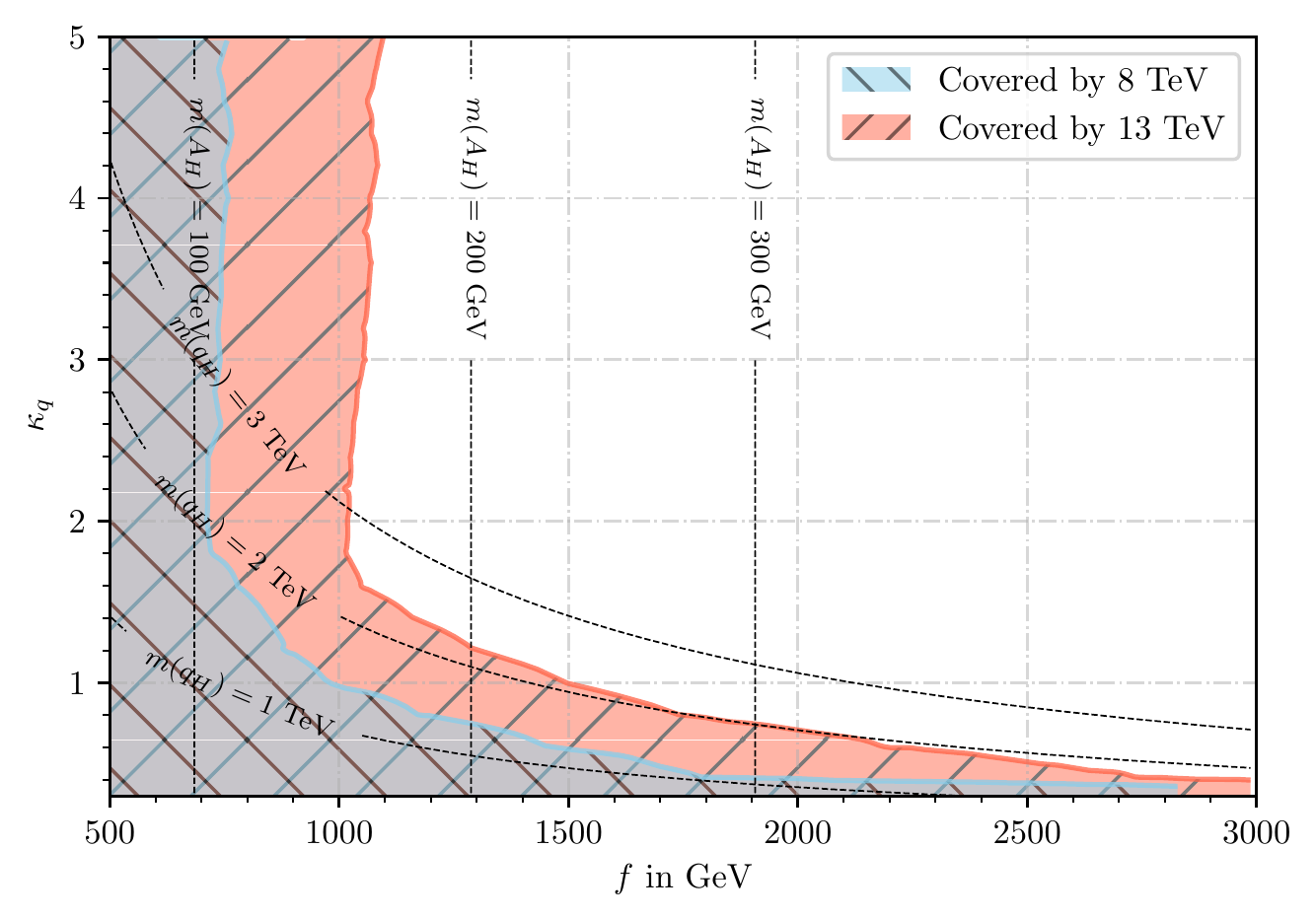} 
\includegraphics[width=0.45\textwidth]
                {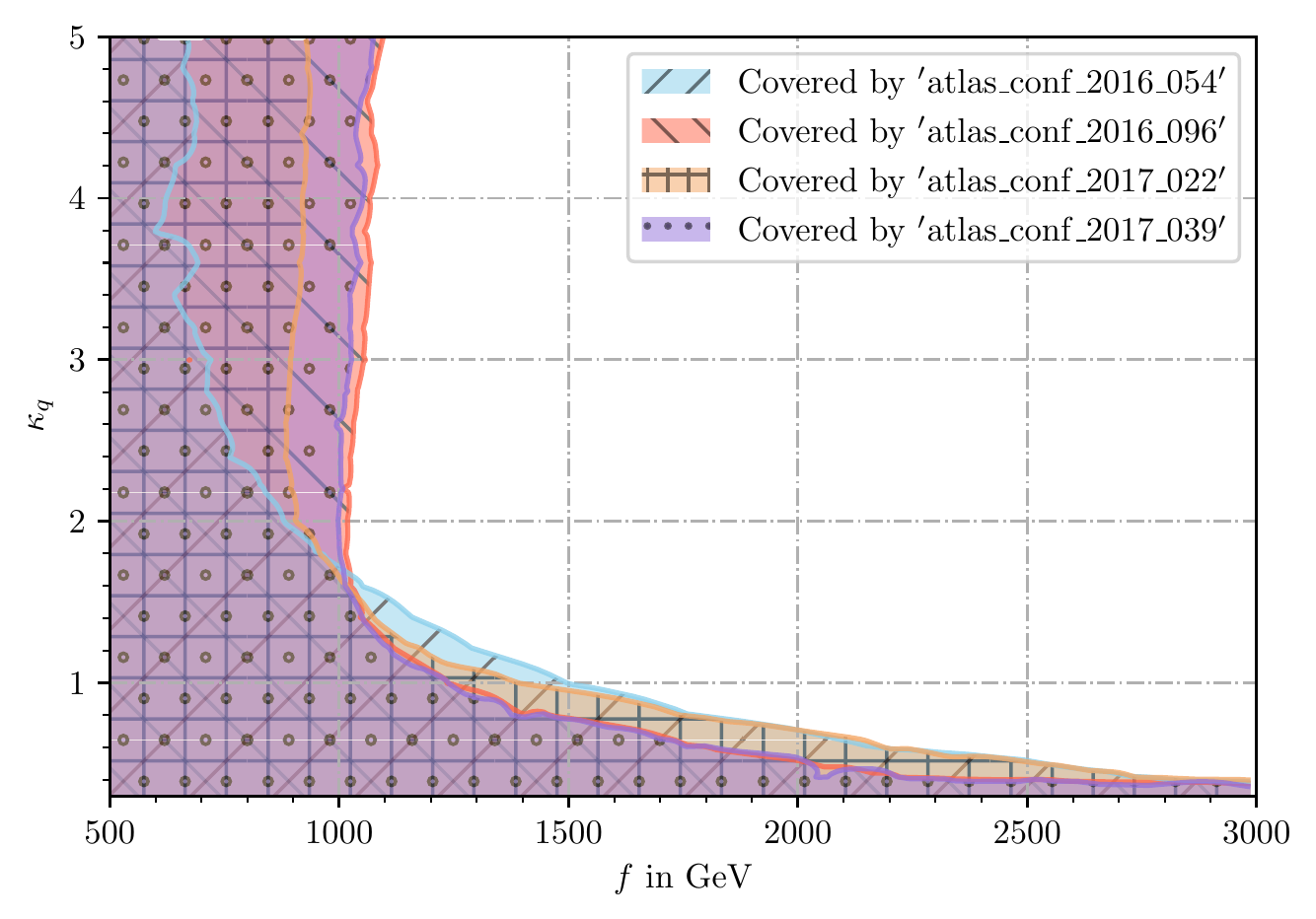}
\caption{Results for scenario (\emph{Fermion
    Universality})$\times$(\emph{Heavy $T^\pm$})$\times$(\emph{TPV})} 
\label{fig:cmresults:univtpvnotop}
\centering
\includegraphics[width=0.45\textwidth]
                {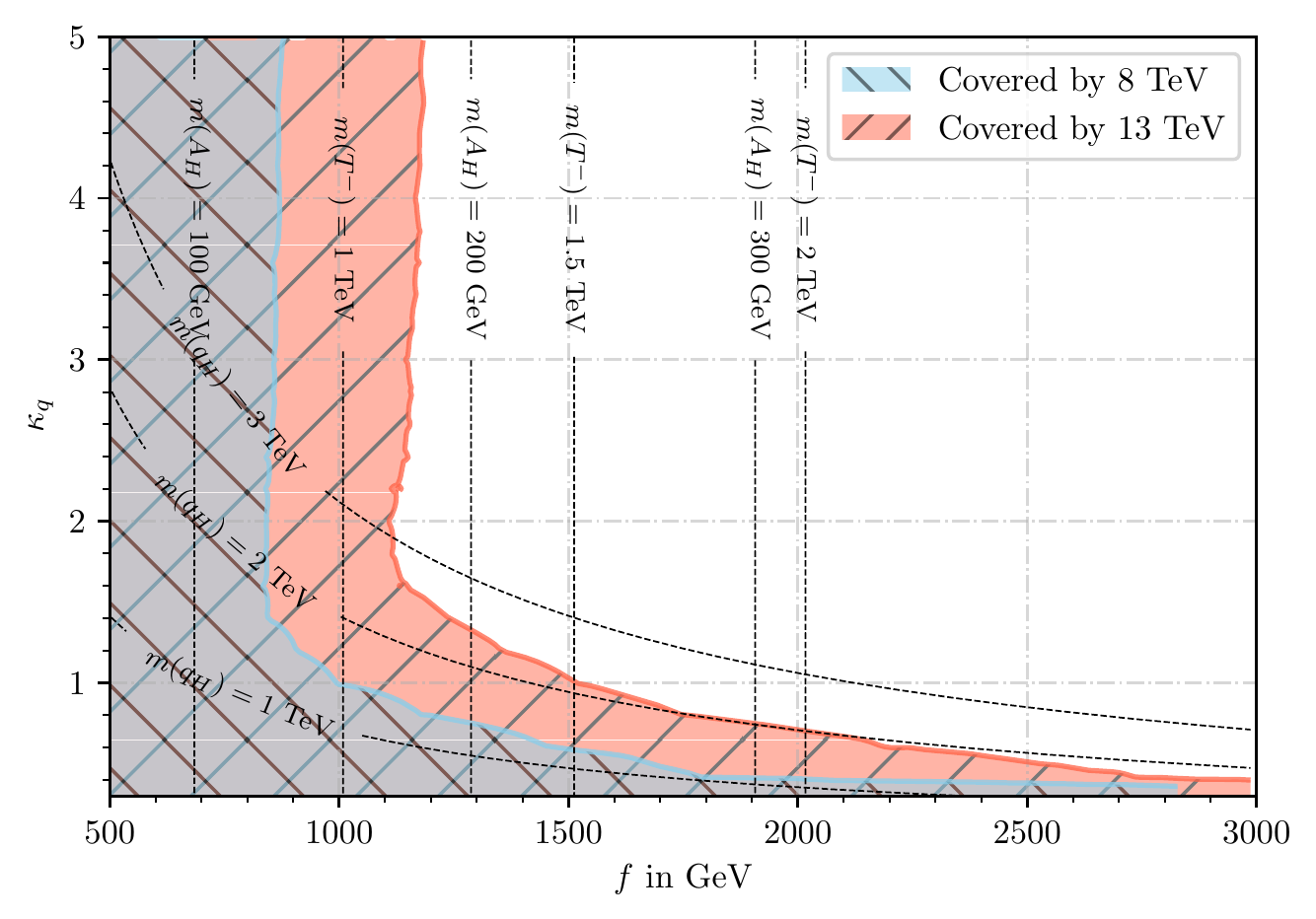} 
\includegraphics[width=0.45\textwidth]
                {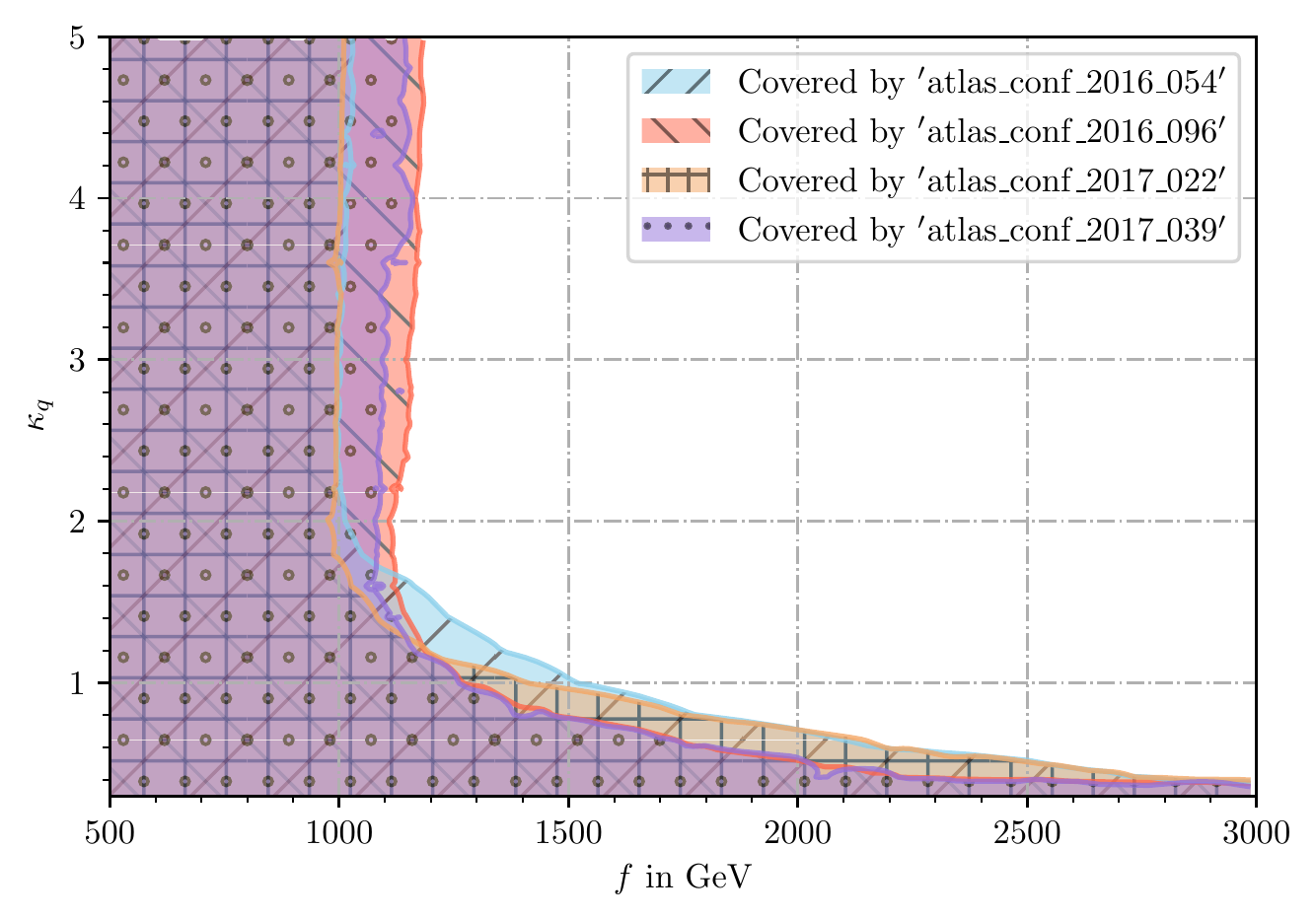}
\caption{Results for scenario (\emph{Fermion
    Universality})$\times$(\emph{Light $T^\pm$})$\times$(\emph{TPV})} 
\label{fig:cmresults:univtpvtop}
\end{figure*}

We now discuss the results of our collider analysis performed with
\Checkmate. Exclusion lines in the $\kappa$-$f$--plane for all $3 \times
2 \times 2$ scenarios are shown in
Figs.~\ref{fig:cmresults:univtpcnotop}-\ref{fig:cmresults:lighttpvtop}. For
each case, we choose two ways to present our results. On the
respective plots in the left column we show the total exclusion line
determined by \Checkmate from LHC analyses at \unit[8]{TeV} and
\unit[13]{TeV}, respectively. The \unit[8]{TeV} results allow direct
comparison to earlier studies,
e.g.\ in~\cite{Reuter:2012sd,Reuter:2013iya}. Drawing 
them in the same plot with the updated \unit[13]{TeV} results
illustrates how the increased energy and the higher integrated
luminosities significantly improve the sensitivity on the Little Higgs
Model with fully or nearly conserved $T$-parity. In the discussion in
the main text of this section we focus on the update from the current
results at $\sqrt{s} = \unit[13]{TeV}$ and will not discuss the
outdated results at \unit[8]{TeV} center-of-mass energy. In the same
set of plots we also show mass contours of the most relevant particles
to understand the bounds. These are  
\begin{itemize}
\item the heavy gauge boson mass $Z_H$ ($=W_H$), 
\item the heavy quarks $q_H$ for all models except \emph{Heavy $q_H$}, 
\item the heavy leptons $\ell_H$ for the model \emph{Heavy $q_H$},
\item the $T$-odd heavy top partner $T^-$ mass for \emph{Light $T^\pm$} benchmarks and
\item the heavy photon mass $A_H$ for \emph{TPV} models.
\end{itemize}
To keep the plots readable we do not show all contours in all
plots. With the exception of $T^\pm$ whose mass values are only
meaningful in the \emph{Light $T^\pm$} scenario, all plots with same
heavy fermion sector scenario (see Table~\ref{tab:benchmarks}) have
the same particle spectrum and therefore, each iso-mass contour can be 
understood to appear in all other plots of the same main benchmark
scenario.   

Alongside the above results we show a second plot each for all benchmark
scenario where we focus on the experimental signature(s) which lead to
the overall bound. For each benchmark study, we show the respective
\Checkmate{} analyses which cover the excluded region at $\sqrt{s} =
\unit[13]{TeV}$. The names in the legend correspond to the
\Checkmate{} analysis identifiers and we provide a small summary of
their respective covered topologies in Tab.~\ref{tab:analysisleg}
for convenience. Note that regions with small $\kappa$ and small $f$
are typically covered by many more LHC analyses but we only show the
minimal set of analyses sufficient to cover the entire excluded
region. A full list of all \checkmate{} analyses that we considered
for this study can be found in Tab.~\ref{tab:app:analyses} in the
appendix~\ref{app:checkmateanal}.

\subsection{Fermion Universality}
\label{sec:res:samekqkl}

We start with a discussion of the \emph{Fermion Universality} model
in which the heavy fermion Yukawa couplings are set to be equal,
$\kappa_q = \kappa_\ell$, and thus features a degenerate spectrum of
heavy quarks and heavy leptons.  

\subsubsection*{$T$-parity conserved and heavy $T^\pm$}

In Fig.~\ref{fig:cmresults:univtpcnotop} we start with the subscenario
of conserved $T$-parity and with the heavy top sector decoupled. The
excluded parameter spaces can be separated into two main regions:  
\begin{itemize} 
\item For large $f \geq \unit[1]{TeV}$, the exclusion line depends
  both on $\kappa$ and $f$ and runs nearly parallel to the iso-mass
  contours of the heavy quarks. It thus nearly follows the inequality
  $f \times \kappa < f\kappa_{\text{max}}$ with $f\kappa_{\text{max}}
  \approx \unit[1.5]{TeV}$ at $\sqrt{s}=\unit[8]{TeV}$ and $\approx
  \unit[2]{TeV}$ at $\sqrt{s}=\unit[13]{TeV}$. The most sensitive
  analysis looks for at least two hard jets and a large amount of
  missing transverse momentum, a topology which in this region appears
  through heavy quark pair production with each heavy quark decaying
  into a quark, an invisible heavy photon and possible additional
  particles via more complicated casscades in the decay, $q_h q_H
  \rightarrow q q A_H A_H + X$. The expected event rate for this
  QCD-induced process mainly depends on the mass of the heavy quarks
  and thus explains why the bound runs nearly parallel to the $q_H$
  iso-mass contours. Still, $t$-channel heavy vector bosons also have
  a small effect on the 
  production cross section and thus the bound drops slightly faster
  with higher $f$, i.e.\ with larger $m(V_H)$, than the $m(q_H$)
  iso-mass contour. The results translate into a bound on $m_{q_H}$ of
  $\geq \unit[3]{TeV}$ for $f \approx \unit[1]{TeV}$ which decreases
  to $m_{q_H} > \unit[2]{TeV}$ for $f \gtrsim \unit[3]{TeV}$. 
\item For smaller values of $f$, the bound becomes nearly independent
  of the specific values of $f$ or $\kappa_q$ and absolutely excludes
  $f > \unit[900]{GeV}$. For large enough values of $\kappa$, the
  heavy quarks are not created abundantly enough and hence we are only
  sensitive to the electroweak production of heavy gauge bosons $V_H$
  whose mass is indepedent of $\kappa_q$. The given limit can then be
  interpreted as an absolute mass bound $m_{Z_H} = m_{W_H} \gtrsim
  \unit[600]{GeV}$. Even though their mass is $\kappa$-independent,
  the bound still becomes stronger for increasing value of
  $\kappa_q$. This is --- see  our discussion in Sec.~\ref{sec:xsects}
  --- due to $\kappa_q$ affecting the mass of the heavy quarks who in
  turn interfere destructively with their contribution to the total
  $V_H V_H$ production cross section. Thus the weakest bound $f >
  \unit[800]{GeV}$ appears for $\kappa_q \approx 2.5$ and improves to
  $f \gtrsim \unit[950]{GeV}$ for $\kappa \gtrsim 5.0$.  

Interestingly, even though the main production channel has changed,
the most sensitive study is the same multijet analysis as before. The
required topology is created from hadronically decaying $W$-bosons in
$W_H \rightarrow W A_H$ and from $b$-jets in the decay to a Higgs
boson of $Z_H \rightarrow h A_H$. 
\end{itemize}

\subsubsection*{$T$-parity conserved and light $T^\pm$}

To see how the sensitivity to the heavy top partners compares
to the previous bound, we show below in
Fig.~\ref{fig:cmresults:univtpctop} the results of the same model, but
now with $R=1.0$ and thus including processes which involve the
production of heavy $T^\pm$. Note that for fixed $R$, the mass of the
$T^\pm$ only depends on $f$ which is why for large $f$ these particles
are not experimentally accessible. Thus, we get the same bound on $(f
\kappa_{\text{max}})$ as explained for the previous benchmark.  

However, if the $T^\pm$ are kinematically accessible they play an
important role for the overall bound. For our special case with
$R=1.0$, we observe that the absolute bound on $f$ increases to $f
\geq \unit[1.3]{TeV}$ and becomes entirely $\kappa$ independent as the
$T^\pm$ production modes, as opposed to the $V_H$ modes discussed
before, do not depend on the heavy quark sector. Again, we observe the
search for multijets plus missing transverse momentum to be most
sensitive for the bound.\footnote{Note that by the time this work was
  completed, the restricted set of analyses implemented in
  \Checkmate{} contained the updated multijet results with an
  integrated luminosity of \unit[36]{fb}$^{-1}$, but had only searches
  for scalar tops implemented which use data from
  \unit[13.3]{fb}$^{-1}$. This may explain why we observe
  multijet final states to be most sensitive even though in
  Sec.~\ref{subsec:topologies} we expected heavy top partners to produce
  distinct decay signatures which mimic scalar top decays in natural
  supersymmetry.} 

Clearly, the precise value of the lower limit on $f$ depends on the
mass of the heavy top partner particles which implicitly depends on
the value of $R$. We emphasize here that the choice $R=1.0$ just
serves as a benchmark case and any other $R$ value would directly
affect the bound, see Eq.~(\ref{eq:heavytopmass}), in either
direction. We chose $R=1$ here for the reason that it is rather
special as it minimizes the LHT contributions to the EWPO,
cf. Sec.~\ref{sec:EWPO}.  Our more
general conclusion from this benchmark study is 
thus that searches for $V_H$ and for  $T^\pm$ can yield competetive
absolute lower bounds on $f$, and while the bound derived from $V_H$
production is nearly independent of the chosen benchmark, the presence
of light top partners may put further constraints on the model. 

\subsubsection*{$T$-parity violated}

\begin{figure*}
\centering
\includegraphics[width=0.45\textwidth]
                {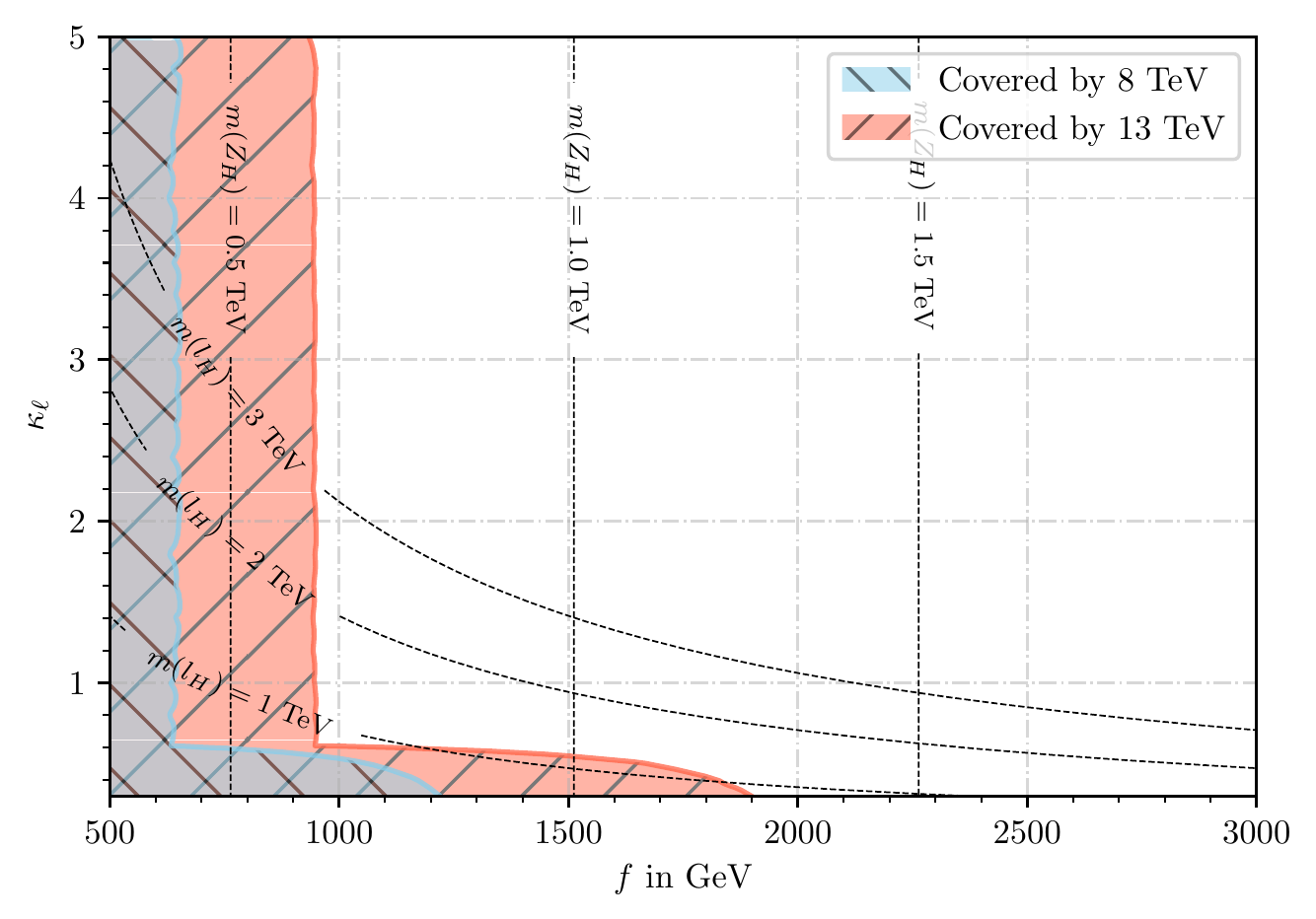} 
\includegraphics[width=0.45\textwidth]
                {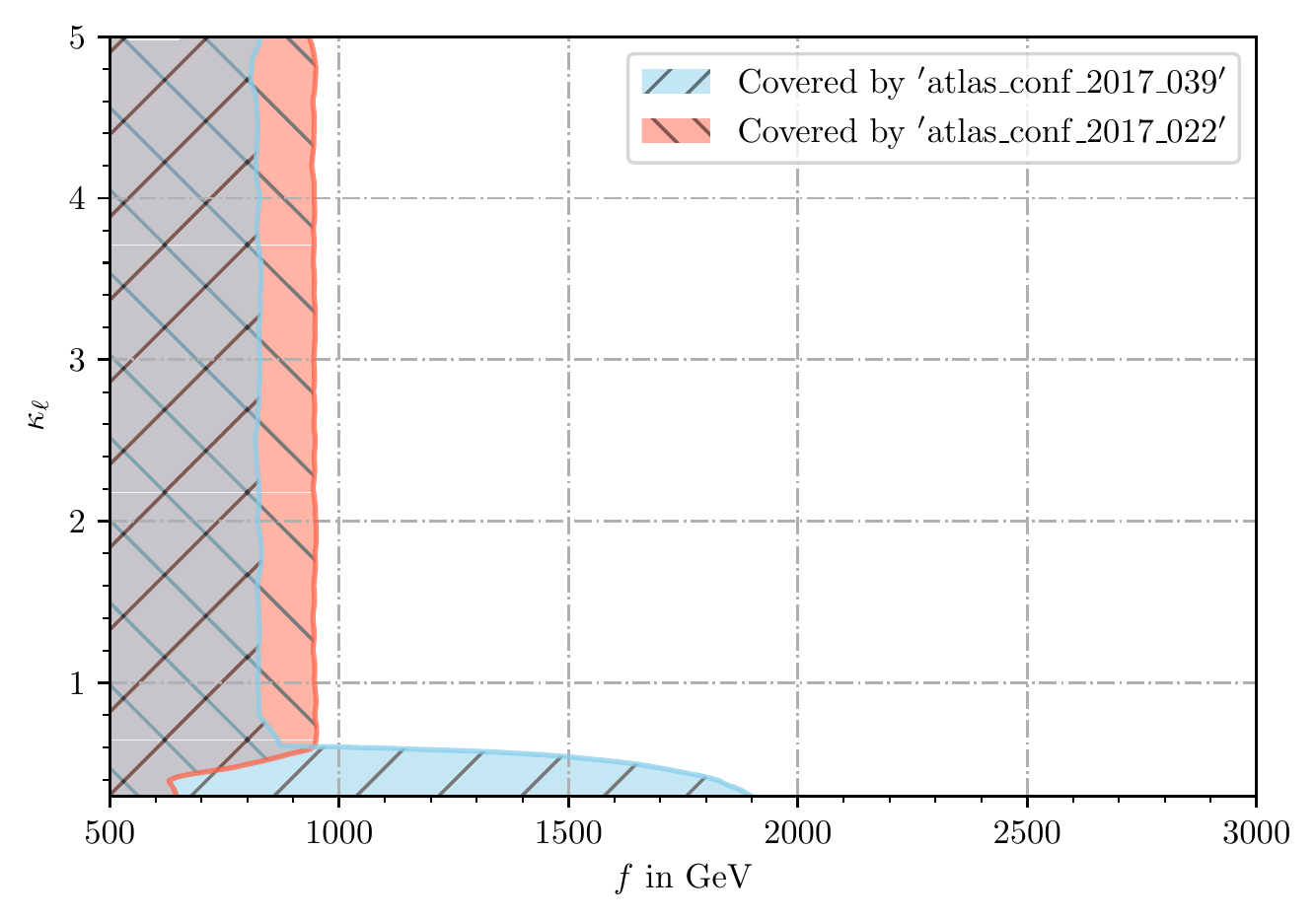}
\caption{Results for scenario (\emph{Heavy $q_H$})$\times$(\emph{Heavy $T^\pm$})$\times$(\emph{TPC})}
\label{fig:cmresults:heavytpcnotop}
\centering
\includegraphics[width=0.45\textwidth]
                {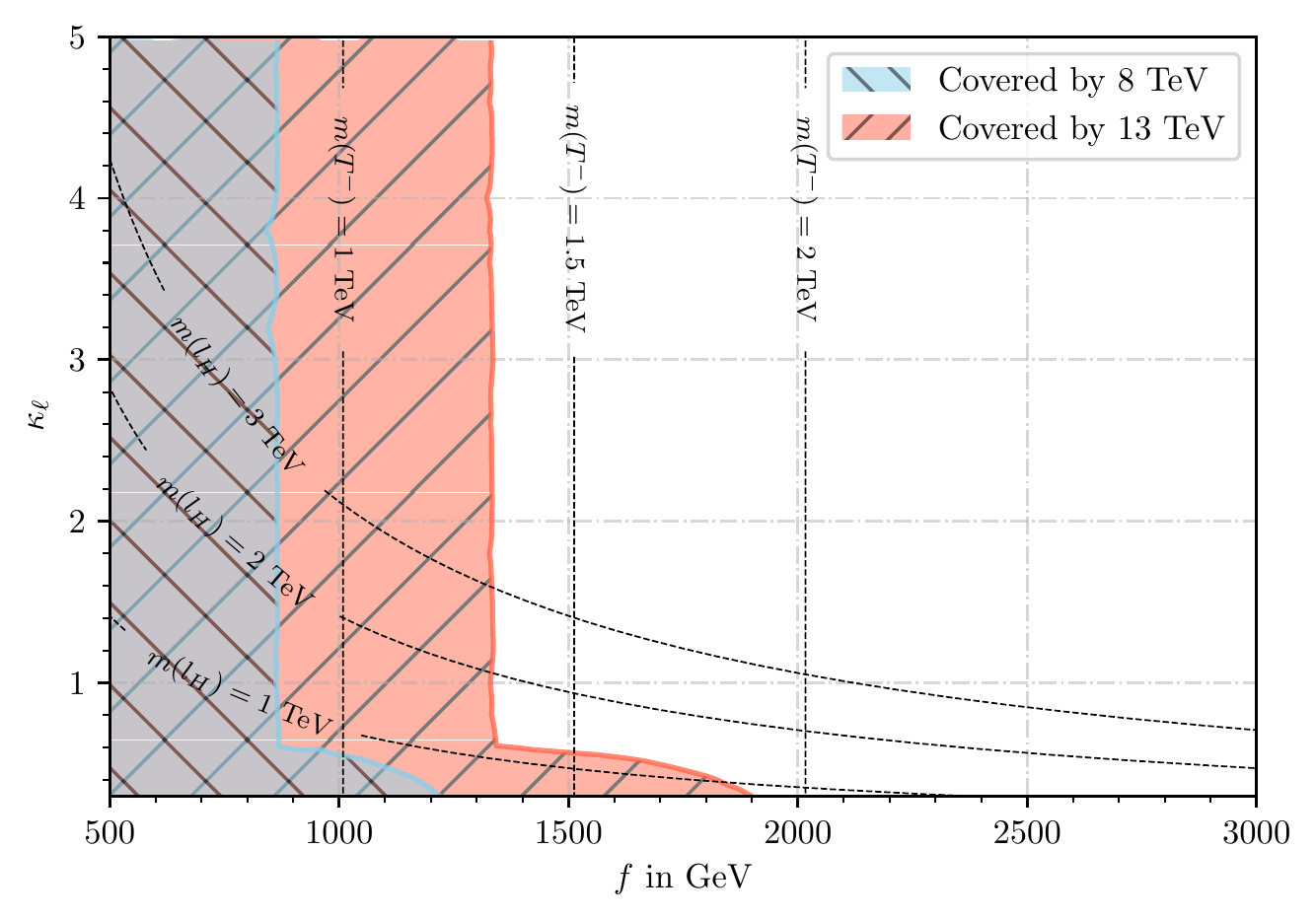} 
\includegraphics[width=0.45\textwidth]
                {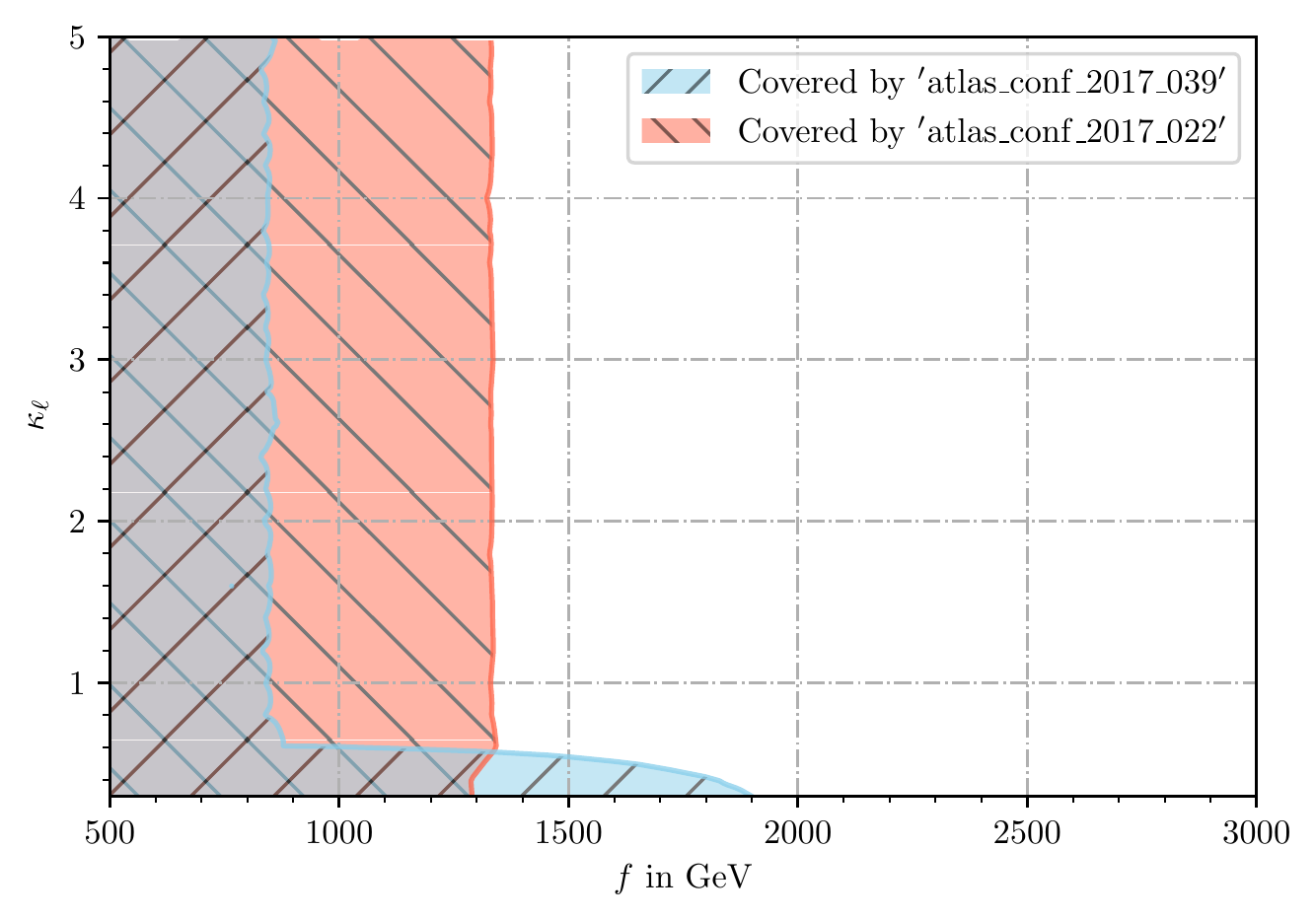}
\caption{Results for scenario (\emph{Heavy $q_H$})$\times$(\emph{Light
    $T^\pm$})$\times$(\emph{TPC})} 
\label{fig:cmresults:heavytpctop}
\includegraphics[width=0.45\textwidth]
                {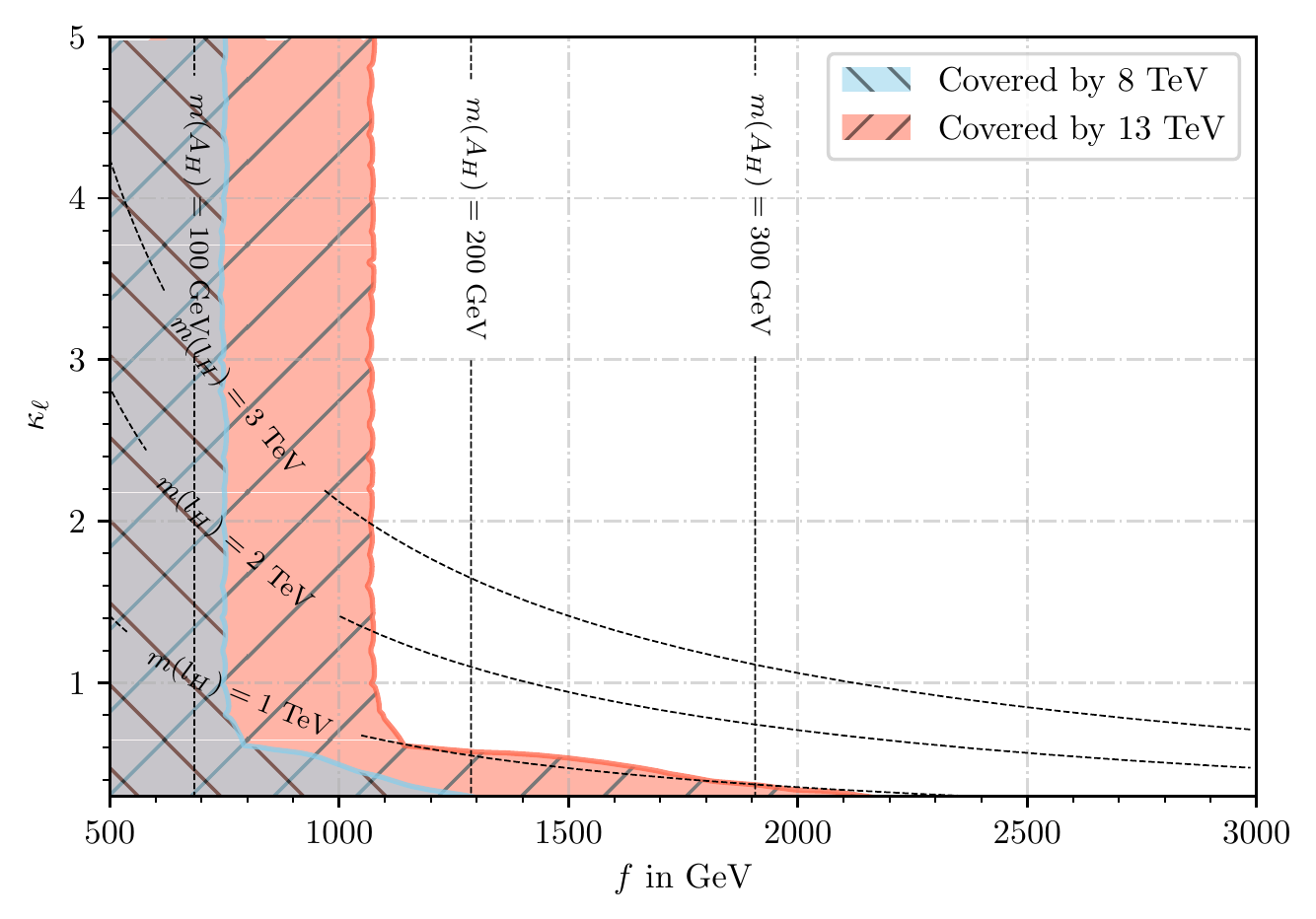} 
\includegraphics[width=0.45\textwidth]
                {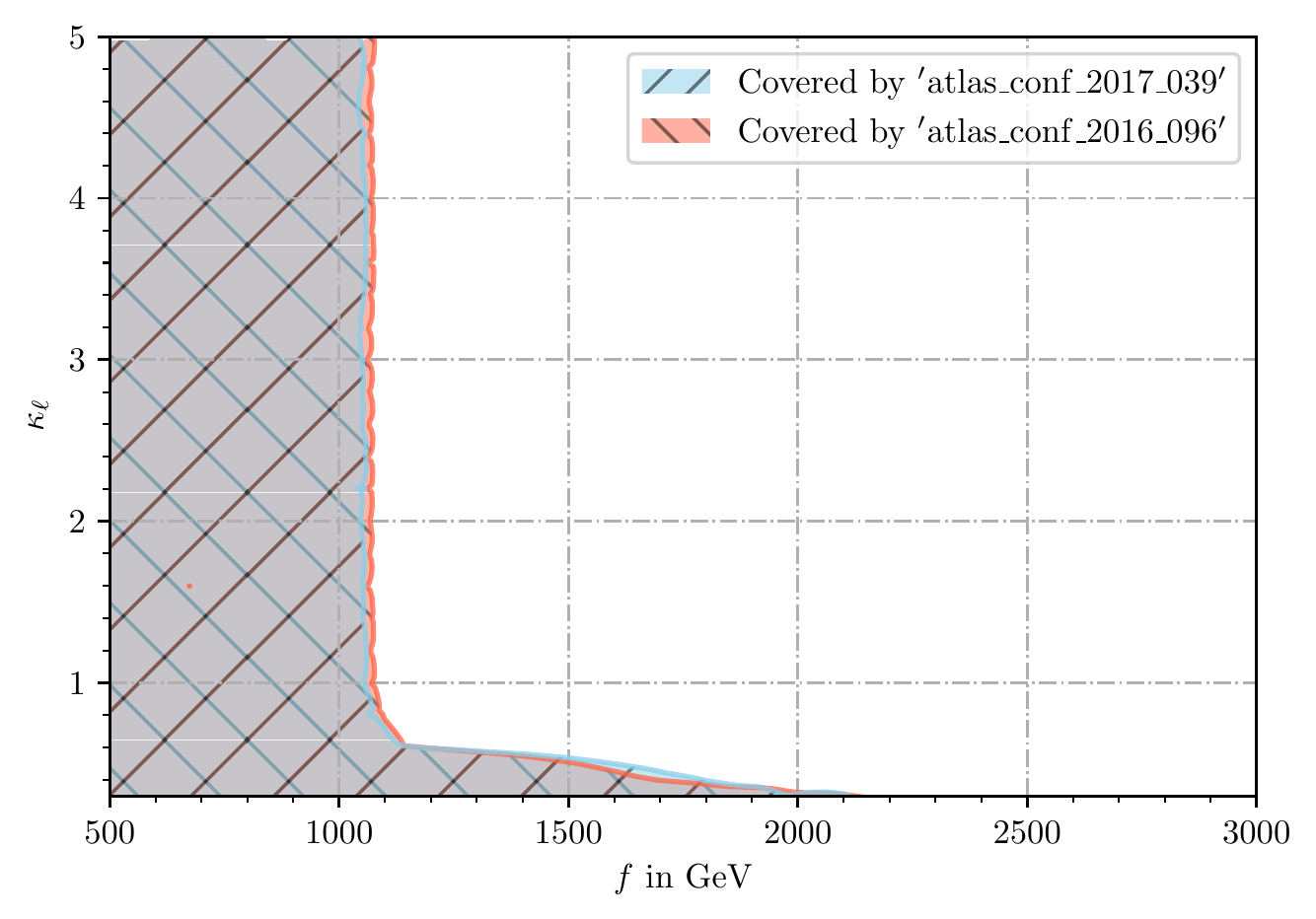}
\caption{Results for scenario (\emph{Heavy $q_H$})$\times$(\emph{Heavy
    $T^\pm$})$\times$(\emph{TPV})} 
\label{fig:cmresults:heavytpvnotop}
\centering
\includegraphics[width=0.45\textwidth]
                {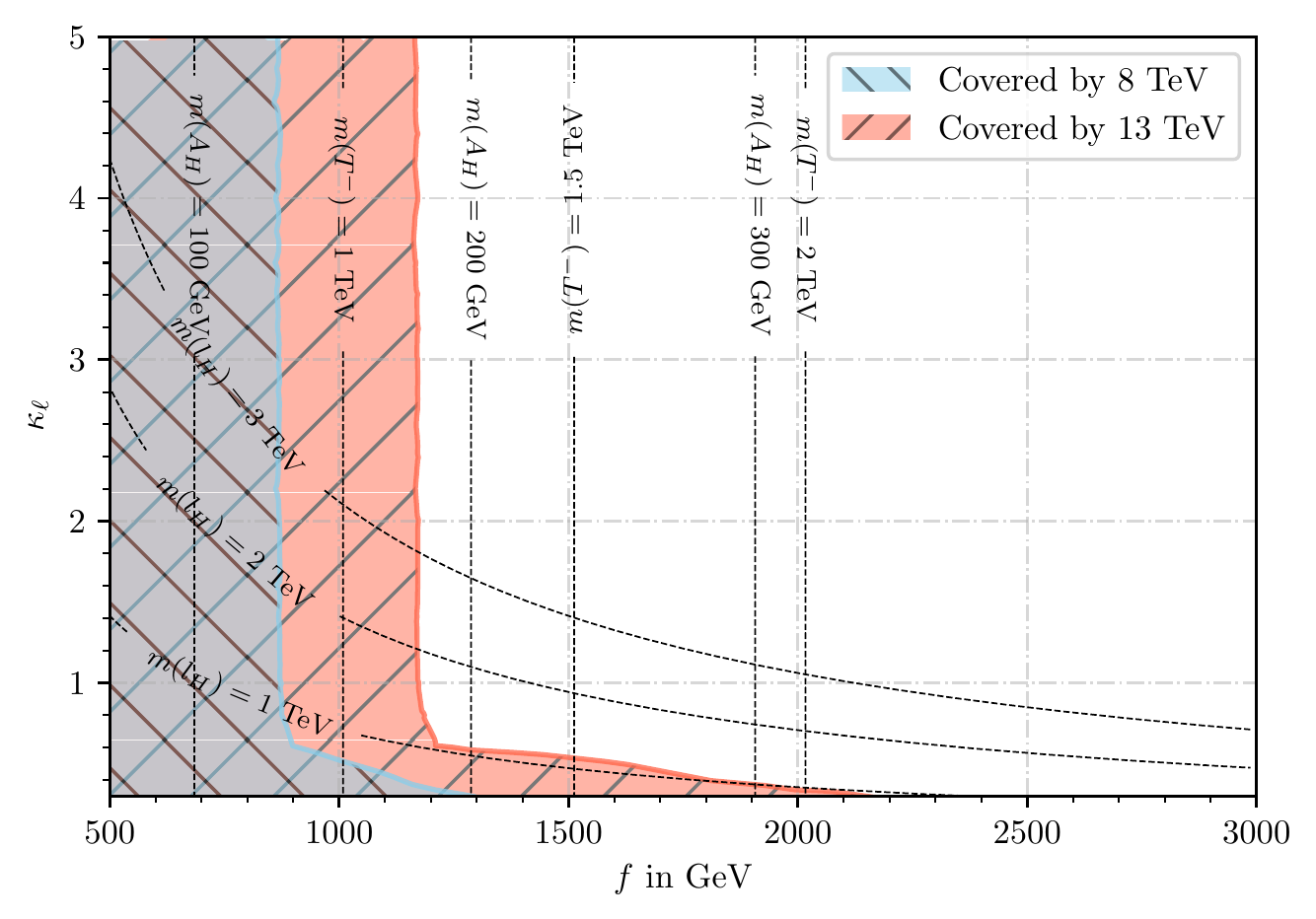} 
\includegraphics[width=0.45\textwidth]
                {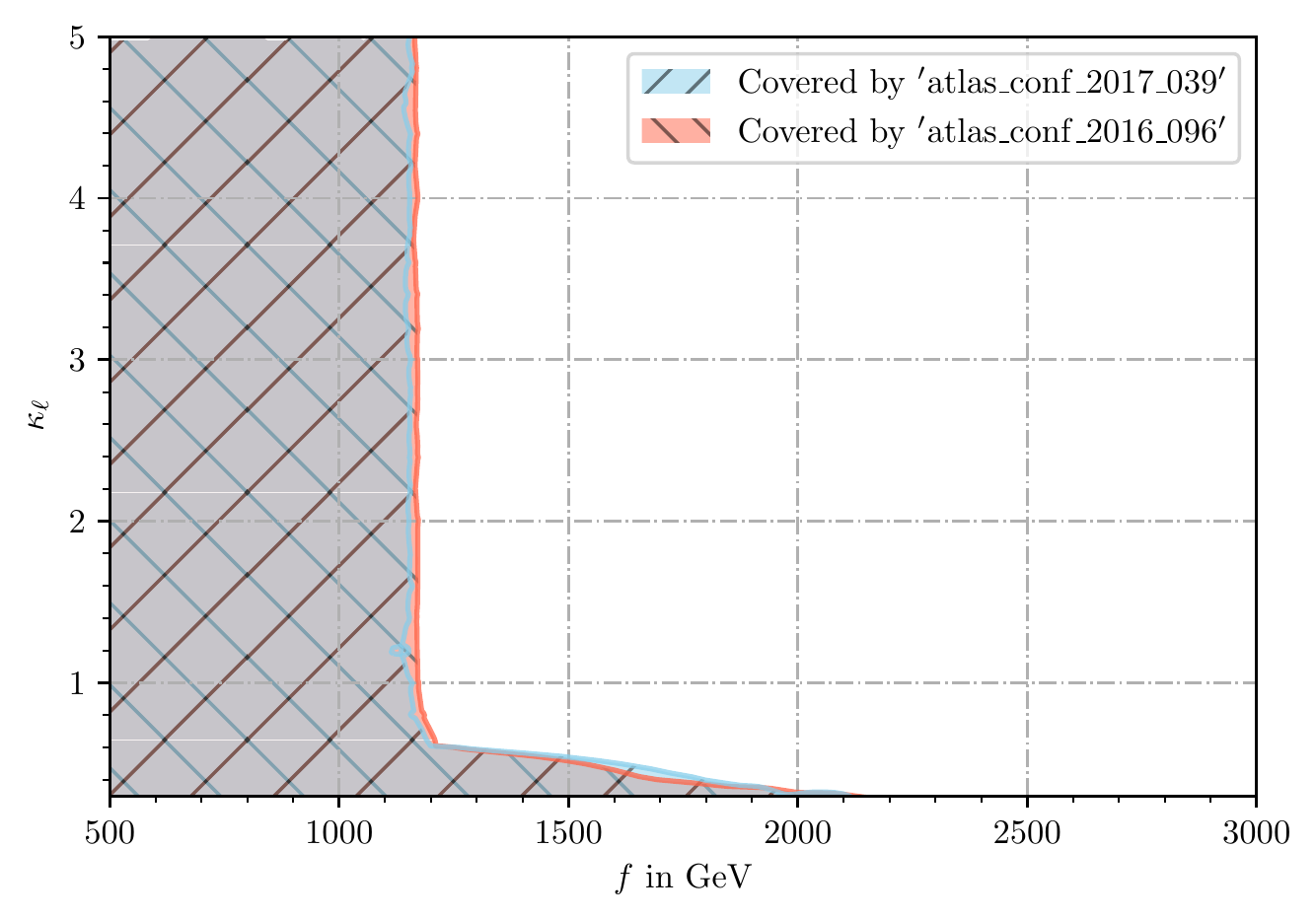}
\caption{Results for scenario (\emph{Heavy $q_H$})$\times$(\emph{Light
    $T^\pm$})$\times$(\emph{TPV})} 
\label{fig:cmresults:heavytpvtop}
\end{figure*}

In Figs.~\ref{fig:cmresults:univtpvnotop},
\ref{fig:cmresults:univtpvtop}  we show the results in case we include
the anomaly-mediated decays of the heavy photon $A_H$ into vector
boson or lepton pairs, both without
(Fig.~\ref{fig:cmresults:univtpvnotop}) and including
(Fig.~\ref{fig:cmresults:univtpvtop}) the heavy top sector. We again
split the discussion into the two main parameter regions already
discussed before: 
\begin{itemize}
\item We again observe a $\kappa$-dependent bound for large values of
  $f$ which follows the iso-mass contour of the heavy quarks. However,
  compared to the $T$-parity conserving case the bound is now slightly
  weaker, $m_{q_h} \geq \unit[2.5]{TeV}$ for $f \approx \unit[1]{TeV}$
  and $m_{q_h} \geq \unit[1.5]{TeV}$ for $f \approx
  \unit[3]{TeV}$. There are two analyses with nearly identical
  sensitivity in this region, namely the already discussed
  zero-lepton--multijet plus \etmiss{} analysis and the related
  multijet analysis which requires one lepton in the final
  state. The fact that their sensitivity is fairly similar
  can be qualitatively understood from the fact that we
  expect many additional final state gauge bosons which produce
  additional leptons and/or jets. Thus, both multijet studies with and
  without leptons become sensitive and we get an overall similar
  signal event rate in the respective signal regions of these two
  studies. In fact, as the branching ratio to $WW$ increases for
  smaller $f$, see Fig.~\ref{fig:cm:br22}, and as $W$-bosons produce
  on average more charged leptons than $Z$-bosons, we expect analyses
  which require a final state lepton to become slightly more sensitive
  for smaller $f$ --- a feature which we exactly observe in our
  results in Fig.~\ref{fig:cmresults:univtpvnotop}, on the right hand
  side.  

At first, it appears unexpected that the bound is not significantly
weakened, even though the originally invisible $A_H$ now decays into
Standard Model particles and thus appears to remove crucial missing
transverse momentum from the event. However, one should bear in mind
that we expect four additional boosted gauge bosons, two from each
$A_H$, in the final state. Thus we expect to pass the \etmiss{}
constraints if at least one of these decays into neutrinos. Even
though on average the branching ratio $V \rightarrow \nu + X$ is only
around \unit[25]{\%}, as we have four gauge bosons the probability of
having an $A_H A_H$ pair decaying into at least one neutrino and thus
producing \etmiss{} is above \unit[70]{\%}. This reduces the \etmiss{}
cut acceptance slightly but not drastically compared to the $T$-parity
conserving case. Furthermore, we get the same visible final state
objects as in the $T$-parity conserving case, together with additional
boosted particles from the  gauge boson decays which may even improve
the final state acceptance. It thus can be understood why the
sensitivity does not drop significantly if $T$-parity violation is
considered.

\item Similarly to before, for a symmetry breaking scale $f$ of the
  order \unit[1]{TeV} we observe a $\kappa$ independent
  bound. Interestingly, the bound has even improved after turning on
  $T$-parity violation and excludes $f \gtrsim \unit[1]{TeV}$ for
  $\kappa \approx 1.5$ and $f \gtrsim \unit[1100]{GeV}$ for $\kappa
  \approx 4.0$. To understand why the limit becomes stronger one needs
  to look at the analysis coverage map on the right of
  Fig.~\ref{fig:cmresults:univtpvnotop}. We see that the bound derived
  from the multijet analysis, which was most sensitive in the
  $T$-parity conserving case, slightly weakened. This can be
  understood with the same arguments as given before for the large-$f$
  region. However, we also observe that the sensitivity is now
  dominated by electroweakino-motivated searches, more specifically by
  analyses which look for final state leptons and missing transverse
  momentum. A more detailed look in the results of that analysis
  reveals that it is in fact the signal region \texttt{SR-Slep-e} which
  produces the bound. This signal region requires 3 high-$p_T$ charged
  leptons which do not originate from a leptonically decaying
  $W$-$Z$-pair and a significant amount of missing transverse
  momentum. Interestingly, such a signature could not be reached in
  the previous $T$-parity conserving benchmark case, because the most
  important topology $p p \rightarrow W_H W_H \rightarrow W W A_H A_H$
  only produces two leptons. Including $T$-parity violation, we can
  get a third, highly energetic lepton if one of the four final state
  gauge bosons is a leptonically decaying $W$. Furthermore, since this
  signal region has no constraints on the final state jet
  multiplicity, the decays of the other three gauge bosons is
  irrelevant. As such, a large signal event rate is expected for this
  analysis if $T$-parity is violated.  

If the top partners are kinematically accessible, see
Fig.~\ref{fig:cmresults:univtpvtop}, the absolute bound on $f$ only
increases slightly by about \unit[100]{GeV}. The electroweak search
stays the most sensitive analysis for this model. The resulting bounds
increase as  more events from the topology $ p p > T^- \bar{T}^-
\rightarrow (bW) (bW) WW VV $ are expected. Again, the impact on the
bound depends on the precise value of $R$ and we only show one example
here which illustrates that the details of the heavy top partner
sector are relevant for the overall LHC limit.  

Interestingly, the multijet analysis does not seem to get a
significant contribution from the presence of the $T^\pm$ even though
it did in the previous case when $T$-parity was conserved,
cf. Figs.~\ref{fig:cmresults:heavytpcnotop},
\ref{fig:cmresults:heavytpctop}. To understand this behavior one
needs to consider the details of the experimental search: this
analysis tries to cover various hierarchies and decay topologies that
can appear in the supersymmetric squark-gluino $\tilde g,\tilde q$
sector and defines many signal regions which target different jet
multiplicities. Different mass scales in the supersymmetric sector are
taken into account by gradually increasing the requirements on the sum
of jet $p_T$ in the event as well as the total amount of \etmiss, more
specifically by using cuts which require minimum values for the ratio
\etmiss/$\sum($jet $p_T)$.  In supersymmetry, jet multiplicity, total
hadronic energy and missing transverse momentum increase
simultaneously as heavier particles on average produce longer decay
chains and give more momentum to the visible jets and the invisible
neutralino and thus a cut on \etmiss/$\sum($jet $p_T)$ has a good
signal acceptance in supersymmetry. 

However, such a cut is disadvantageous for our most important topology
$T^\pm \rightarrow t A_H$ if $A_H$ decays via TPV: the additional
decay of $A_H$ into gauge bosons is expected to produce a
significantly larger amout of jets and hadronic energy while
reducing the amount of missing transverse momentum, resulting
in a large drop in the signal acceptance. Therefore adding the $T^\pm$
to the experimentally accessible spectrum hardly increases the amount
of signal events in this case and the bound only improves little. 

\end{itemize}

\subsection{Heavy $q_H$}
We continue with the discussion of the results for the \emph{Heavy
  $q_H$} scenario which fixes $\kappa_q$ to 3.0 and thus effectively
decouples the $q_H$ from the experimental reach. The results for all
subscenarios (with/without $T$-parity violation and ex-/including the
heavy top partner sector) are shown in
Figs.~\ref{fig:cmresults:heavytpcnotop}-\ref{fig:cmresults:heavytpvtop}. The
plots show the same information as in the previous section
\ref{sec:res:samekqkl}, however note that the ordinate is  now chosen
to be the free parameter $\kappa_\ell$ and the iso-mass contours are given
for the $\ell_H$ instead of the $q_H$ now.\footnote{As the mass of the
  $\ell_H$ and $q_H$ are identical for $\kappa_q = \kappa_\ell$, see
  Eq.~(\ref{eq:qHmasses}), the iso-mass contours for $\ell_H$ appear
  at the same position as those for $q_H$ in the previous benchmark.} 

To understand how the bounds change compared to the previous benchmark
scenario, it is worth repeating the two main phenomenological
consequences of this benchmark case: 
\begin{enumerate}
\item $q_H \rightarrow q V_H$ topologies are replaced by $\ell_H
  \rightarrow \ell V_H$. Multijet final states are thus replaced by
  multilepton final states. As the production cross section for
  $\ell_H \ell_H$ is 2 to 3 orders of magnitude smaller than the
  corresponding cross section for $q_H q_H$, we expect a far weaker
  sensitivity in the heavy fermion dominated region (i.e. large $f$,
  small $\kappa$).
\item $\sigma(p p \rightarrow V_H V_H)$ was dependent on $\kappa$ but
  is independent of $\kappa_\ell$ as no contributions from $t$-channel
  $q_H$ exist in this benchmark case. Thus we expect the bounds
  produced from $V_H$ pair production to be entirely $\kappa_\ell$
  independent and very similar to the case $\kappa = 3.0$ of the
  previous benchmark. 
\end{enumerate}

With these pieces of information in mind, the results in
Figs.~\ref{fig:cmresults:heavytpcnotop}-\ref{fig:cmresults:heavytpvtop}
compare straightforwardly to the bounds of the earlier benchmark
scenario in
Figs.~\ref{fig:cmresults:univtpcnotop}-\ref{fig:cmresults:univtpvtop}:
\begin{itemize}
\item For $f \approx \unit[1]{TeV}$, vector boson production and
  potential heavy top partner production are the most sensitive
  channels and they produce $\kappa_\ell$ independent bounds of $f
  \gtrsim \unit[950]{GeV}$ (TPC, no $T^\pm$), $f \gtrsim
  \unit[1350]{GeV}$ (TPC, with $T^\pm$), $f \gtrsim \unit[1100]{GeV}$
  (TPV, no $T^\pm$) and $f \gtrsim \unit[1200]{GeV}$ (TPC, with
  $T^\pm$). The bounds correspond to those for the previous benchmark
  for large values of $\kappa \gtrsim 4.0$. The most dominant
  topologies also do not change: we observe multijet final states to
  be the most sensitive ones in case $T$-parity is conserved while
  multilepton final states become more important if $T$-parity is
  violated. 
\item For $\kappa_\ell \lesssim 0.5$, the mass of the $\ell_H$ drops
  below the mass of the heavy vector bosons and thus decays of type
  $V_H \rightarrow \ell_H \ell$ can happen, see
  Fig.~\ref{fig:cm:br2324}. The boosted final-state leptons of this
  decay can be observed via a multilepton analysis as can be seen in
  the right of Fig.~\ref{fig:cmresults:heavytpcnotop}. This
  significantly improves the sensitivity and improves the bound on $f$
  to up to \unit[1.9]{TeV}. As the branching ratios depend on
  $\kappa$, this bound is now slighly dependent on $\kappa$. 
\item The ``$f \kappa_{\text{max}}$''-bound which we were able to set
  in the previous benchmark almost disappears for this scenario where
  the $q_H$ are decoupled. The expected event rates from $\ell_H
  \ell_H$ pair production are so small that no feasible bound can be
  set from this topology in case of $T$-parity conservation, even with
  the newest $\sqrt{s}=\unit[13]{TeV}$ results. It is only in the case
  of $T$-parity violation that we can observe an exclusion for very
  small values of $\kappa$ which follows the $m(\ell_H) =
  \unit[1]{TeV}$ mass contour, caused by a slight increase of the
  expected multilepton event rates from leptonic gauge boson decays,
  see our discussion above. 
\end{itemize}

All in all we observe that the presence or absence of the $q_H$
partner particles plays a very important role for determining the LHC
limits in the low $\kappa$ region, i.e.\ for $\kappa \lesssim
1.5$. However, the heavy gauge boson sector also puts very important
constraints on $f$ and as the collider phenomenology of this sector is
almost, but not completely, independent of the heavy fermion sector, the
absolute bounds on $f$ are very robust against choices for the heavy
quark sector. In fact, they tend to become stronger as the presence of
light $q_H$ decreases the $V_H V_H$ production cross section. 
\begin{figure*}
\centering
\includegraphics[width=0.45\textwidth]
                {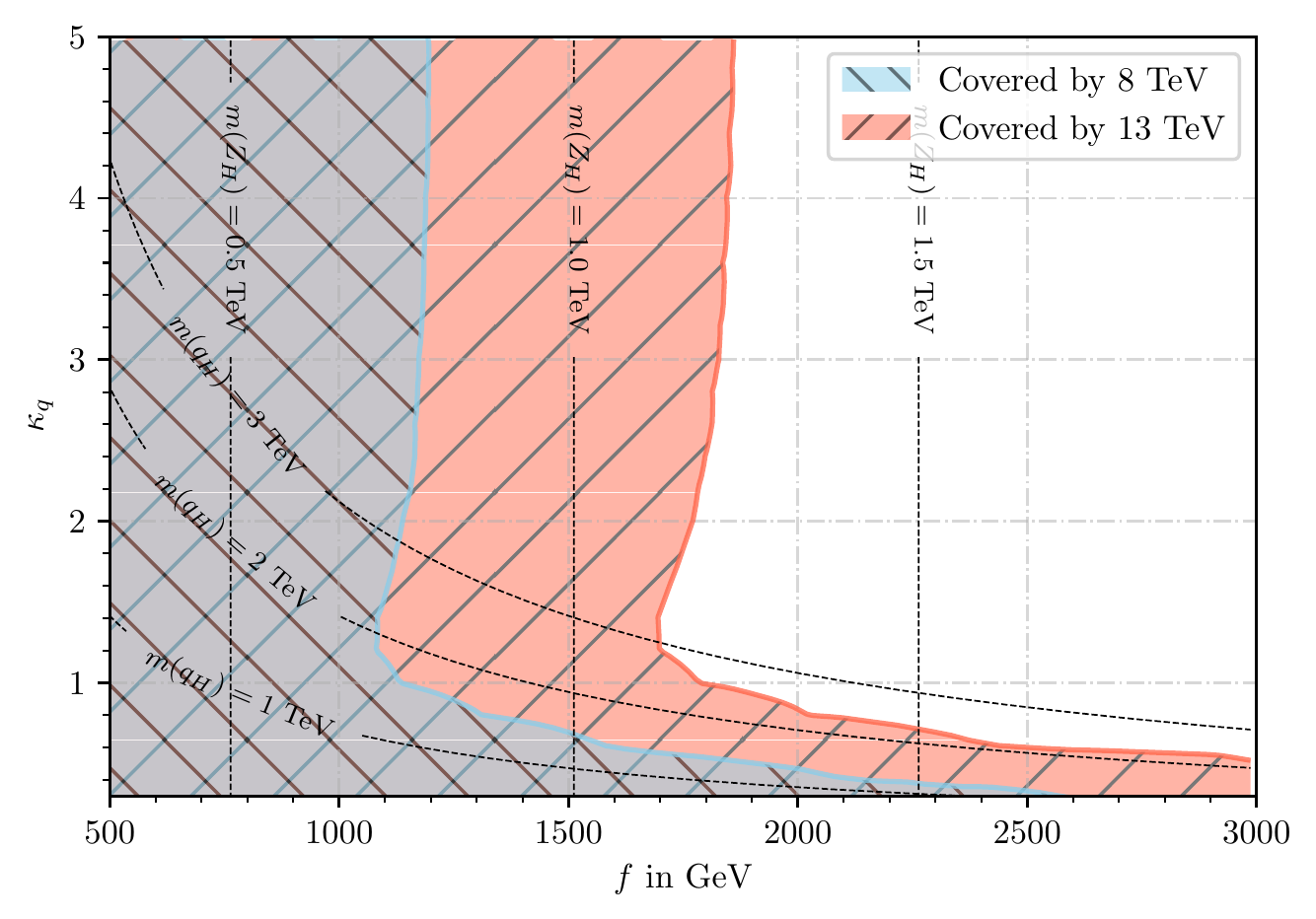} 
\includegraphics[width=0.45\textwidth]
                {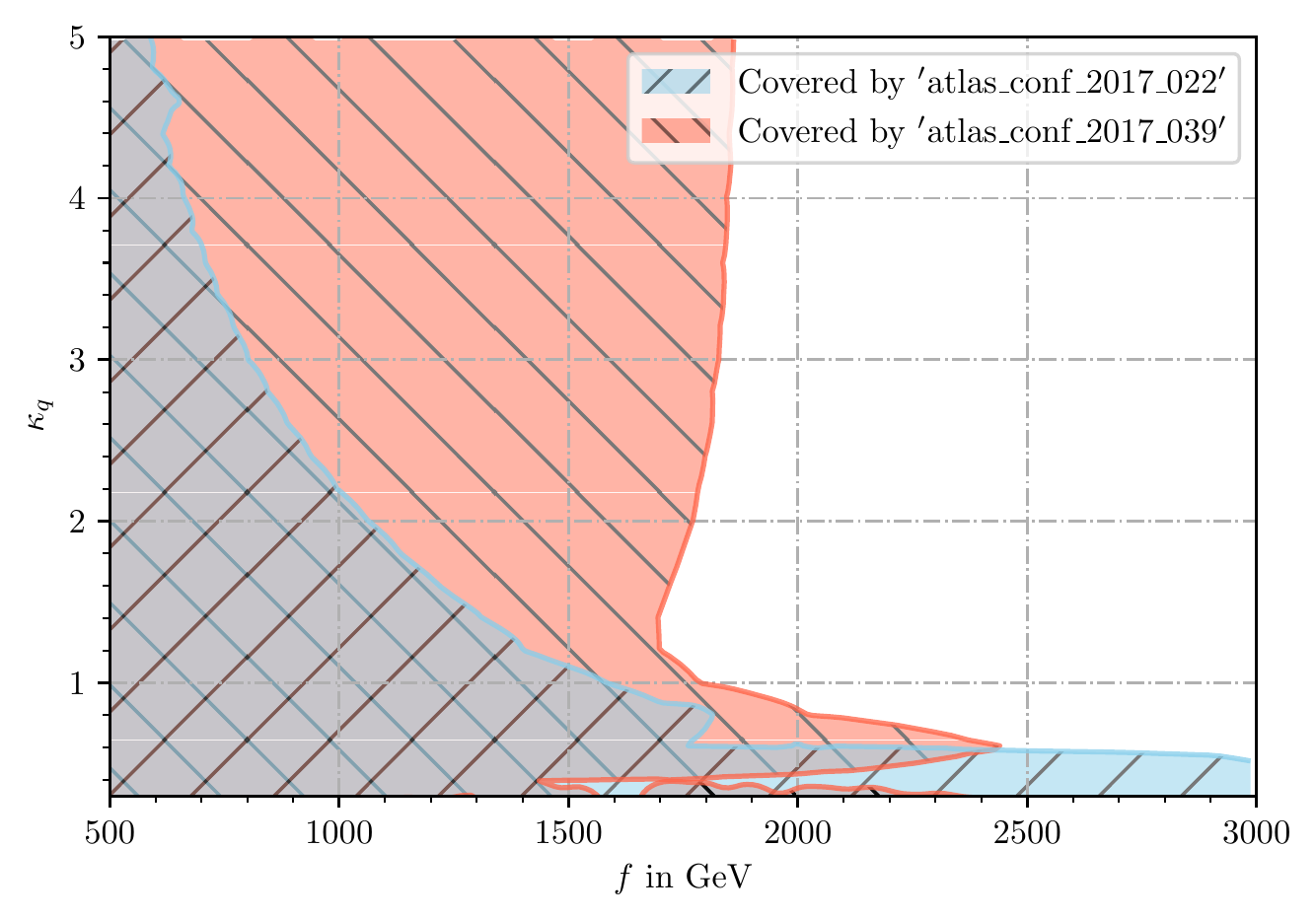}
\caption{Results for scenario (\emph{Light $\ell_H$})$\times$(\emph{Heavy $T^\pm$})$\times$(\emph{TPC})}
\label{fig:cmresults:lighttpcnotop}
\includegraphics[width=0.45\textwidth]
                {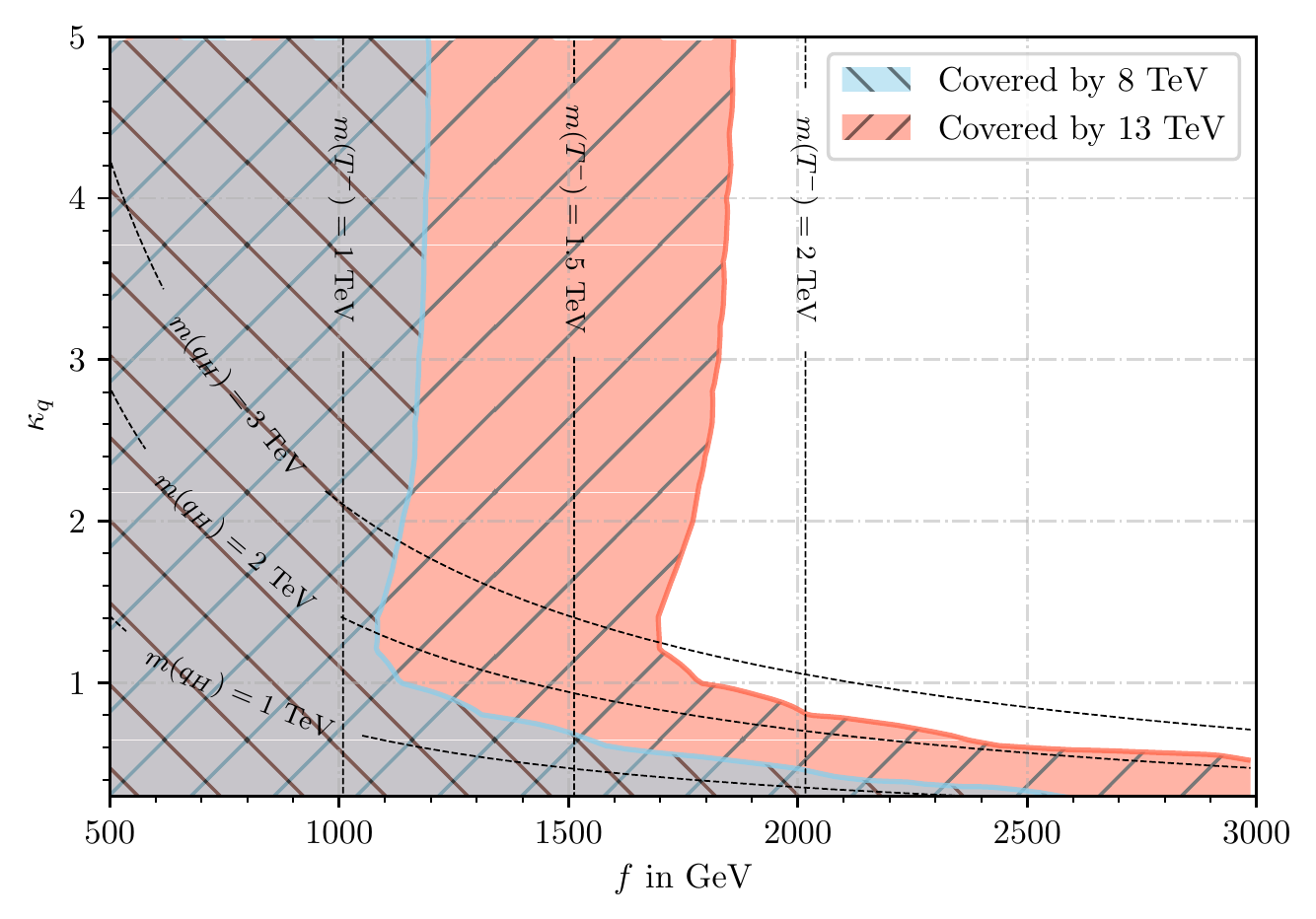} 
\includegraphics[width=0.45\textwidth]
                {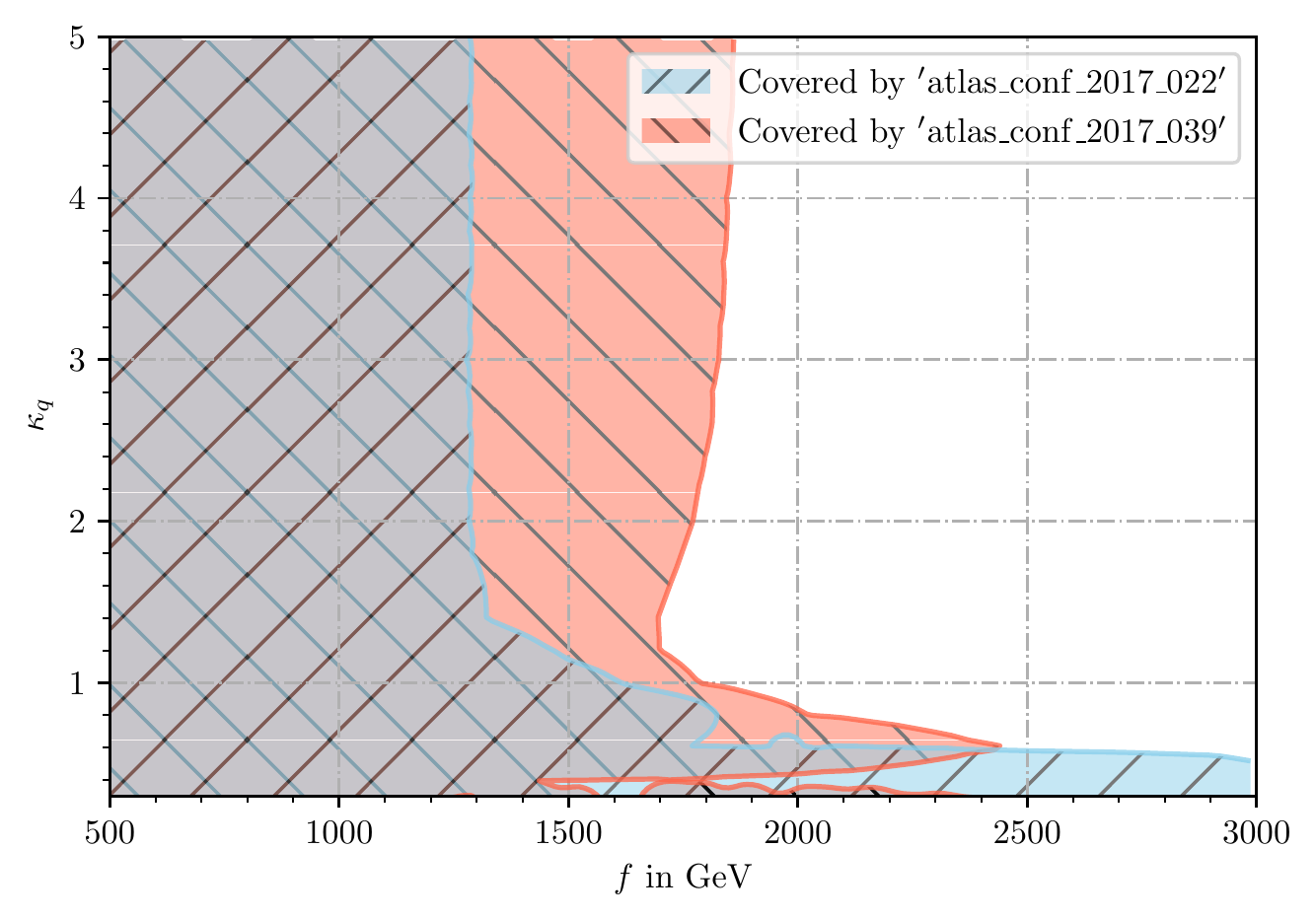}
\caption{Results for scenario (\emph{Light
    $\ell_H$})$\times$(\emph{Light $T^\pm$})$\times$(\emph{TPC})} 
\label{fig:cmresults:lighttpctop}
\centering
\includegraphics[width=0.45\textwidth]
                {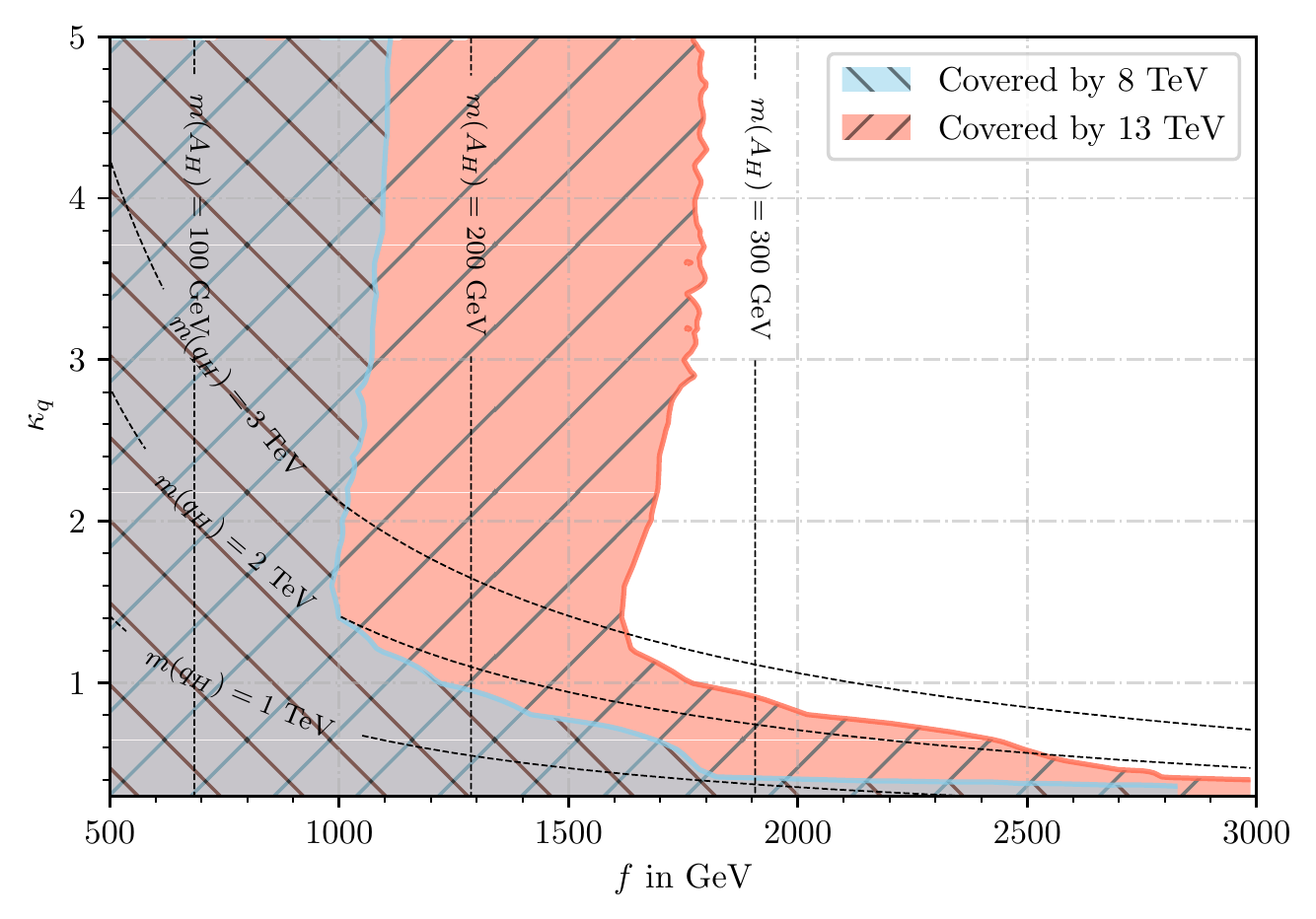} 
\includegraphics[width=0.45\textwidth]
                {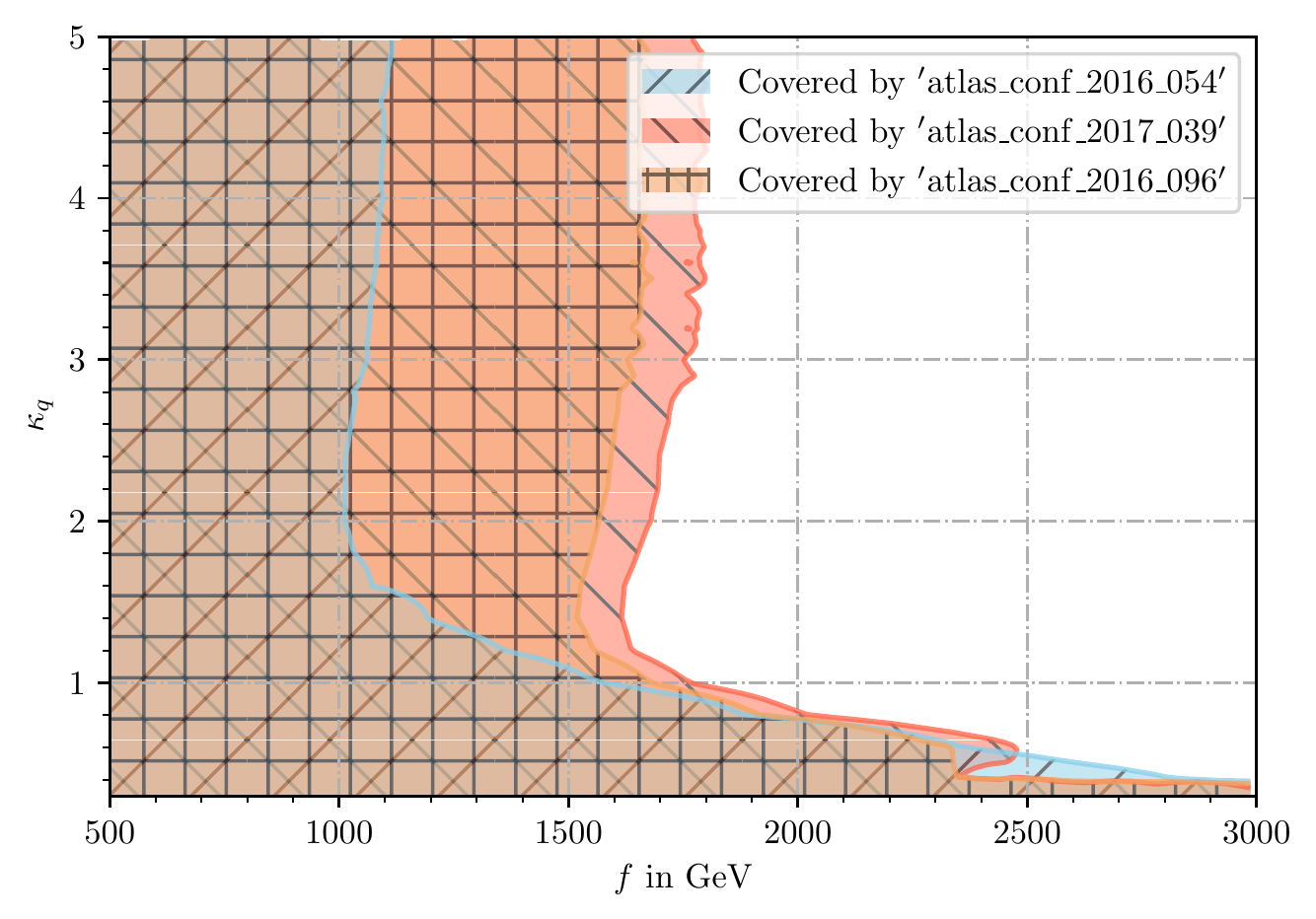}
\caption{Results for scenario (\emph{Light
    $\ell_H$})$\times$(\emph{Heavy $T^\pm$})$\times$(\emph{TPV})} 
\label{fig:cmresults:lighttpvnotop}
\includegraphics[width=0.45\textwidth]
                {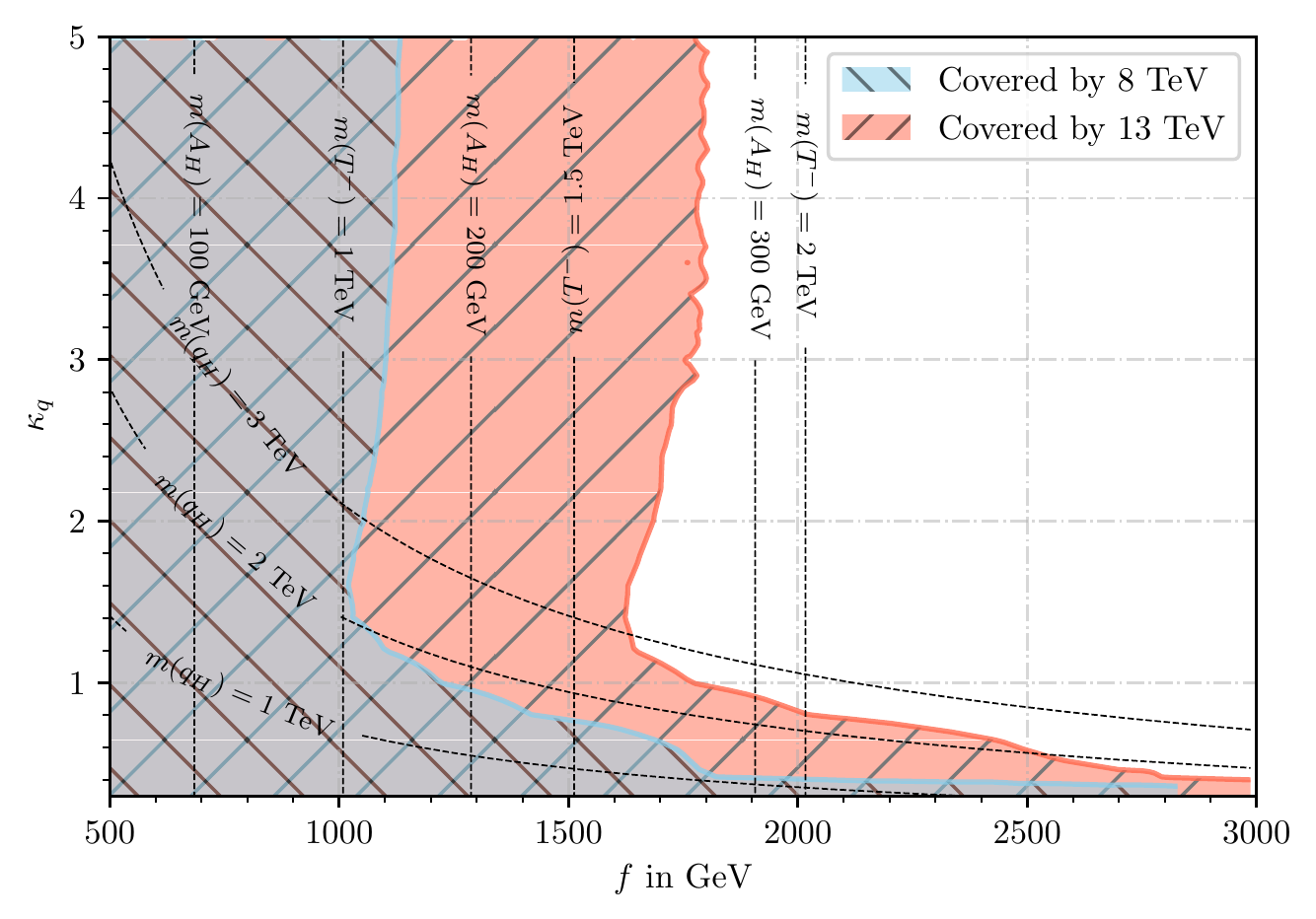} 
\includegraphics[width=0.45\textwidth]
                {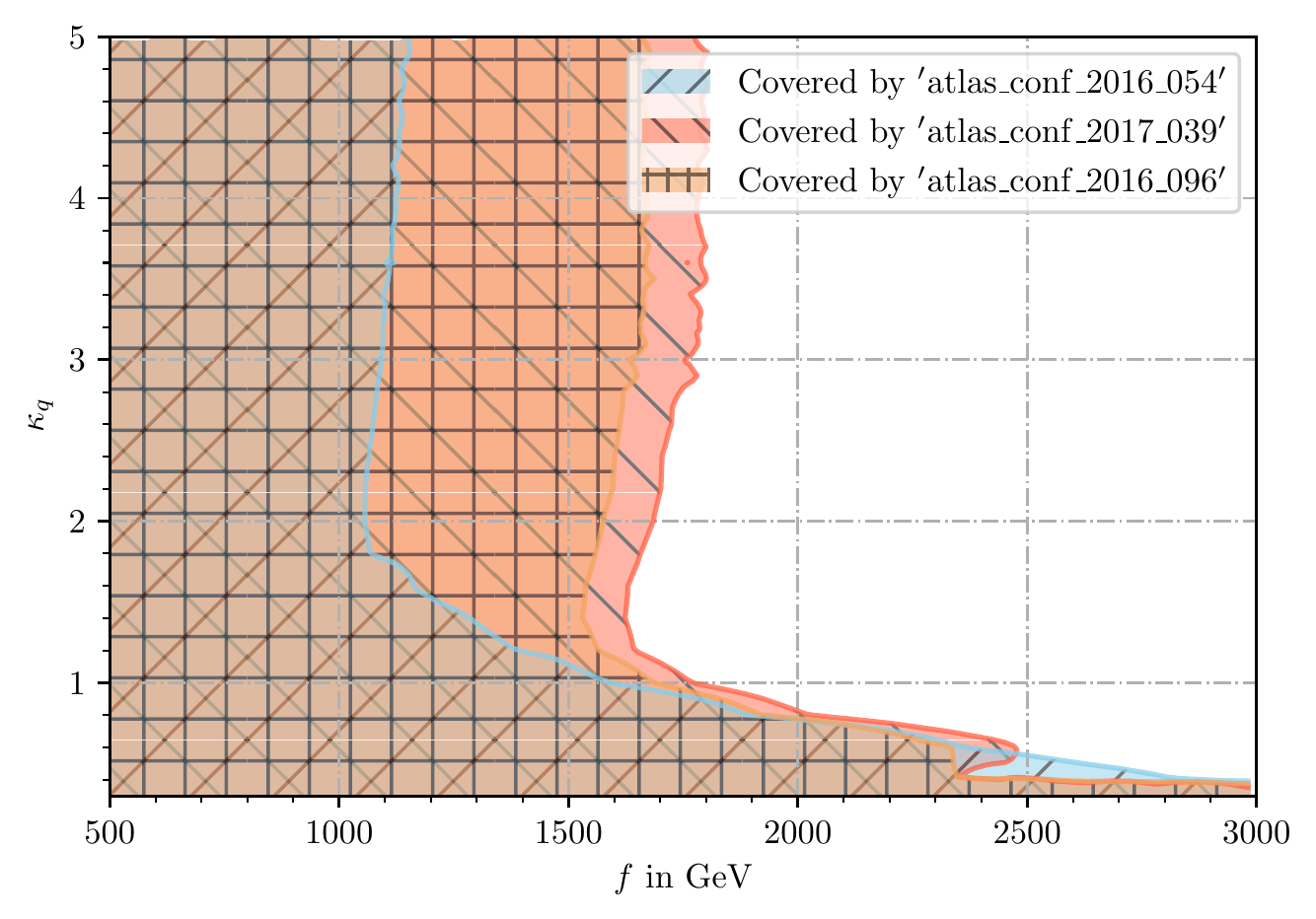}
\caption{Results for scenario (\emph{Light
    $\ell_H$})$\times$(\emph{Light $T^\pm$})$\times$(\emph{TPV})} 
\label{fig:cmresults:lighttpvtop}
\end{figure*}

\subsection{Light $\ell_H$}

In our third main benchmark scenario we again scan $\kappa_q$ and thereby
the mass of the heavy quarks. However, the degeneracy with the heavy
lepton sector is now lifted by fixing $\kappa_\ell = 0.2$. Since we
know from the results of the previous benchmark that no bound can be
set if we search for the direct production of $\ell_H$ alone, we \emph{only}
consider the effects of the light $\ell_H$ with respect to the
branching ratio of the heavy gauge bosons. Note again from our results
in Sec.~\ref{subsec:topologies} that while in the \emph{Fermion
  Universality} model we dominantly expect bosonic decays $Z_H
\rightarrow h A_H, W_H \rightarrow W A_H$, the \emph{Light Leptons}
benchmark mainly produces leptonic decays $Z_H \rightarrow \ell_H
\ell, W_H \rightarrow \ell_H \nu / \nu_H \ell$ with subsequent
$\ell_H/\nu_H \rightarrow \ell/\nu A_H$.  

In
Figs.~\ref{fig:cmresults:lighttpcnotop}-\ref{fig:cmresults:lighttpvtop}
we show the results of this benchmark, again for all four
subscenarios.  
\begin{itemize}
\item As in the \emph{Fermion Universality} scenario, we observe two
  main regions of exclusion which intersect at $f \approx
  \unit[1.6]{TeV}$ and $\kappa_q \approx 1.2$.  
\item For small $\kappa_q$ and large $f$, we again observe a $q_H$
  dominated bound similar to the one seen in the \emph{Fermion
    Universality} scenario. The analysis coverage map reveals that for
  $\kappa_q > 0.5$, the bound is set by a multilepton analysis. The
  already mentioned 3$\ell$ signal region is
  very sensitive to the final state topology $q_H q_H \rightarrow q q
  W_H W_H \rightarrow q q \ell \ell \ell \ell A_H A_H$ with one of the
  leptons not being identified and the other three leptons being
  highly boosted due to the large $q_H-W_H$ mass splitting. For
  $\kappa_q < 0.5$, the heavy vector bosons start predominantly
  decaying into hadronic final states --- see
  Fig.~\ref{fig:cm:br2324kl} --- in which case multijet final states
  start becoming more sensitive and reproduce the same bound as in the
  \emph{Fermion Universality} scenario.  
\item The $q_H$ dominated bound is again insensitive to the presence
  of the heavy top partners. Furthermore, it again slightly weakens in
  the presence of $T$-parity violation as the most sensitive final
  state stays identical but the  \etmiss{} cut efficiency drops due to
  the $A_H$ decaying.  
\item For larger values of $\kappa_q$, we again observe a nearly
  $\kappa_q$-independent absolute bound on $f$. This bound is again
  produced from direct production of heavy vector bosons and shows a
  small $\kappa_q$ dependence due to the cross section dependence of
  this parameter, see our discussion before. Compared to the
  \emph{Fermion Universality} scenario, the limit has become
  tremendously stronger due to the presence of light $\ell_H$ and
  improves to $f \gtrsim \unit[1.6]{TeV}$ for $\kappa_q \approx 1.5$
  and to $f \gtrsim \unit[2]{TeV}$ for $\kappa_q \gtrsim 5.0$. As the
  analysis coverage map on the right of
  Fig.~\ref{fig:cmresults:lighttpcnotop} shows, the vector-boson
  dominated region is now tested by the multilepton analysis which
  identifies the boosted leptons from the $V_H \rightarrow \ell
  \ell/\nu A_H$ decays. As this final state has small Standard Model
  background contamination --- most importantly since the leptons do
  not originate from $W$ or $Z$ decays --- it produces a very clean
  signal and thus leads to a very strong exclusion. 
\item In this scenario, the presence of the heavy top partners does
  not improve the bound derived from heavy vector boson production at
  all: the bound derived from $T^-$ production --- see the
  \emph{Fermion Universality} benchmark discussion --- is only
  sensitive to scales $f \lesssim \unit[1350]{GeV}$ and thus cannot
  compete with the much stronger bound set from the vector boson
  sector. Furthermore, the multilepton final state produced from the
  $V_H$ decays do not get any contributions from any of the expected
  $T^-$ decays. The limit is therefore unaffected. 
\item As the final state leptons from the $V_H$ decays already produce
  a very clean signal, a possible decay of the $A_H$ induced by $T$-parity
  violation  only results in a smaller \etmiss{} cut efficiency as
  explained before. Thus, we only observe that the bound is slightly
  weakened in models with $T$-parity violation. 
\end{itemize}

To summarize the results of this benchmark, we observe that a lighter
$\ell_H$ sector changes the decay patterns of the heavy vector bosons
and this globally leads to a significant improvement on the
bounds. This improvement even overcomes possible contributions from
the heavy top partner sector and is only slightly weakened by the
presence of $T$-parity violation. Therefore we again conclude that the
lower limits on $f$ derived in the \emph{Fermion Universality}
benchmark from searches for heavy vector bosons are very robust
regarding changes in the heavy fermion sector. 

Note that for this benchmark we chose a specific value of
$\kappa_\ell$ and thus in fact only analyzed the impact of light
$\ell_H$ for a particular assumption for their masses. It is thus
worthwhile discussing how changing $\kappa_\ell$ would affect our
results: 
\begin{itemize}
\item In our benchmark, the branching ratio $V_H \rightarrow \ell_H
  \ell$ was nearly \unit[100]{\%}. Clearly, the partial decay width
  $V_H \rightarrow \ell_H \ell$ depends on the $\ell_H$ mass and thus
  the leptonic branching ratio may drop if we increase the heavy
  lepton mass. The resulting bounds would then gradually shift from
  those derived in the \emph{Light $\ell_H$} to those in the
  \emph{Fermion Universality} benchmark.  
\item The kinematic configuration of the $V_H \rightarrow \ell
  \ell^{(\prime)} A_H$ decay depends on the mass of the intermediate
  on-shell $\ell_H$. Changing the mass results in different expected
  energy distributions for the signal leptons and can therefore affect
  the signal acceptance after applying the cuts in analysis
  \texttt{atlas\_conf\_2017\_039}. However, as the mass splitting
  $V_H-A_H$ is of order \unit[750]{GeV} for $f \approx
  \unit[1.5]{TeV}$ and is independent of the benchmark model,
  the final state leptons are always expected to be high-energetic
  enough to pass the constraints.  
\end{itemize}

\subsection{Prospects for $\sqrt{s}=\unit[14]{TeV}$}
\label{sec:results14tev}
\begin{table}
\footnotesize
    \begin{tabularx}{\columnwidth}{XXXl}
      \toprule \midrule
      CM identifier & Final State & Designed for & Ref.\\
\midrule
      \texttt{atlas\_2014\_010\_hl\_3l} &       \etmiss{} + 3 $\ell$  & $\tilde \chi^\pm, \tilde \chi^0$ & \cite{ATL-PHYS-PUB-2014-010} \\               
      \texttt{atlas\_phys\_2014\_010\_sq\_hl} &    \etmiss{} + 0 $\ell$ + 2-6 j           & $\tilde q, \tilde g$ & \cite{ATL-PHYS-PUB-2014-010} \\
      \texttt{dilepton\_hl} &    \etmiss{} + 2 $\ell$          & $\tilde \chi^\pm, \tilde \ell$ & \cite{dileptonhl} \\
      \bottomrule
    \end{tabularx}
\caption{Small summary of all $\sqrt{s} = \unit[14]{TeV}$ analyses
  which appear in the discussion of our results. More details, also on
  other tested analyses, are given in Tab.~\ref{tab:app:analyses} in
  the appendix.} 
\label{tab:analysisleg14}
\end{table}
As we observed in our results, the update from a center-of-mass energy
of $\sqrt{s} = \unit[8]{TeV}$ to $\sqrt{s} = \unit[13]{TeV}$ and the
increase of integrated luminosity between LHC Run 1 and Run 2 yielded
significantly stronger bounds for all of the considered benchmark
scenarios. In that context, the interesting question arises to which
extent the sensitivity is expected to further improve at a high
luminosity LHC running at $\sqrt{s} = \unit[14]{TeV}$. For that
purpose, we used the ATLAS high luminosity studies implemented in
\Checkmate{} to determine the expected bounds at very high statistics,
$\int \mathcal{L} = \unit[3000]{fb}^{-1}$. This gives a rough estimate
for the overall sensitivity range of the Large Hadron Collider to the
Littlest Higgs Model in general. The corresponding cross sections are
shown in Figs.~\ref{fig:xs:univ14},\ref{fig:xs:univ142} in the
appendix~\ref{app:figures}. 

Again, all analyses which have been used by this study are listed in
Tab.~\ref{tab:app:analyses} in the appendix and we provide a shortened
version in Tab.~\ref{tab:analysisleg14} which only lists those analyses
which appear in our discussion of the most sensitive analyses. As one
can see in the full table in Tab.~\ref{tab:app:analyses}, at this
stage the list of high luminosity analyses is very limited as only few
official, experimental and some phenomenological high performance
studies have been implemented so far. These cover the most important
topologies, i.e.\ missing transverse momentum with either a monojet,
multijet or multileptons final state, however these old experimental
studies use far fewer, less optimized signal regions compared to their
counterparts at lower center-of-mass energies. Hence, our results
should only be understood as rough approximations and much more
sophisticated studies, especially on the experimental side, would be
required to get results which are qualitatively at the same level as
our earlier, detailed re-interpretation of existing experimental data.  

Since the number of tested topologies is fairly small and is not
expected to cover all the various final states we discussed before, we
do not consider the full set of benchmark models introduced previously
at this stage. Instead, we concentrate on the results for \emph{TPC}
$\times$ \emph{Heavy $T^\pm$} for the three scenarios \emph{Fermion
  Universality}, \emph{Heavy $q_H$} and \emph{Light $\ell_H$}. These
give a good overview to the general expected sensitivity at high
statistics. As can be seen from the results discussed above, the
macroscopic structure of the excluded parameter areas are very similar
for cases with and without $T$-parity violation and with the heavy top
partners included or not. Hence, one can apply the phenomenological
discussions of the previous sections to appoximately determine the
excluded areas for the other benchmark cases which we do not
explicitly discuss in the following.  

\begin{figure*}
\centering
\includegraphics[width=0.45\textwidth]
                {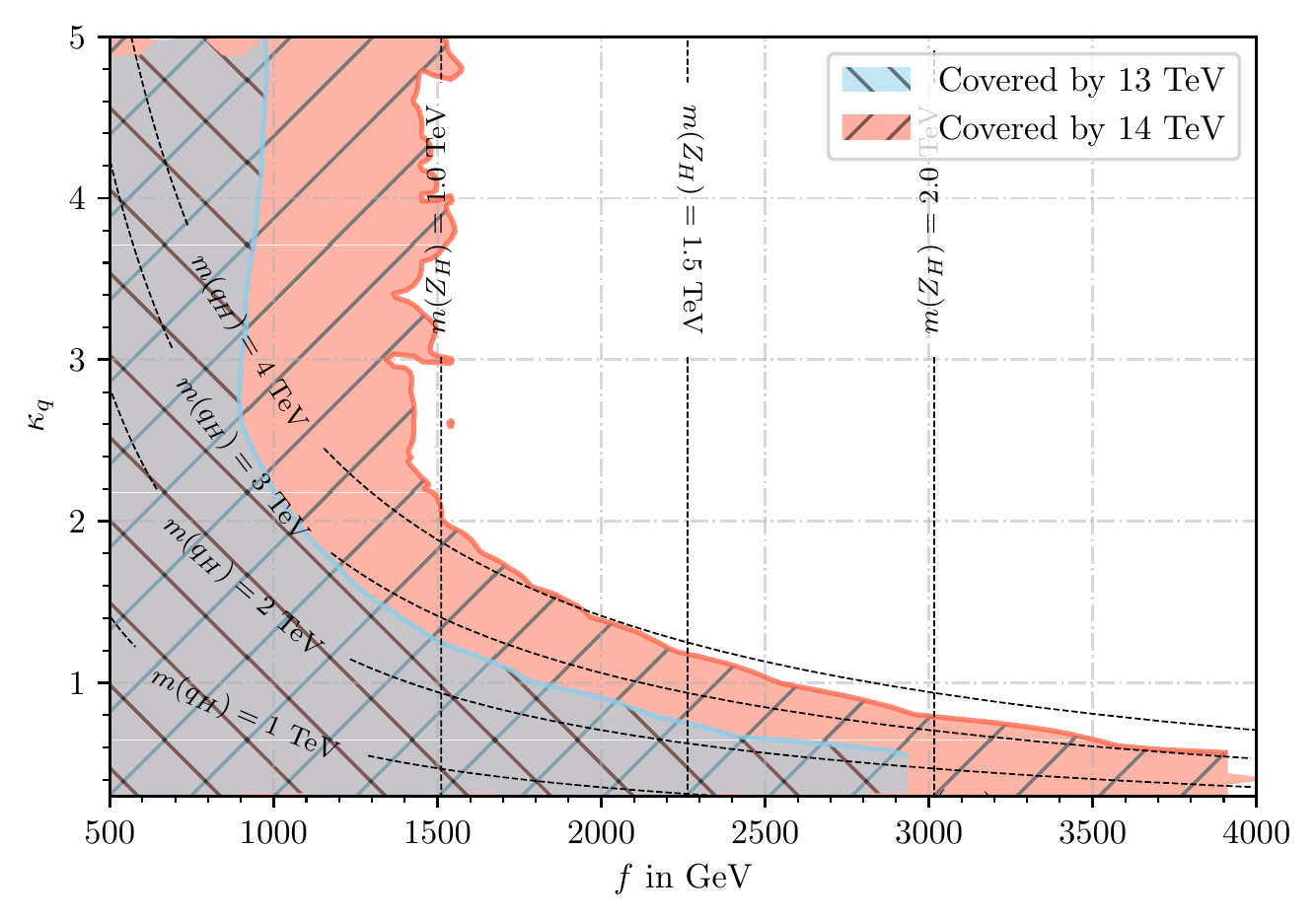} 
\includegraphics[width=0.45\textwidth]
                {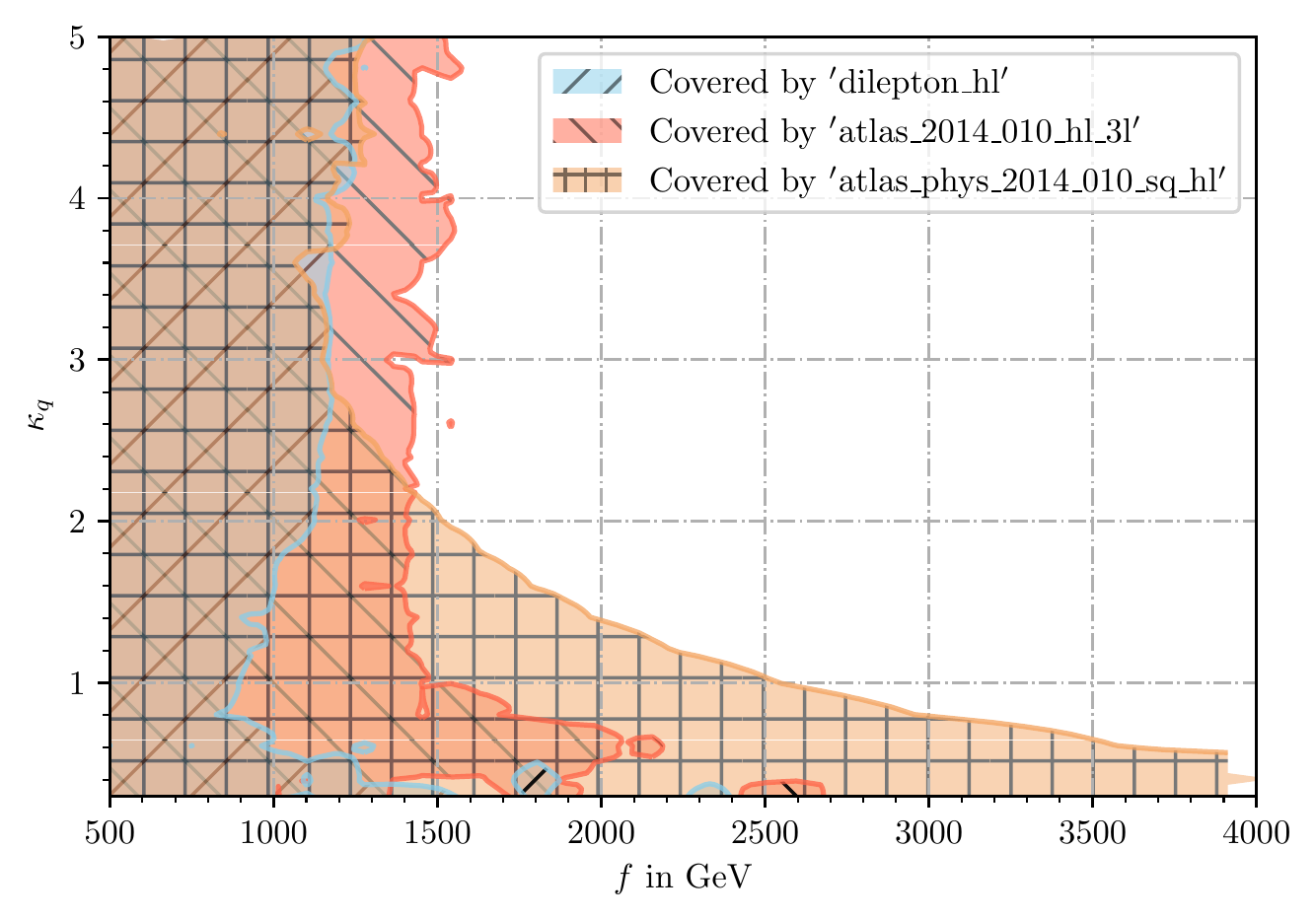}
\caption{Expected results at $\sqrt{s}=\unit[14]{TeV}, \int
  \mathcal{L} = \unit[3000]{fb}^{-1}$ for scenario (\emph{Fermion
    Universality})$\times$(\emph{Heavy
    $T^\pm$})$\times$(\emph{TPC}). } 
\label{fig:cmresults:14tev:samekqkl}
\includegraphics[width=0.45\textwidth]
                {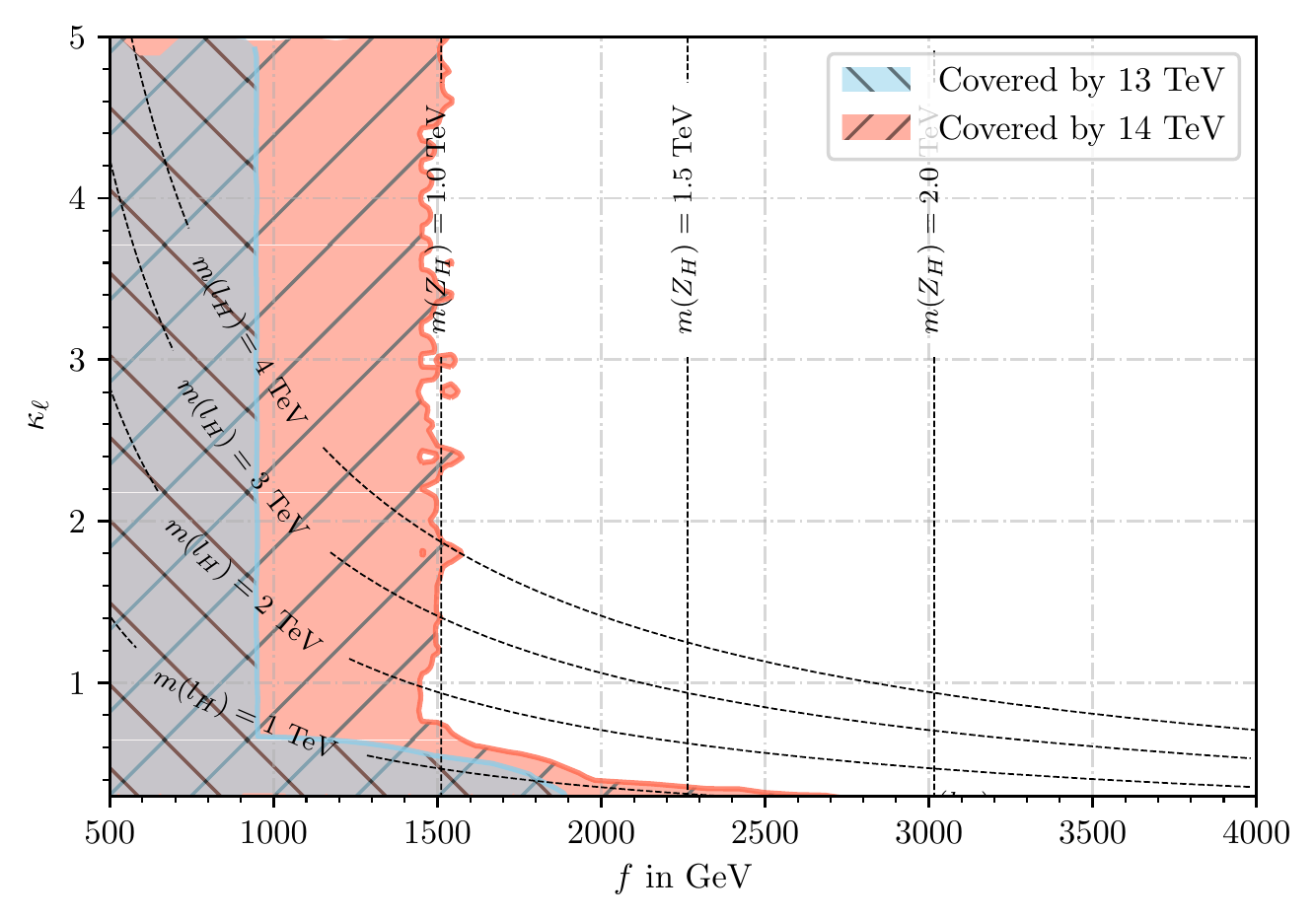} 
\includegraphics[width=0.45\textwidth]
                {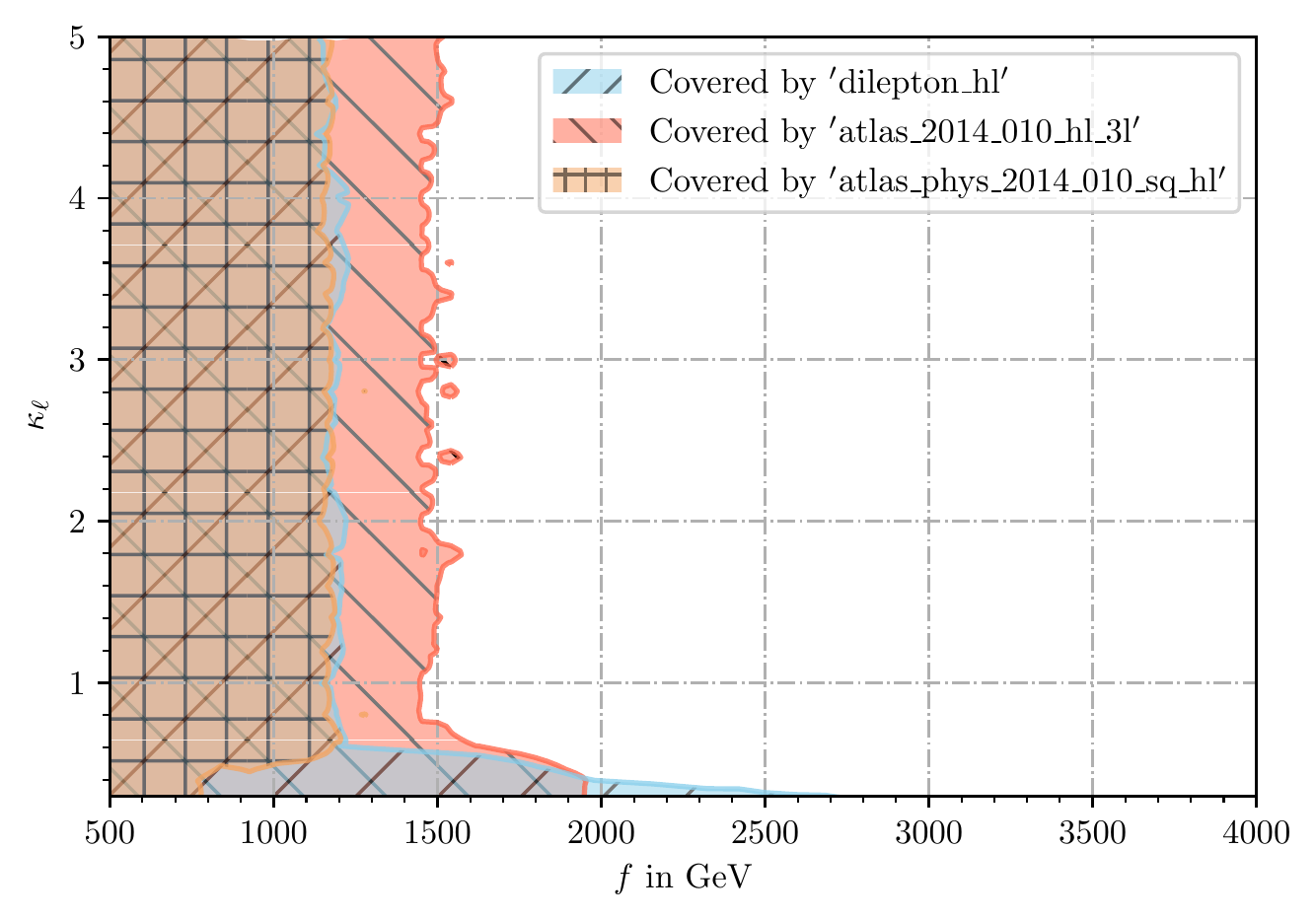}
\caption{Expected results at $\sqrt{s}=\unit[14]{TeV}, \int
  \mathcal{L} = \unit[3000]{fb}^{-1}$ for scenario (\emph{Heavy
    $q_H$})$\times$(\emph{Heavy $T^\pm$})$\times$(\emph{TPC}).} 
\label{fig:cmresults:14tev:kq4}
\includegraphics[width=0.45\textwidth]
                {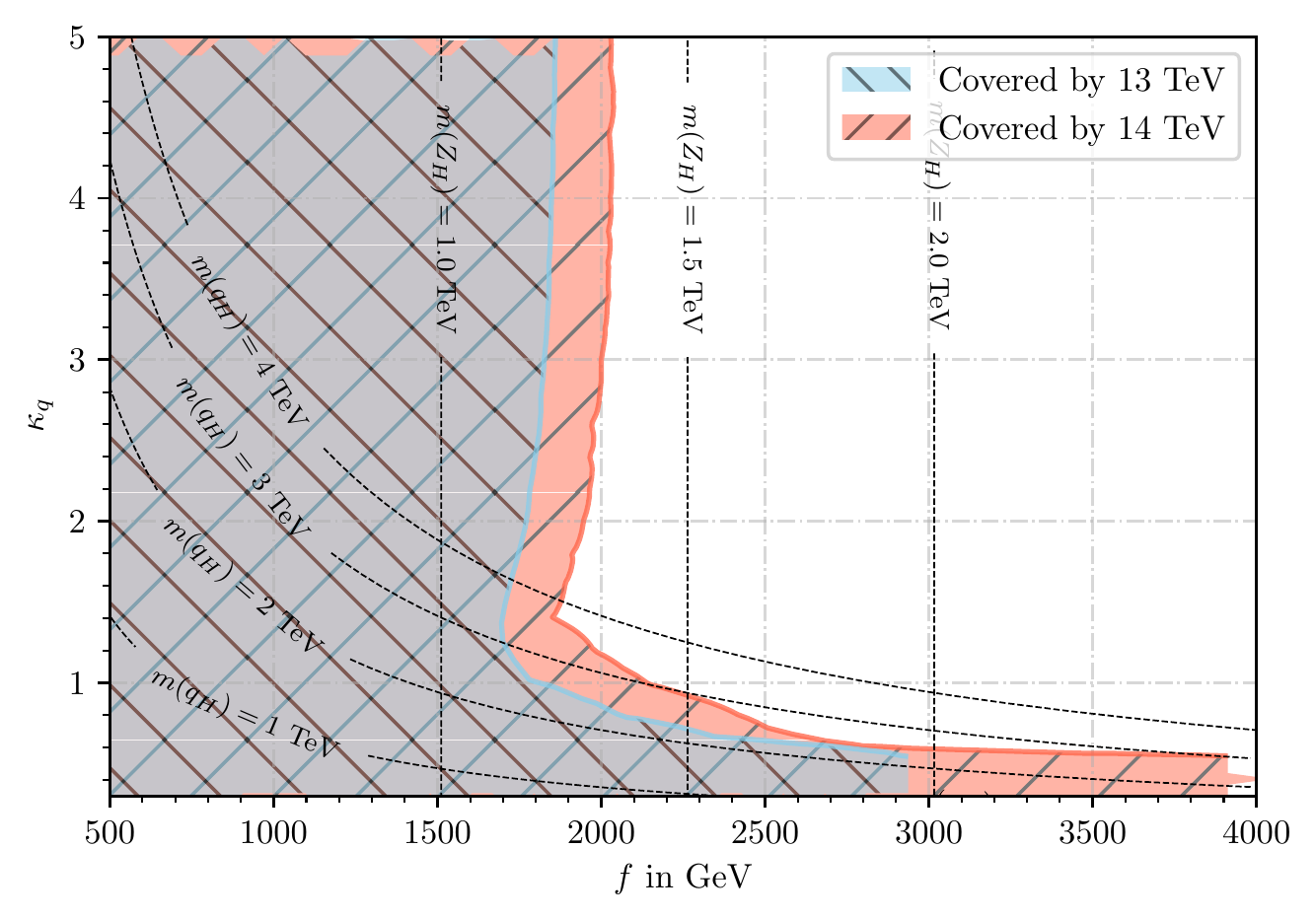} 
\includegraphics[width=0.45\textwidth]
                {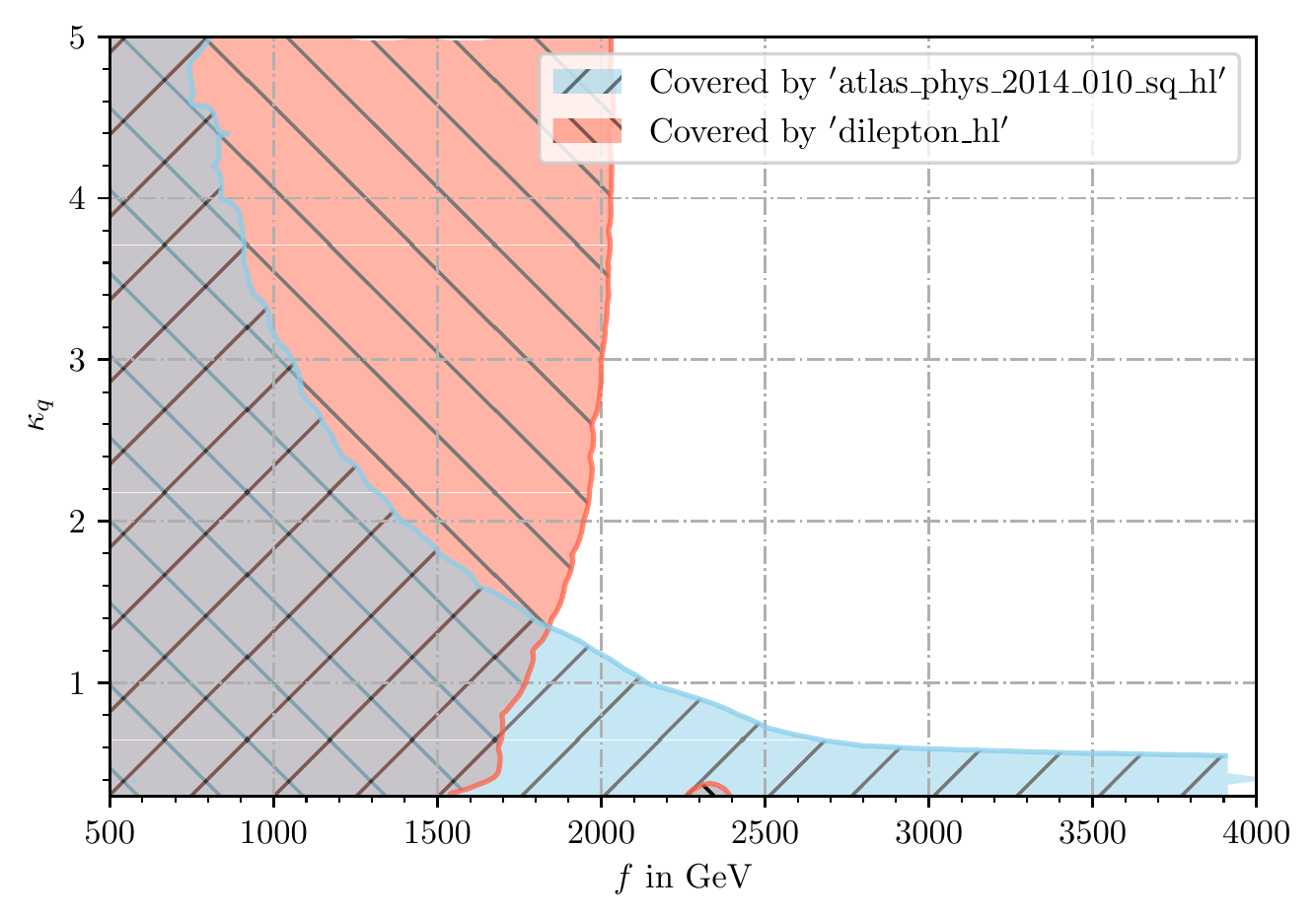}
\caption{Expected results at $\sqrt{s}=\unit[14]{TeV}, \int
  \mathcal{L} = \unit[3000]{fb}^{-1}$ for scenario (\emph{Light
    $\ell_H$})$\times$(\emph{Heavy $T^\pm$})$\times$(\emph{TPC}).} 
\label{fig:cmresults:14tev:kl2}
\end{figure*}

The results of our scans are shown in
Figs.~\ref{fig:cmresults:14tev:samekqkl}-\ref{fig:cmresults:14tev:kl2}. Note
the extended $f$-axis range compared to
Figs.~\ref{fig:cmresults:univtpcnotop}-\ref{fig:cmresults:lighttpvtop}
to better illustrate the even higher $f$-reach at high center-of-mass
energies. The figures on the left column again show the overall
expected experimental reach at \unit[14]{TeV} and compare to current
results from \unit[13]{TeV} data which corresponds to the results
discussed in the previous sections. The figures in the right column,
similarly to before, show the most sensitive analyses in different
regions of parameter space. Fluctuations in the contours originate
from sizable statistical uncertainties in our Monte Carlo
description\footnote{We use the same sample size as in our previous
  studies, however due to the 100-fold integrated luminosity the
  statistical uncertainty near the boundary increases approximately by
  a factor of 10.}, however do not affect the qualitative description
of the overall bound. 

In general, the structure of the bounds is kept, i.e.\ there is a
(nearly) $\kappa$-independent bound for small $f$ and larger values of
$\kappa$ while there is a bound which follows the iso-mass contours
for large values of $f$.  

\begin{itemize}
\item In the \emph{Fermion Universality} scenario, the $q_H$ mass
  bound for large values of $f$ increases by 1 to \unit[1.5]{TeV} and
  excludes heavy quarks with masses $m(q_H) \gtrsim \unit[4]{TeV}$ for
  $f \approx \unit[2]{TeV}$ and $m(q_H) \gtrsim \unit[3]{TeV}$ for $f
  \approx \unit[4]{TeV}$. As before, this bound originates from the
  high luminosity version of a multijet plus \etmiss{} search designed
  to find heavy squarks or gluinos in supersymmetry. The $V_H$
  dominated bound for large values of $\kappa$ probes heavy vector
  boson masses of order \unit[1]{TeV}. Compared to the previous result
  determined at \unit[13]{TeV}, the most sensitive analysis is now
  quoted to be the multilepton instead of the multijet final state. To
  reduce the contamination from pileup which is expected to become an
  important issue for the high luminosity LHC, the multijet final
  states require the scalar sum of the transverse momenta of all
  reconstructed objects to exceed \unit[3]{TeV}. In the $V_H$
  dominated region, the expected signal $V_H \rightarrow A_H V, V
  \rightarrow \text{hadrons}$ with $m(V_H) \approx \unit[1]{TeV}$
  typically does not pass this constraint and for example requires a
  boosted final state due to  a high $p_T$ jet from initial-state
  radiation (ISR) whose requirement
  significantly reduces the expected event rate.  
\item The \emph{Heavy $q_H$} scenario at $\sqrt{s} = \unit[14]{TeV}$
  does not significantly improve the $\ell_H$-induced bound for small
  values of $\kappa$. We expect a weak bound which follows the
  $\ell_H$ mass contour and excludes masses of order $m(\ell_H) \approx
  \unit[1-1.5]{TeV}$. This bound originates from an extrapolated search for
  dilepton final states. This is however only a minor improvement to
  the bound which can be set already from today's result. As in the
  previous benchmark scenario, the $V_H$ produces a
  $\kappa$-independent bound of $m(V_H) \gtrsim \unit[1]{TeV}$.  
\item Lastly, the bound in the \emph{Light $\ell_H$} scenario only
  improves little compared to  the current \unit[13]{TeV} results. In
  the large $f$ region, the most sensitive analysis channel at LHC Run
  2 is a multijet final state with one additional lepton which has a
  particularly small Standard Model contamination. Unfortunately, we
  do not have a high luminosity version of this analysis available and
  can only consider final states with many jets but no final state
  lepton. As the characteristic feature of the the \emph{Light
    $\ell_H$} scenario is the appearance of at least one lepton in all
  relevant final state decay chains, we lose sensitivity due to our
  restricted amount of available analyses. For larger values of
  $\kappa$, the bound on $m(V_H)$ only increases by about
  \unit[100]{GeV}, determined from a search which requires two leptons
  in the final state. This analysis is designed to target either of
  the two supersymmetric topologies $\tilde \ell \tilde \ell
  \rightarrow \ell \ell \tilde \chi \tilde \chi$ or $\tilde \chi^+
  \tilde \chi^- \rightarrow W W \tilde \chi \chi$ followed by leptonic
  $W$ decays. Though some of the final states produced by our
  benchmark scenario pass the constraints set for these particular
  topologies, none of the signal regions are specifically designed for
  our topology. Thus, again our bound does not represent the full
  sensitivity which can be expected from the high luminosity LHC but
  significant aditional effort would be required to determine the
  necessary experimental predictions for our desired topologies. 
\end{itemize}

%% file: sections/06_summary.tex
\section{Comparison of LHC limits with Bounds from Electroweak
  Precision Observables}
\label{sec:comp}

\begin{figure*}
\centering
\includegraphics[width=0.45\textwidth]
                {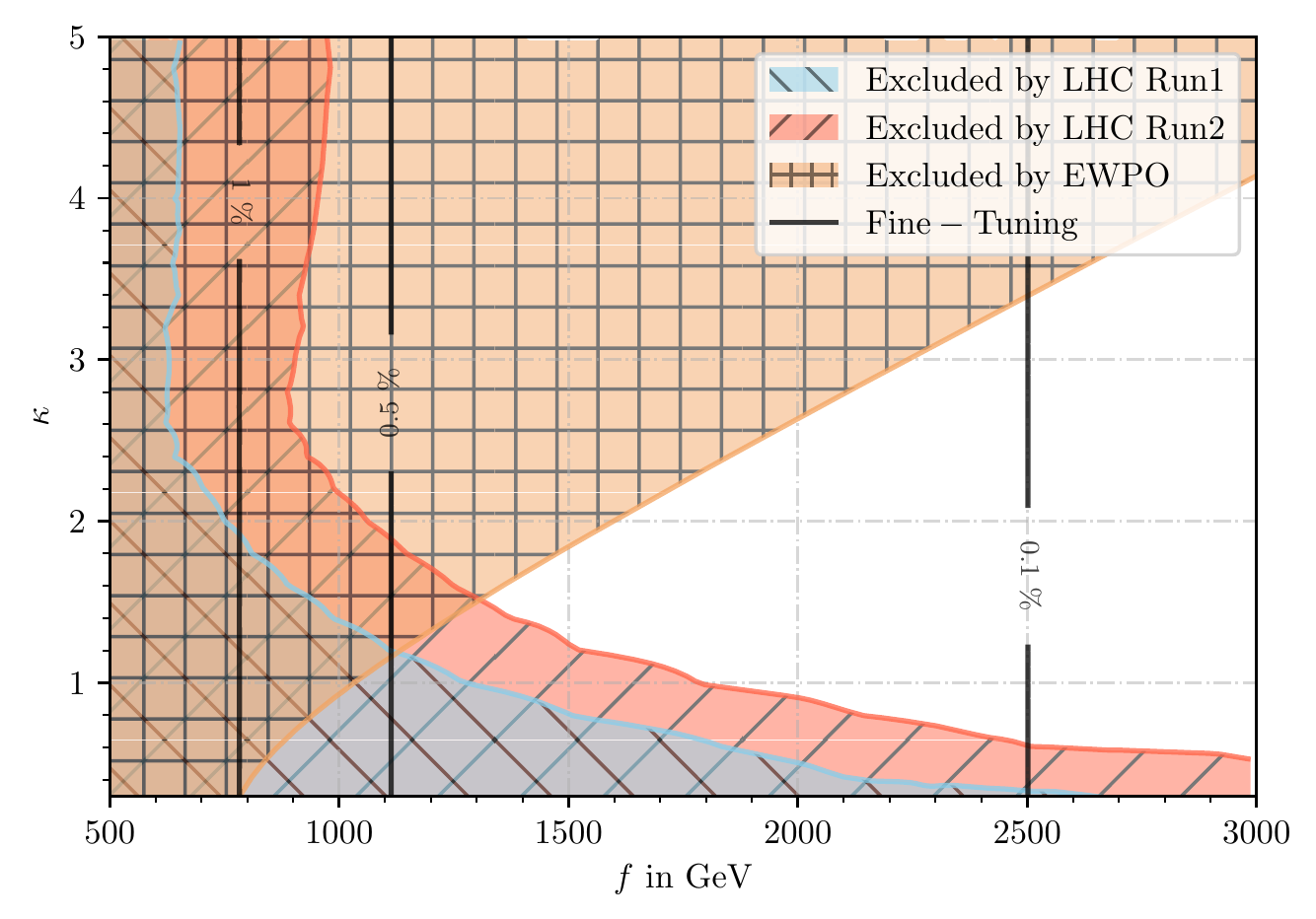} 
\includegraphics[width=0.45\textwidth]
                {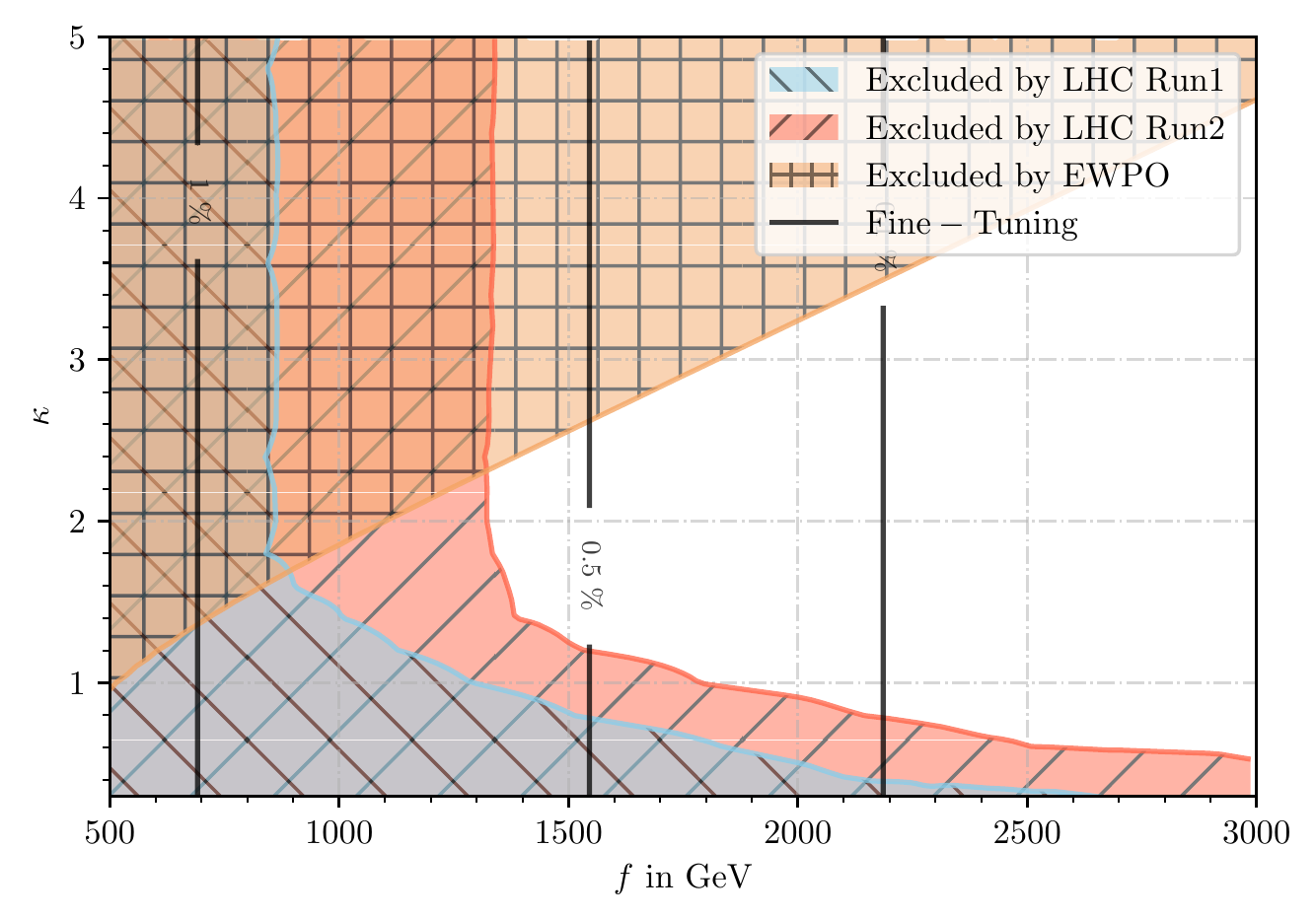}
\caption{Combined results for scenario (\emph{Fermion
    Universality})$\times$(\emph{TPC}), left: \emph{Heavy $T^\pm$},
  right: $\times$(\emph{Light $T^\pm$})} 
\label{fig:cmewporesults:kqkl}
\includegraphics[width=0.45\textwidth]
                {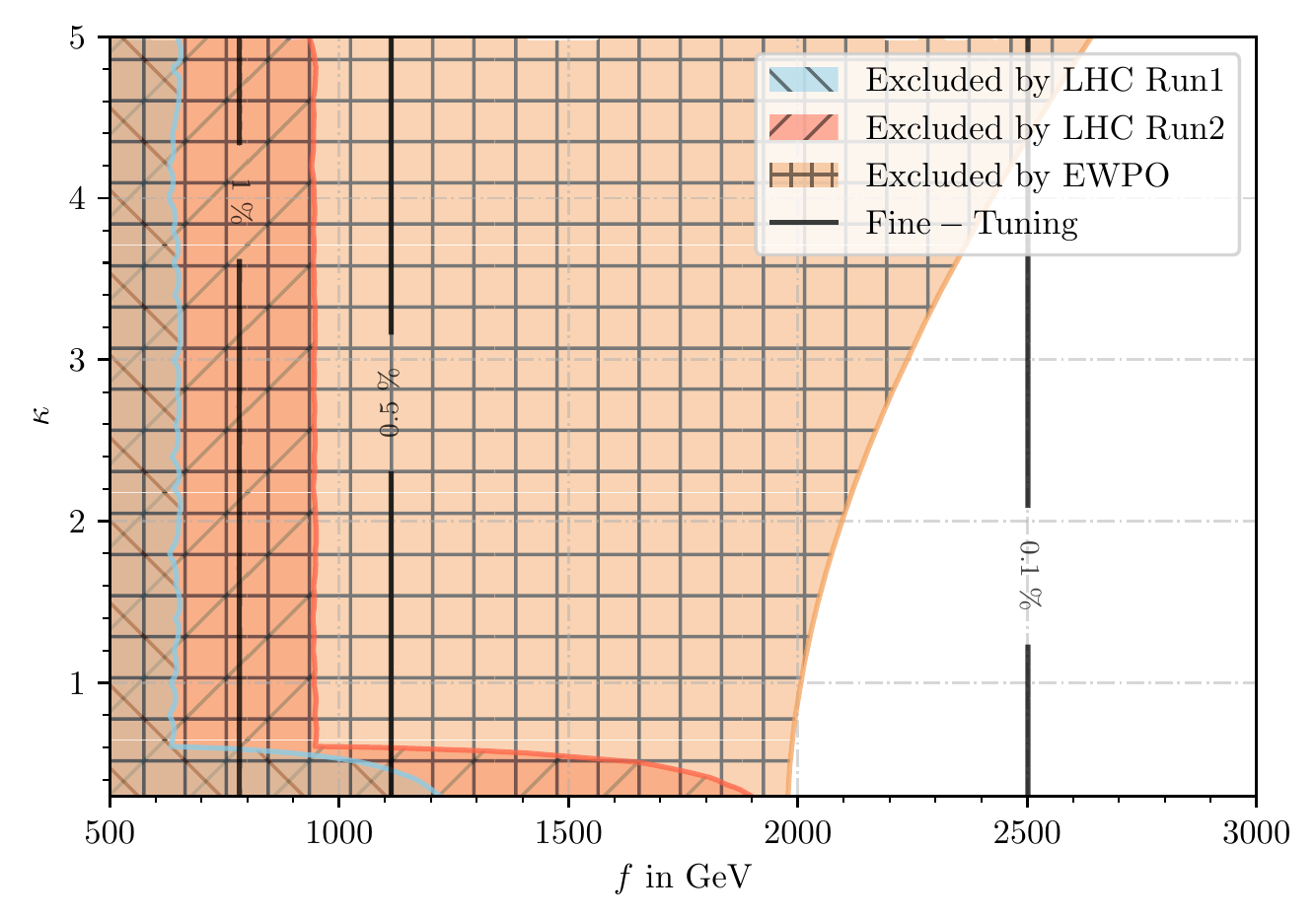} 
\includegraphics[width=0.45\textwidth]
                {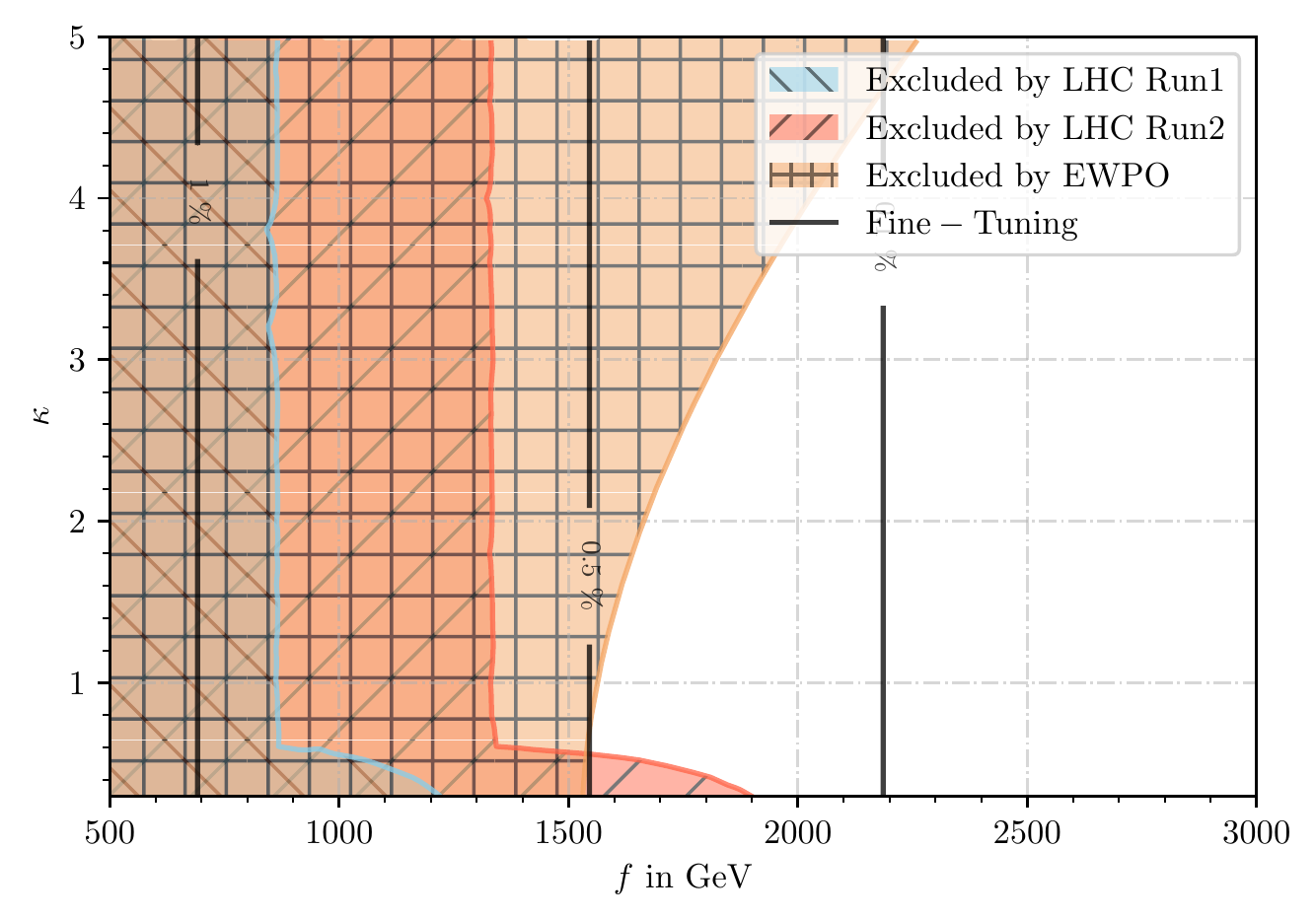}
\caption{Combined results for scenario (\emph{Heavy
    $q_H$})$\times$(\emph{TPC}), left: \emph{Heavy $T^\pm$}, right:
  $\times$(\emph{Light $T^\pm$})} 
\label{fig:cmewporesults:kq4}
\includegraphics[width=0.45\textwidth]
                {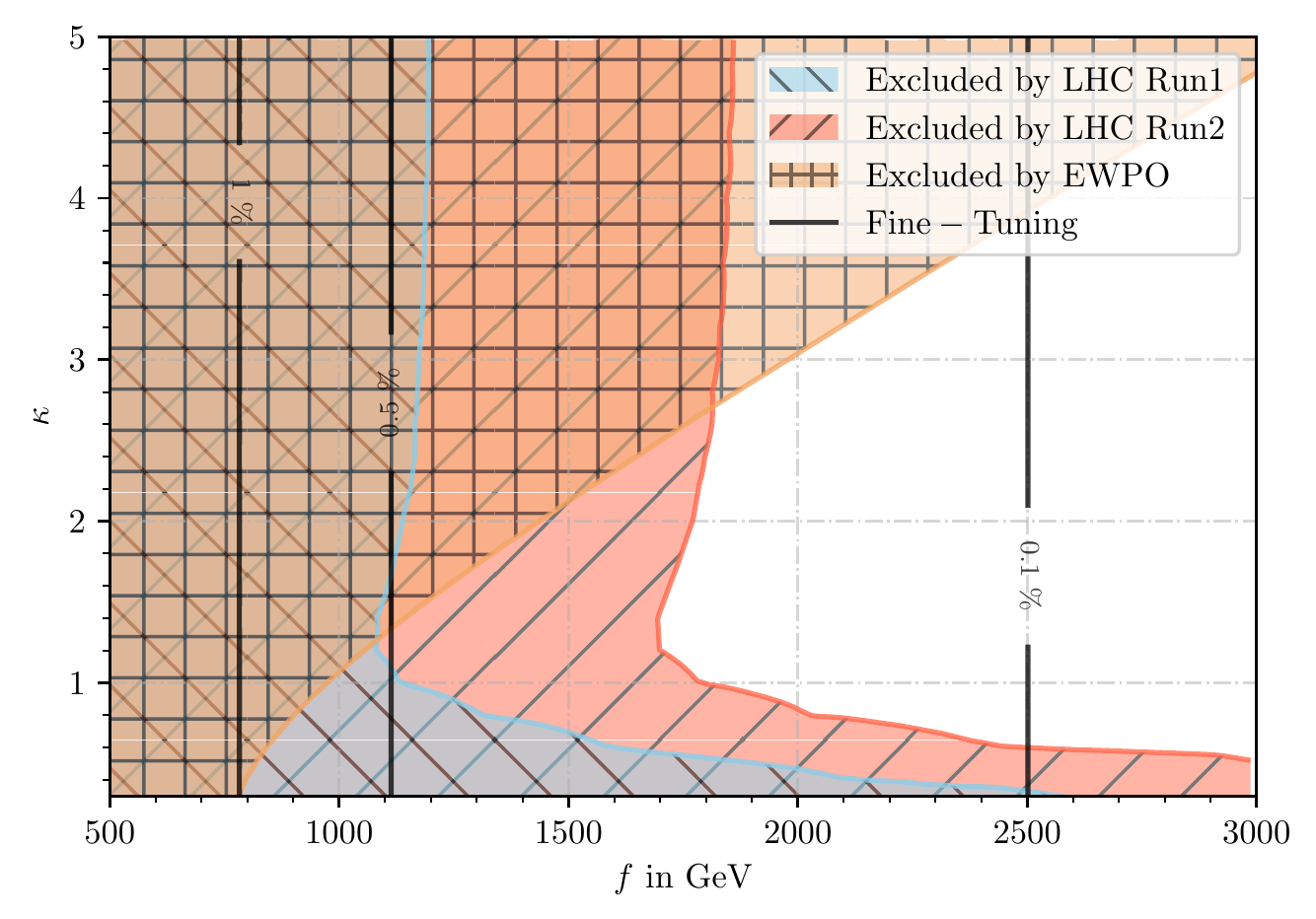} 
\includegraphics[width=0.45\textwidth]
                {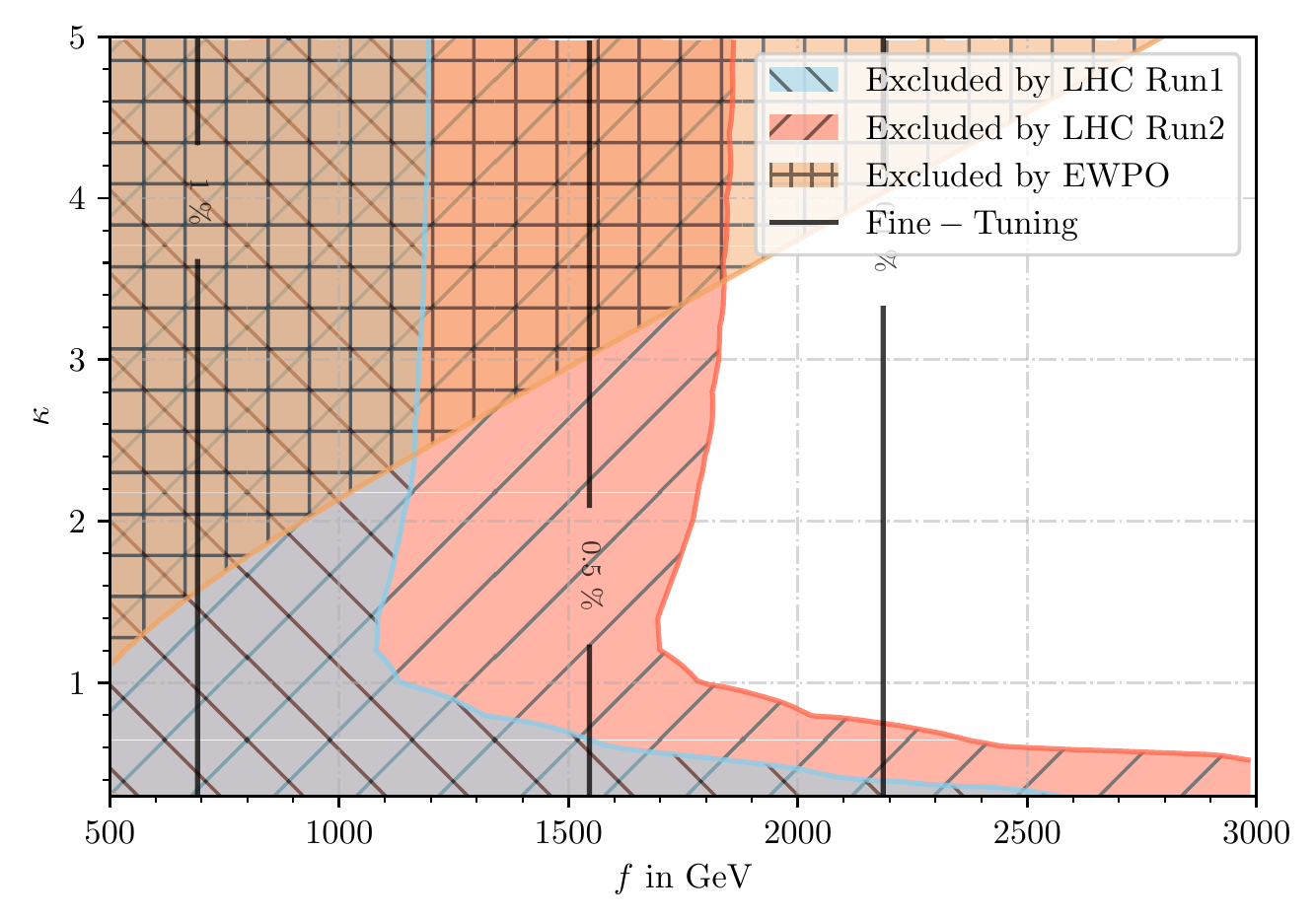}
\caption{Combined results for scenario (\emph{Light $\ell_H$})$\times$(\emph{TPC}), left: \emph{Heavy $T^\pm$}, right: $\times$(\emph{Light $T^\pm$})}
\label{fig:cmewporesults:kl2}
\end{figure*}

In the previous section we discussed the bounds which can be put on
various benchmark scenarios of the Littlest Higgs Model with
$T$-parity (and its possible violation). As explained in
Sec.~\ref{sec:littlehiggstheory}, an 
appealing property of this model is its considerably small amount of
fine tuning in the Higgs sector. Moreover, not only do the null
results of searches for these new $T$-odd particles set bounds on this
model but also, see Sec.~\ref{sec:EWPO}, electroweak precision
observables (EWPO) put tight constraints on $f$ and $\kappa$. In the
following we want to combine these three pieces of information,
putting a particular focus on the relevance of the newest LHC results
for the total combined bound on the model. 

In Figs.~\ref{fig:cmewporesults:kqkl}-\ref{fig:cmewporesults:kl2} we
show compilations of bounds from electroweak precision observables,
see Sec.~\ref{sec:EWPO}, the amount of fine tuning in the Higgs sector
according to Eq.~(\ref{eq:finetuning}) and the 8 and \unit[13]{TeV}
LHC bounds discussed in Sec.~\ref{sec:results}. We only show results
for the case of $T$-parity conservation as electroweak precision
observables are not affected by the presence of $T$-parity violating
operators and the respective TPV collider bounds are very similar, see
our results of the previous section. In each figure we show the
results for \emph{Heavy $T^\pm$} scenario, i.e. $R=0.2$, and \emph{Light
  $T^\pm$} scenario, i.e. $R=1.0$. Note that the choice of this parameter has
an important impact on the fine tuning measure $\Delta$.  

In general, we observe that LHC results produce an absolute lower
bound on $f$ for large $\kappa$ and a lower bound which approximately
follows $f \cdot \kappa$ for small $\kappa$. Electroweak precision
data, however, tend to produce upper bounds which
approximately follow the ratio $f / \kappa$. Therefore, we have two
very complementary bounds 
which together exclude a considerably large region of parameter
space. This complementarity mostly originates from the opposite
dependence of the respective bounds on $\kappa$ and $R$: the collider
data produce stronger bounds for lighter particles and therefore show their
largest sensitivity for small values of $\kappa$ and/or $R =
1.0$. Loop corrections to precision observables, however, increase if
the corresponding coupling constants increase and therefore show their
strictest bounds for \emph{large} values of $\kappa$ and
$R=0.2$.\footnote{Note that the free parameter $R$ defines the Yukawa
  coupling $\lambda_2$ via Eq.~(\ref{eq:topmass}) which
  \emph{increases} if $R$ \emph{decreases}.} 

We now move the general discussion to some indiviual results of particular benchmark models:
\begin{itemize}
\item In the case of \emph{Lepton Universality}, we observe that the
  updated collider results from \unit[13]{TeV} are only relevant in
  the regions dominated by $q_H$ and $T^\pm$ production. Most
  importantly, bounds derived from $V_H V_H$ production only cover the
  region $f < \unit[1]{TeV}, \kappa > 2$ and are not competitive with
  the  limits from electroweak precision data which cover the same
  region in the \emph{Light $T^\pm$} scenario and an even much larger
  region $f < \unit[1.3]{TeV}, \kappa > 1.5$  in the case of
  \emph{Heavy $T^\pm$}.  

In the case of \emph{Heavy $T^\pm$}, the combined bound from
electroweak precision observables and $q_H$ production excludes
symmetry breaking scales $f$ below \unit[1.3]{TeV}, independent of
$\kappa$, and by that requires a fine tuning below \unit[0.5]{\%}. If
the heavy top partners $T^\pm$ are lighter, the EWPO bounds
weaken. However, at the same time the collider bounds increase,
resulting in approximately the same bound of $f > \unit[1.3]{TeV}$ as
before which however corresponds to a slightly smaller fine tuning of
approximately \unit[0.6]{\%}. Judging from the two benchmark scenarios
for $T^\pm$, we conclude that the combination of electroweak precision
data and newest LHC results does not allow for values of $f <
\unit[1.3]{TeV}$ for values of $R \in [0.2, 1.0]$. As the EWPO bounds
become stronger for heavier $T^\pm$  and the collider result becomes
stricter for lighter $T^\pm$, the lower bound on $f$ should become
even stricter for any value of $R$ outside this range. 

\item The combined results of the \emph{Heavy $q_H$} scenario show a
  similar complementarity effect as in the previous model: whilst the
  LHC results are significantly weakened if the heavy quarks are
  decoupled, the bounds from electroweak precision observables become
  even stricter due to their dependence on $\kappa^2$, see
  Sec. \ref{sec:EWPO}, and thus become stronger if $\kappa_q = 3.0$ is
  fixed. Here, the bounds implicitly depend on the value of $R$ and
  exclude values of $f$ below \unit[1.5]{TeV} for $R=1.0$ (\emph{Light
    $T^\pm$}), and values below \unit[2]{TeV} for $R=0.2$ (\emph{Heavy
    $T^\pm$}). Even in the case of \emph{Light $T^\pm$} the LHC result
  cannot compete. Still, the bounds are already very close to the EWPO
  limit such that we again conclude that any other value of $R$ should
  not produce a significantly weaker but potentially an even stronger
  bound on $f$ if the mass of the $T^\pm$ is chosen even lighter. Note
  that for very small values of $\kappa_\ell$, the LHC bound derived
  from $V_H \rightarrow \ell_H \ell$ pushes the lower bound on $f$ by
  a few hundred \unit{GeV}, but not considerably. The minimal allowed
  fine tuning is  around \unit[0.5]{\%} for the \emph{Light $T^\pm$}
  scenario and reduces to  approximately \unit[0.25]{\%} for
  the \emph{Heavy $T^\pm$} scenario.   

\item For the \emph{Light $\ell_H$} scenario, the complementarity
  between LHC and EWPO results appears in the opposite direction as
  before: Due to the small value of $\kappa_\ell$, electroweak
  precision observables are slightly weaker than in the previous
  benchmark cases. However, at the same time the collider bounds
  improve significantly due to the very distinctive decay topology
  which produces sevaral hard leptons, see our discussion in the
  previous section. In this benchmark, the lower bound $f >
  \unit[1.7]{TeV}$ originates solely from the collider result and is
  independent of the details of the heavy top partner sector. It is
  only the region with large values of $f \gtrsim \unit[1.8]{TeV},
  \kappa \gtrsim 2.5$ where the EWPO bound may become more relevant
  --- depending on the chosen value of $R$. The minimal allowed fine
  tuning is around \unit[0.35]{\%} in the \emph{Heavy $T^\pm$} and
  \unit[0.4]{\%} in the \emph{Light $T^\pm$} scenario, respectively. 
\end{itemize}

All in all, we observe that without taking the LHC data into account, fine
tuning above \unit[1]{\%} would still be allowed in regions with light
$q_H$ and light $T^\pm$. These regions, however, are nowadays testable
at collider experiments and results from the first LCH run at
\unit[8]{TeV} already pushed the fine-tuning to the sub-percent
level. Using the updated results acquired during the $\sqrt{s} =
\unit[13]{TeV}$ period, limits derived from the Large Hadron Collider
become more and more severe. Though the precise position of the total
bound depends on the details of the heavy fermion sector, the heavy
top partner masses, and the presence or absence of
$T$-parity violation, we observe that due to their complementary
behavior regarding the EWPO bounds, values of $f$ below
\unit[1.3]{TeV} and fine-tuning above \unit[0.6]{\%} seems to be
excluded by now. Within our considered benchmark scenarios we
observe that \emph{Fermion Universality} is the most weakly constrained
model. However, the newest \unit[13]{TeV} results show a significant
improvement already when put in comparison with the earlier
\unit[8]{TeV} bounds. Furthermore, our approximate future sensitivity
study in Sec.\ref{sec:results14tev} gives us reason to expect an even
further improvement by LHC results in the near and far future, putting
the Littlest Higgs Model with $T$ parity more and more to the test.

\section{Summary}
\label{sec:summary}

In this study we reinterpreted null results from LHC searches for
physics beyond the Standard Model in the context of the Littest Higgs
Model with conserved and broken $T$-parity. This model is an elegant
implementation of global collective symmetry breaking combined with a
discrete symmetry to explain the natural lightness of the Higgs boson
as a (pseudo-)Nambu-Goldstone boson. Bounds on the symmetry-breaking
scale $f$ from data until 2013 were still as low as roughly 600 GeV.
This model predicts heavy partners for the Standard Model quarks
$q_H$, leptons $\ell_H$, gauge bosons $W_H, Z_H, A_H$ and special
partners for the top quark $T^\pm$. The mass hierarchies and the
presence of the discrete $T$-parity result in a model which shares
many phenomenlogical similarities with supersymmetric extensions of
the Standard Model, most importantly it features a stable,
invisible $A_H$ if $T$-parity is conserved, similar to the lightest
neutralino in supersymmetry with conserved $R$-parity. 

 Using the degrees of freedom for the full theory, we defined a set of
 benchmark scenarios which make different assumptions about the mass
 hierarchies in the heavy fermion sector, the masses of the heavy top
 partners and the possible presence of small $T$-parity violating
 operators.  By making use of the collider phenomenology tool
 \Checkmate{}, we systematically analyzed all relevant topologies at
 the LHC and derived bounds for all benchmark scenarios, excluding
 those regions which would have predicted a signal in any of the many
 considered search channels. We also give rough estimates for the
 bounds expected from a high luminosity LHC running with $\sqrt{s} =
 \unit[14]{TeV}$ and $\unit[3]{ab}^{-1}$ of integreated luminosity.  

Our results show that $q_H$ pair production, $V_H$ pair production and
$T^-$ pair production, respectively, produce strong bounds in the
model parameter space due to null results in searches dedicated for
squarks and electroweakinos in supersymmetry. Most importantly,
searches which require a large amount of hard jets and a significant
amount of missing transverse momentum produce the strongest results in
regions where $q_H$ and $T^-$ production is important whilst searches
for final states with multilepton and missing energy become more relevant
as soon as  heavy vector boson production is the dominant
channel. Color-neutral heavy leptons are mostly irrelevant for the
LHC, unless they are light enough to appear in decay topologies like
$V_H \rightarrow \ell_H \ell$ in which case they are again largely
constrained by searches for multileptons and missing energy. Allowing
for a  small amount of $T$-parity violation surprisingly only has a
minor impact on the result if compared to the case where $T$-parity is
exactly conserved. This can by explained by the fact that in the case
of $T$-parity violation via anomalous WZW-terms, $A_H$ decays
predominantly into the Standard Model gauge bosons whose leptonic decays
can produce the required missing energy plus additional hard particles
which improve the signal-to-background ratio.  

As the masses of the particles $q_H, \ell_H,$ $V_H$ and $T^\pm$ depend
differently on the Yukawa-like parameters $\kappa_q, \kappa_\ell$ and
$R$, precise LHC bounds depend on the particular values of these three
parameters. On the other hand, all particle masses grow linearly with the
symmetry breaking scale $f$ and we conclude that LHC results from the
$\sqrt{s} = \unit[13]{TeV}$ run  exclude any value of $f$ below
\unit[950]{GeV} at 95\% confidence level. The weakest bound appears in
a scenario where only the heavy gauge bosons are kinematically
accessible and all Yukawa parameters are such that the other particles
are decoupled from LHC observability. 

Our LHC bounds are complementary to those derived from electroweak
precision observables as the former constrain light particles with
small Yukawa couplings while the latter put limits on sizable
contributions from large Yukawa couplings. This complementarity
strongly removes the dependence of the bound on the details of the
heavy fermion sector as weaker limits from the LHC are compensated by
corresponding stronger bounds from EWPO and vice versa. All in all,
from our benchmark results we conclude that the symmetry
breaking scale $f$ must be larger than \unit[1.3]{TeV} and the fine
tuning cannot be better than \unit[0.4]{\%}.  Even stronger bounds
are possible if more details about the heavy fermion sector are known
and these limits can easily be derived from our exhaustive set of
results for the various benchmark scenarios. This constitutes an
improvement of more than 700 GeV compared to LHC run 1 data.

Though the Littlest Higgs model with $T$-parity has been constrained
much stronger by LHC run 2 data, it is still a rather natural solution
to the shortcomings of the electroweak and scalar sector, and we will
need full high-luminosity data from the LHC to decide whether
naturalness is actually an issue of the electroweak sector or not. A
qualitative improvement of all bounds on the model, particularly in
the Higgs sector and the heavy lepton sector, might need the running
of a high-energy lepton collider (or a hadron collider at much higher
energy).

%% file: sections/07_additionalplots.tex
\section{Supplementary Figures for the Collider Analysis}
\label{app:figures}

In this section we provide additional figures which are useful to
better understand and/or reproduce our results but which are not
necessarily needed for the discussion of the main text.

\begin{figure*}
\centering
\includegraphics[width=0.45\textwidth]
                {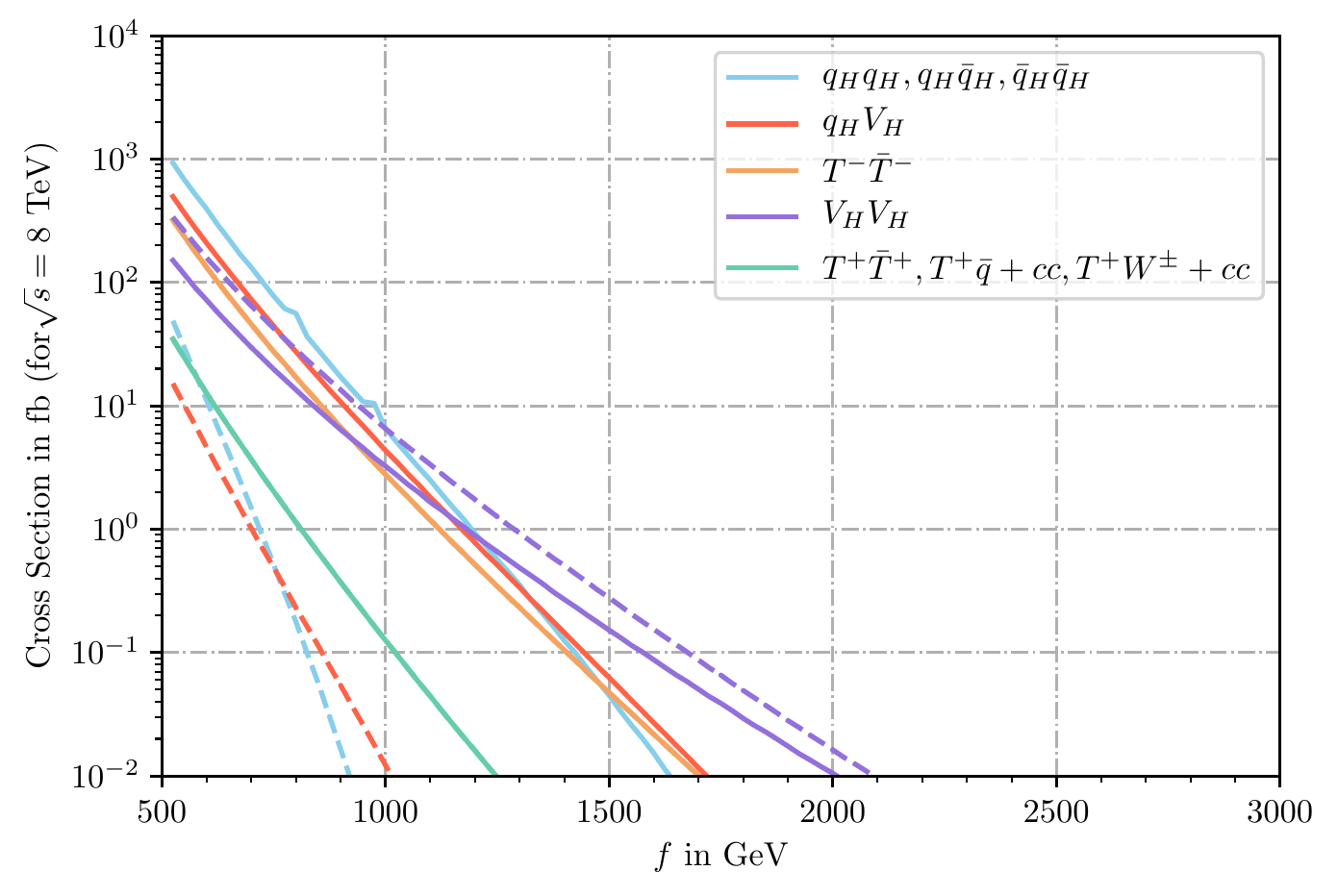}  \quad
\includegraphics[width=0.45\textwidth]
                {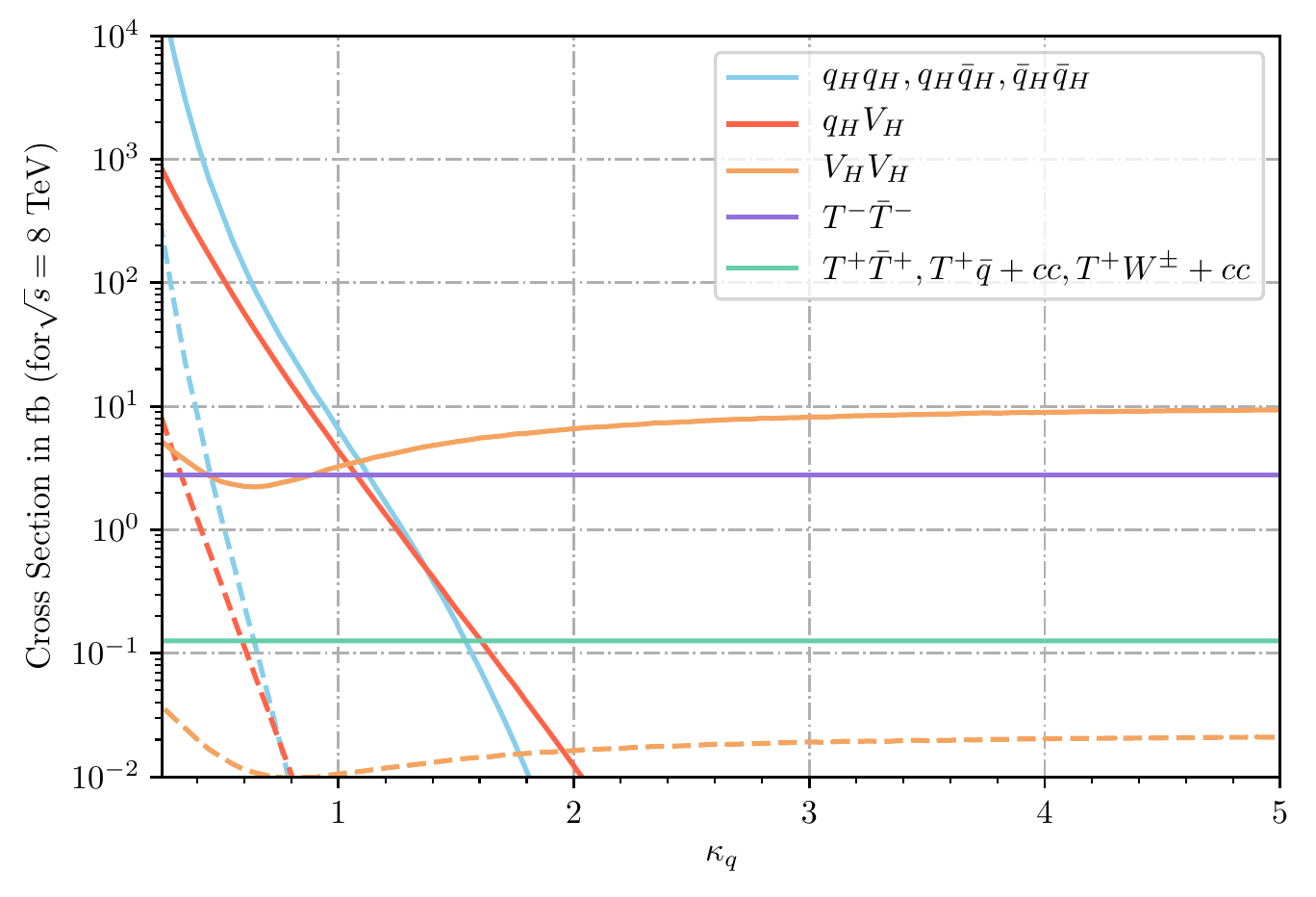}
\caption{Same as Fig.~\ref{fig:xs:univ} for $\sqrt{s} = \unit[8]{TeV}$.}
\label{fig:xs:univ8}
\includegraphics[width=0.45\textwidth]
                {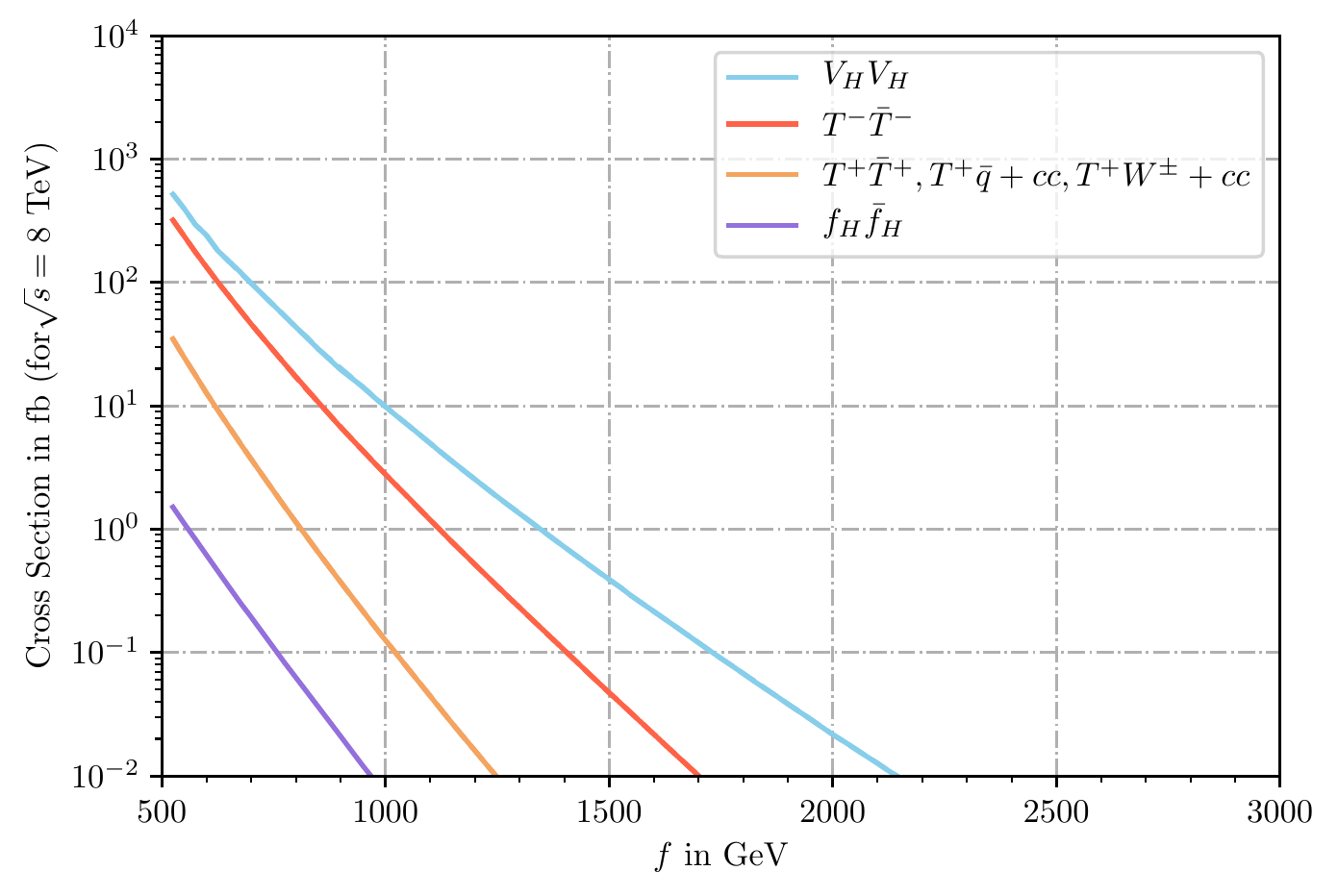} \quad
\includegraphics[width=0.45\textwidth]
                {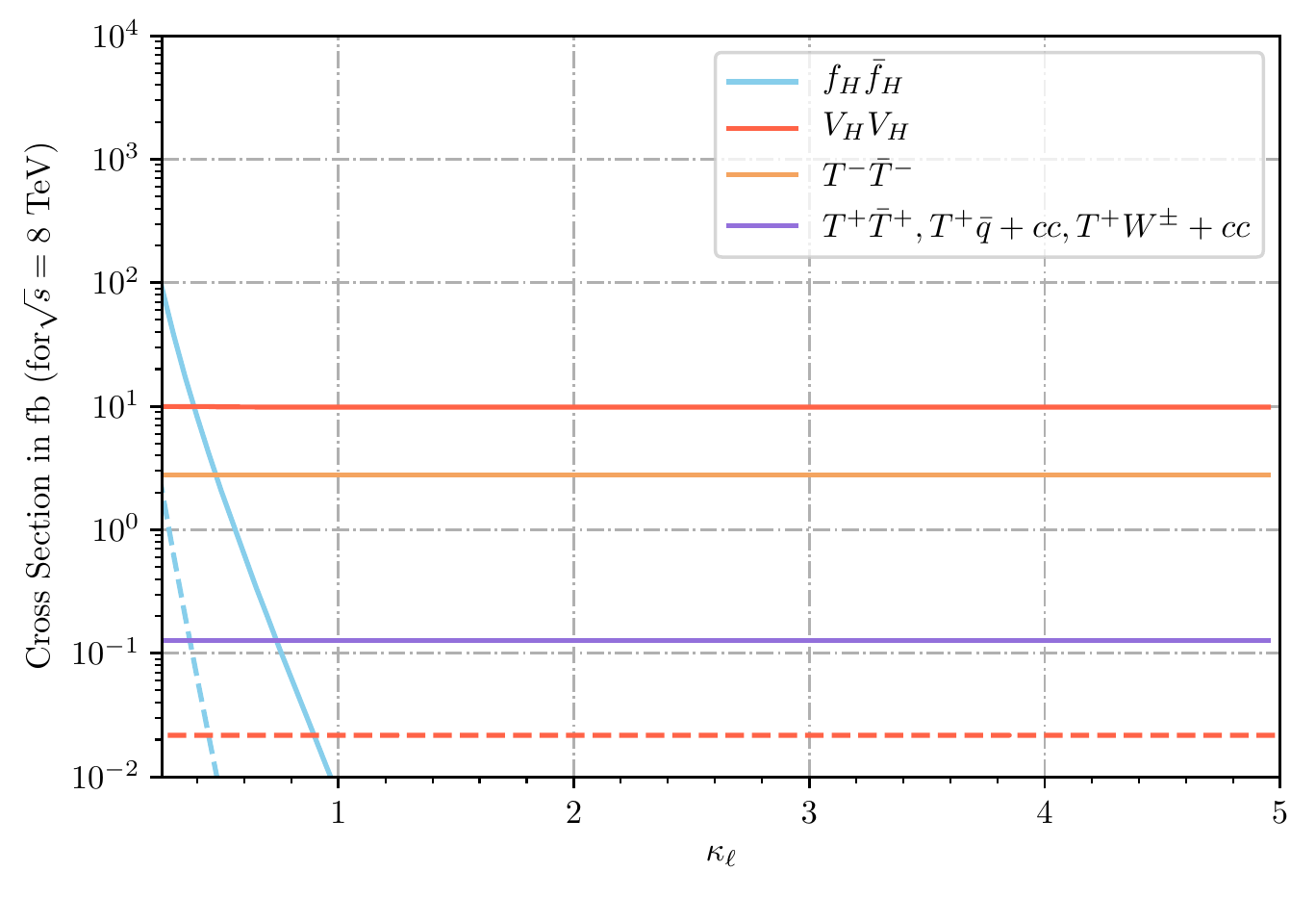}                
\caption{Same as Fig.~\ref{fig:xs:univ8} for benchmark model
  \emph{Heavy $q_H$}+\emph{Light $T^\pm$}.}\label{fig:cm:xs28} 
\centering
\includegraphics[width=0.45\textwidth]
                {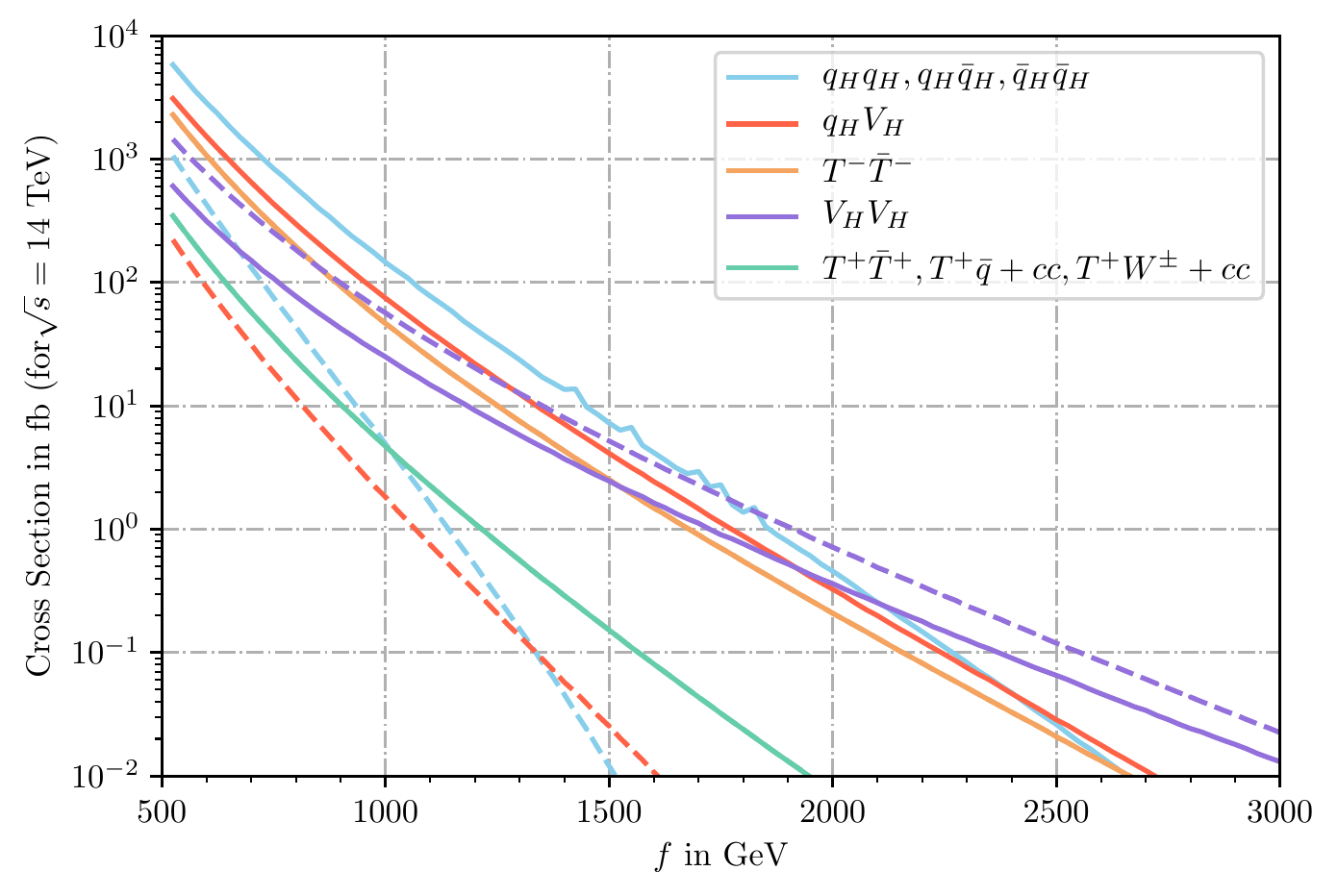}  \quad
\includegraphics[width=0.45\textwidth]
                {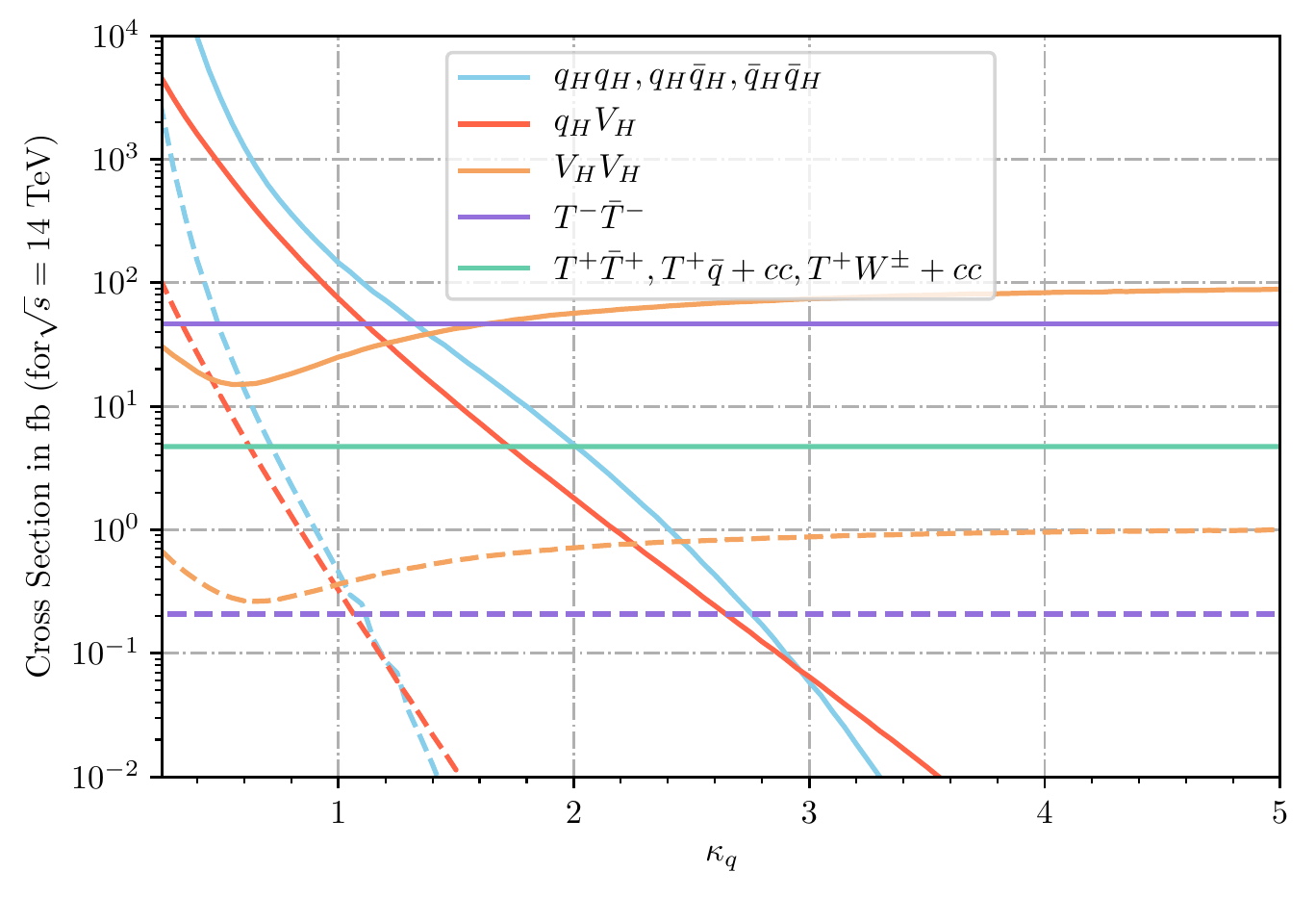}
\caption{Same as Fig.~\ref{fig:xs:univ} for $\sqrt{s} = \unit[14]{TeV}$.}
\label{fig:xs:univ14}
\includegraphics[width=0.45\textwidth]
                {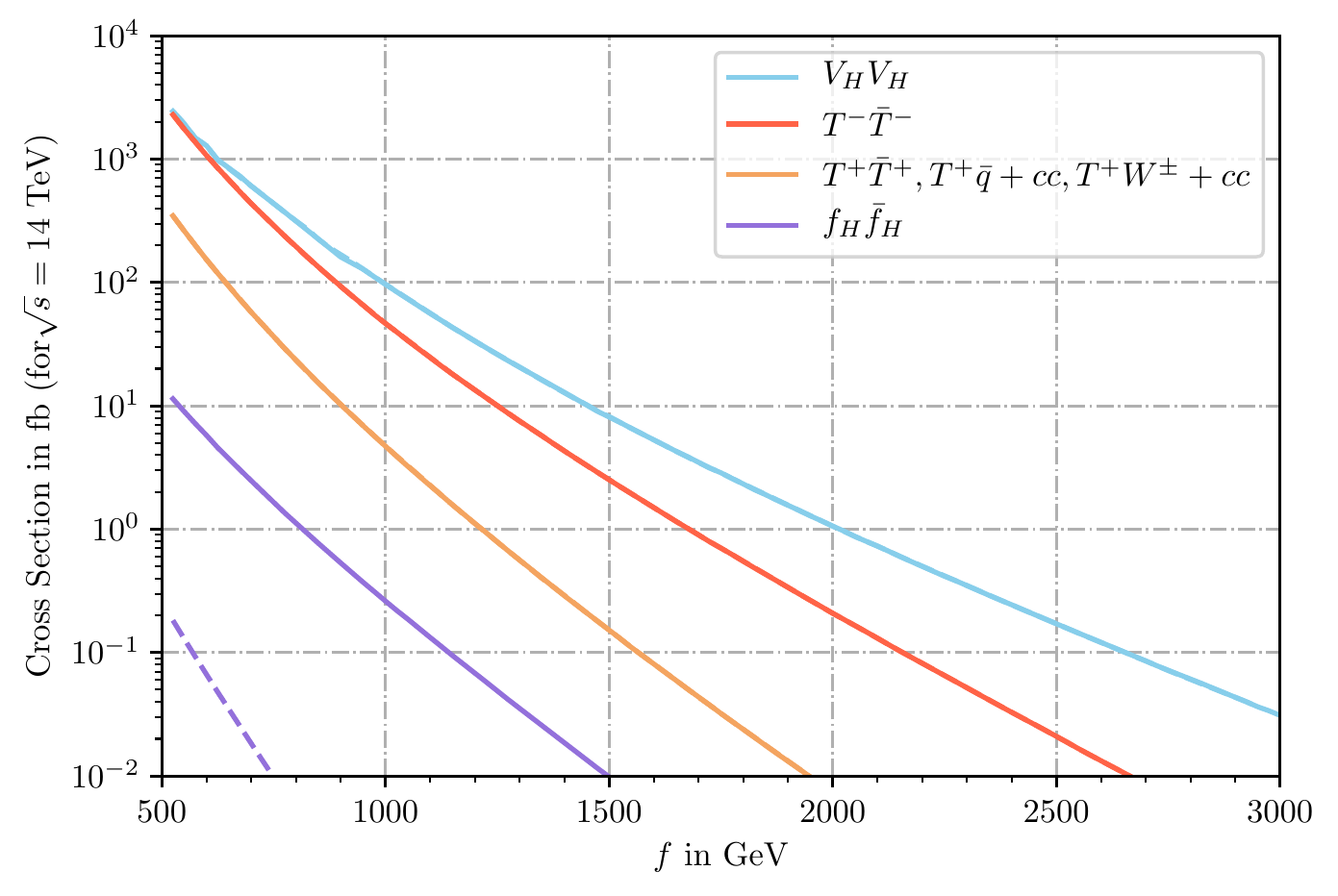} \quad 
\includegraphics[width=0.45\textwidth]
                {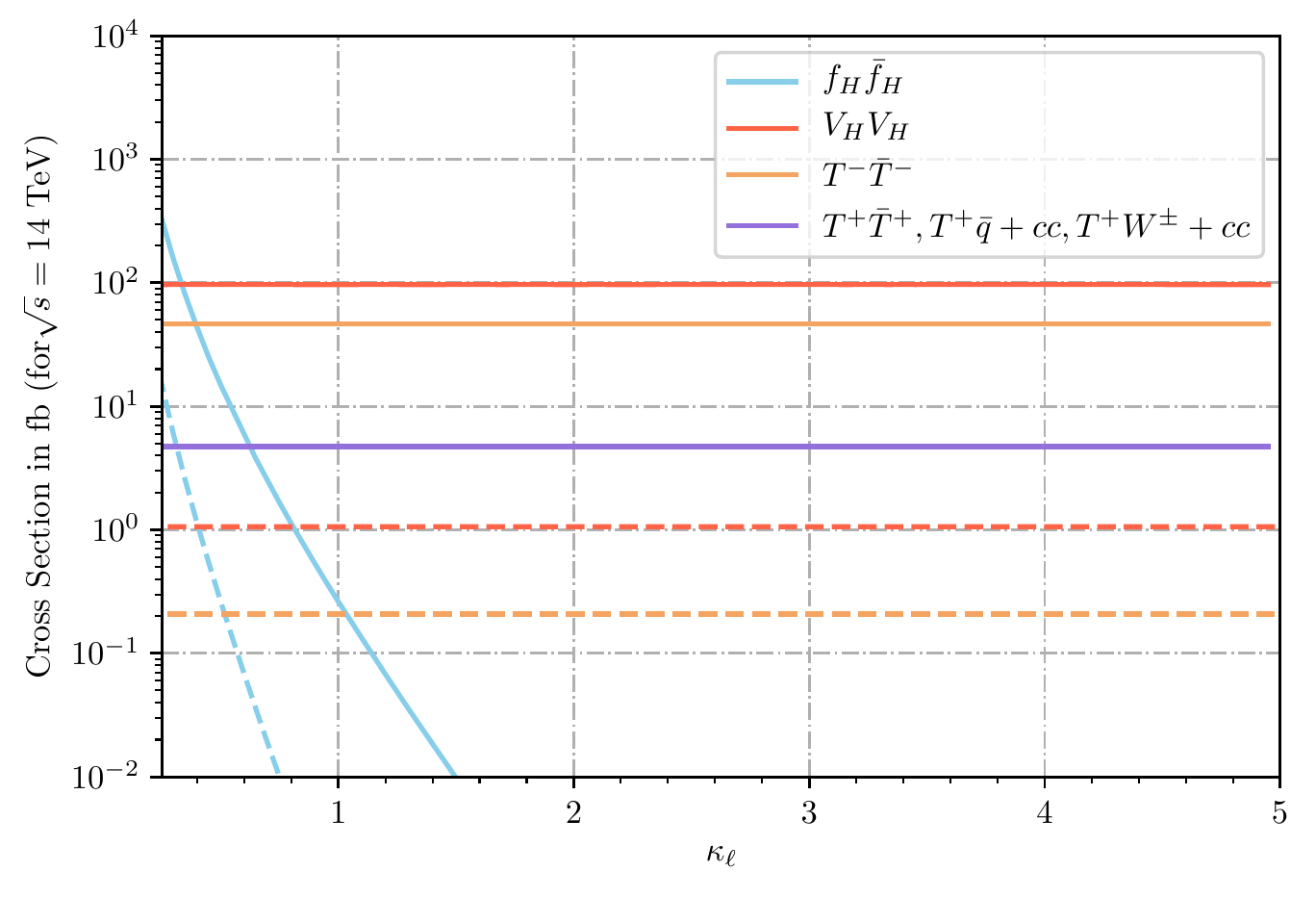}
\caption{Same as Fig.~\ref{fig:xs:univ14} for benchmark model
  \emph{Heavy $q_H$}+\emph{Light $T^\pm$}.}\label{fig:cm:xs214} 
\label{fig:xs:univ142}
\end{figure*}

This includes the cross sections for a center-of-mass energy of 8 TeV
in Fig.~\ref{fig:xs:univ8} for the {\em Fermion Universality/Light
$\ell_H$ + Light $T^\pm$} scenario as a function of the Little Higgs
scale $f$ and $\kappa_q$, respectively. In Fig.~\ref{fig:cm:xs28} we show
the 8 TeV cross sections for the {\em Heavy $q_H$ + Light $T^\pm$}
scenario. The lower figures, Fig.~\ref{fig:xs:univ14} and
Fig.~\ref{fig:xs:univ142}, show the cross sections for the same
scenarios, but now for 14 TeV full LHC center-of-mass energies.

%%%%%
%\newpage

\section{Full List of CheckMATE Analyses}
\label{app:checkmateanal}
\nopagebreak
\begin{table*}
\scriptsize
\setlength{\tabcolsep}{1.1pt}
\def\arraystretch{0.70}%  1 is the default, change whatever you need
\begin{tabularx}{\textwidth}{lXlll}
\toprule
\Checkmate{} identifier &  Search designed for&  \#SR & $L_{\text{int}}$ & Ref.\\
\midrule
\midrule
\multicolumn{5}{c}{$\sqrt{s} = \unit[8]{TeV}$} \\
\midrule
\texttt{atlas\_1308\_1841}           &  New phenomena in final states with large jet multiplicities and \etmiss{}                   &     13 &          20.3 & \cite{Aad:2013wta} \\
\texttt{atlas\_1308\_2631}           &  Direct $\tilde t/\tilde b$ pair production in final states with \etmiss{} and two $b$-jets                    &     6  &         20.1 & \cite{Aad:2013ija}\\
\texttt{atlas\_1402\_7029}           &  Direct production of $\tilde \chi^\pm/\tilde \chi^0$ in events with 3 $\ell$ and \etmiss{}          &     20 &        20.3  & \cite{Aad:2014nua}\\
\texttt{atlas\_1403\_4853}           &  Direct $\tilde t$ pair production in final states with 2$\ell$                   &     12 &       20.3  & \cite{Aad:2014qaa}\\
\texttt{atlas\_1403\_5222}           &  Direct $\tilde t$ pair production in events with a $Z$, $b$-jets and \etmiss{}                    &     5  &      20.3  & \cite{Aad:2014mha}\\
\texttt{atlas\_1404\_2500}           &  Supersymmetry in final states with jets and 2 SS $\ell$ or 3$\ell$                             &     5  &     20.3   & \cite{Aad:2014pda}\\
\texttt{atlas\_1405\_7875}           &  Search for $\tilde q$ and $\tilde g$ in final states with jets and \etmiss{}                       &     15 &    20.3   & \cite{Aad:2014wea}\\
\texttt{atlas\_1407\_0583}           &  $\tilde t$ pair production in final states with 1 isol. $\ell$, jets and \etmiss{}                    &     27 &   20.3   & \cite{Aad:2014kra} \\
\texttt{atlas\_1407\_0608}           &  Pair produced 3rd gen. squarks decaying via $c$ or compressed scenarios                             &     3  &  20.3   & \cite{Aad:2014nra}\\
\texttt{atlas\_1411\_1559}           &  New phenomena in events with a photon and \etmiss{}                          &     1  &              20.3  & \cite{Aad:2014tda} \\
\texttt{atlas\_1501\_07110}          &  Direct production of $\tilde \chi^\pm/\tilde \chi^0$ decaying into a Higgs boson   &     12 &             20.3   & \cite{Aad:2015jqa} \\
\texttt{atlas\_1502\_01518}          &  New phenena in final states with an energetic jet and large \etmiss{}                   &     9  &            20.3   & \cite{Aad:2015zva} \\   
\texttt{atlas\_1503\_03290}          &  Supersymmetry in events with an SFOS $\ell$ pair, jets and large \etmiss{}                   &     1  &           20.3   & \cite{Aad:2015wqa} \\   
\texttt{atlas\_1506\_08616}          &  Direct Pair production third generation squarks           &     12 &          20.0  &  \cite{Aad:2015pfx} \\   
\texttt{atlas\_conf\_2012\_104}       &  Supersymmetry in final states with jets, 1 isolated lepton and \etmiss{}                   &     2  &          5.8    & \cite{ATLAS:2012tna} \\  
\texttt{atlas\_conf\_2013\_024}       &  Directo $\tilde t$ pair production in the all-hadronic $t \bar t$ + \etmiss{} final state      &     3  &         20.5 & \cite{ATLAS:2013cma}    \\  
\texttt{atlas\_conf\_2013\_049}      &  Direct $\tilde \ell/\tilde \chi^\pm$ production in final states with 2 OS $\ell$, no jets and \etmiss{}                                &      9  &        20.3 &   \cite{TheATLAScollaboration:2013hha}  \\ 
\texttt{atlas\_conf\_2013\_061}      &  Strongly produced Supersymmetric particles with $\geq 3$ $b$-jets and \etmiss{}                 &      9  &       20.1     & \cite{TheATLAScollaboration:2013tha} \\  
\texttt{atlas\_conf\_2013\_089}      &  Strongly produced Supersymmetric particles decaying into 2 leptons                                  &      12 &      20.3   & \cite{TheATLAScollaboration:2013via} \\  
\texttt{atlas\_conf\_2015\_004}      &  Invisibly decaying Higgs bosons produced in vector boson fusion                         &      1  &     20.3   & \cite{ATLAS:2015yda} \\ 
\texttt{atlas\_conf\_2012\_147}      &  New phenomena in monojets plus \etmiss{}        &     4  &                     10.0   & \cite{ATLAS:2012zim} \\  
\texttt{atlas\_conf\_2013\_035}      &  Direct production of $\tilde \chi^\pm / \tilde \chi^0$ in events with 3 leptons and \etmiss{}         &     6  &                    20.7 & \cite{ATLAS:2013rla}   \\   
\texttt{atlas\_conf\_2013\_037}      &  Direct $\tilde t$ pair production in final states with 1 isolated $\ell$, jets and \etmiss{}          &     6  &                   20.7 & \cite{ATLAS:2013pla}    \\   
\texttt{atlas\_conf\_2013\_047}      &  $\tilde q$ and $\tilde g$ in final states with jets and \etmiss{}           &     10 &                  20.3 & \cite{TheATLAScollaboration:2013fha}    \\  
\texttt{cms\_1303\_2985}             &  Supersymmetry in hadronic final states with $b$-jets and \etmiss{} using $\alpha_T$                            &     59 &                11.7   & \cite{Chatrchyan:2013mys} \\  
\texttt{cms\_1408\_3583}            &  Dark Matter, Extra Dimensions and Unparticles in monojet events                                 &      7  &              19.7   & \cite{Khachatryan:2014rra}  \\   
\texttt{cms\_1502\_06031}            &  New Physics in events with 2$\ell$, jets and \etmiss{}         &     6  &             19.4   & \cite{Khachatryan:2015lwa} \\   
\texttt{cms\_1504\_03198}             &  Dark Matter produced in association with $t \bar{t}$ in final states with 1$\ell$        &    1  &            19.7 & \cite{Khachatryan:2015nua}   \\   
\texttt{cms\_sus\_13\_016}             &   Supersymmetry in events with 2 OS $\ell$, many jets, $b$-jets and large  \etmiss{}                                &    1  &            19.5 & \cite{CMS:2013ija}    \\ 
\texttt{cms\_exo\_14\_014}              &  Heavy Majorana neutrinos in events with SS dileptons and jets              &   16 &           19.7 & \cite{Khachatryan:2016olu}    \\
\midrule
\multicolumn{4}{c}{$\sqrt{s} = \unit[13]{TeV}$} \\
\midrule
\texttt{atlas\_1602\_09058} & Supersymmetry in final states with jets and two SS leptons or 3 leptons & 4 & 3.2 & \cite{Aad:2016tuk}\\   
\texttt{atlas\_1604\_01306} & New phenomena in events with a photon and \etmiss{} & 1 & 3.2 & \cite{Aaboud:2016uro}  \\   
\texttt{atlas\_1604\_07773} & New phenomena in final states with an energetic jet and large \etmiss{} & 13 & 3.2 & \cite{Aaboud:2016tnv}  \\   
\texttt{atlas\_1605\_03814} & $\tilde q$ and $\tilde g$ in final states with jets and \etmiss{} & 7 & 3.2  & \cite{Aaboud:2016zdn} \\   
\texttt{atlas\_1605\_04285} & Gluinos in events with an isolated lepton, jets and \etmiss{} & 7 & 3.3  & \cite{Aad:2016qqk} \\   
\texttt{atlas\_1605\_09318} & Pair production of $\tilde g$ decaying via $\tilde t$ or $\tilde b$ in events with $b$-jets and \etmiss{} & 8 & 3.3 & \cite{Aad:2016eki}   \\   
\texttt{atlas\_1606\_03903} & $\tilde t$ in final states with one isolated lepton, jets and \etmiss{} & 3 & 3.2  & \cite{Aaboud:2016lwz} \\   
\texttt{atlas\_1609\_01599} & Measurement of $ttV$ cross sections in multilepton final states & 9 & 3.2  & \cite{Aaboud:2016xve} \\   
\texttt{atlas\_conf\_2015\_082} & Supersymmety in events with leptonically decaying $Z$, jets and \etmiss{} & 1 & 3.2  & \cite{TheATLAScollaboration:2015nxu} \\   
\texttt{atlas\_conf\_2016\_013} & Vector-like $t$ pairs or 4 $t$ in final states with leptons and jets & 10 & 3.2 & \cite{TheATLAScollaboration:2016gxs}   \\   
\texttt{atlas\_conf\_2016\_050} & $\tilde t$ in final states with one isolated lepton, jets and \etmiss{} & 5 & 13.3 & \cite{ATLAS:2016ljb} \\   
\texttt{atlas\_conf\_2016\_054} & $\tilde q$, $\tilde g$ in events with an isolated lepton, jets and \etmiss{} & 10 & 14.8 & \cite{ATLAS:2016lsr} \\   
\texttt{atlas\_conf\_2016\_076} & Direct $\tilde t$ pair production and DM production in final states with 2$\ell$ & 6 & 13.3 & \cite{ATLAS:2016xcm}  \\   
\texttt{atlas\_conf\_2016\_078} & Further searches for $\tilde q$ and $\tilde g$ in final states with jets and \etmiss{} & 13 & 13.3 & \cite{ATLAS:2016kts}  \\   
\texttt{atlas\_conf\_2016\_096} & Supersymmetry in events with 2$\ell$ or $3\ell$ and \etmiss{} & 8 & 13.3 & \cite{ATLAS:2016uwq} \\   
\texttt{atlas\_conf\_2017\_022} & $\tilde q$, $\tilde g$ in final states with jets and \etmiss{} & 24 & 36.1 & \cite{ATLAS:2017cjl}  \\   
\texttt{atlas\_conf\_2017\_039} & Electroweakino production in final states with 2 or 3 leptons & 37 & 36.1 & \cite{ATLAS:2017uun}  \\   
\texttt{atlas\_conf\_2017\_040} & Dark Matter or invisibly decaying $h$, produced in associated with a $Z$ & 2 & 36.1  & \cite{ATLAS:2017zkz}\\   
\texttt{cms\_pas\_sus\_15\_011} & New physics in final states with an OSSF lepton pair, jets and \etmiss{} & 47 & 2.2 &  \cite{CMS:2015bsf}    \\   
\midrule
\multicolumn{4}{c}{$\sqrt{s} = \unit[14]{TeV}$} \\
\midrule
\texttt{atlas\_phys\_pub\_2013\_011} &  Search for Supersymmetry at the high luminosity LHC ($\tilde t$ sector) & 4 & 3000  & \cite{ATL-PHYS-PUB-2013-011}\\   
\texttt{atlas\_2014\_010\_hl\_3l} & Search for Supersymmetry at the high luminosity LHC  ($\tilde \chi^\pm/\tilde \chi^0$ sector) & 1 & 3000 & \cite{ATL-PHYS-PUB-2014-010}  \\   
\texttt{atlas\_phys\_2014\_010\_sq\_hl} & Search for Supersymmetry at the high luminosity LHC ($\tilde q/\tilde g$ sector) & 10 & 3000 & \cite{ATL-PHYS-PUB-2014-010} \\   
\texttt{dilepton\_hl*} & Custom Search for $\tilde \ell/\tilde \chi^\pm$ in final states with 2 leptons and \etmiss{} & 9 & 3000  & \cite{dileptonhl} \\   
\texttt{atlas\_14tev\_monojet*} & Custom Search for DM in final states with an energetic jet and \etmiss{} & 5 & 3000 & \cite{Kim:2015hda}   \\   
\bottomrule
\end{tabularx}
\caption{Full list of all \Checkmate{} analyses used for this
  study. The column labelled \#SR yields the number of signal
  regions. Entries for the integrated luminosities $L_{\text{int}}$
  are given in fb$^{-1}$.} 
\label{tab:app:analyses}
\end{table*}

Table~\ref{tab:app:analyses} gives the full list of used \Checkmate{}
analyses. The first column shows the \Checkmate{} idenitifer, the
second the purpose for which the analysis was designed for. The last
three columns show the number of signal regions in the corresponding
analysis (marked \#SR), the integrated luminosity for that analysis
and the reference to the publication or conference notes from the
experimental collaborations. More details on the respective analyses
and corresponding validation material can be found on
\url{http://checkmate.hepforge.org}. High luminosity analyses marked
with * do not correspond to official experimental studies but have
been implemented by the \Checkmate{} collaboration. More information
can be found in the respective references.